\documentclass[a4paper,11pt]{article}
\pdfoutput=1 

\usepackage{jheppub} 

\usepackage{mathtools}

\usepackage[T1]{fontenc} 
\usepackage{amsmath}
\usepackage{tensor}
\usepackage{import}
\usepackage{braket}
\usepackage{slashed}
\usepackage[force]{feynmp-auto}
\allowdisplaybreaks

\newcommand{\eq}[1]{eq.~\eqref{eq:#1}}
\newcommand{\eqs}[2]{eqs.~\eqref{eq:#1} and \eqref{eq:#2}}
\renewcommand{\sec}[1]{sec.~\ref{sec:#1}}
\newcommand{\secs}[2]{secs.~\ref{sec:#1} and \ref{sec:#2}}
\newcommand{\fig}[1]{fig.~\ref{fig:#1}}
\newcommand{\figs}[2]{figs.~\ref{fig:#1} and \ref{fig:#2}}
\newcommand{\app}[1]{app.~\ref{app:#1}}

\newcommand{\tr}{\text{tr}}
\newcommand{\fsl}[1]{\slashed{#1}}
\def\df{{\rm d}}
\def\img{{\rm i}}
\def\nn{\nonumber}

\def\al{\alpha}
\def\bt{\beta}
\def\ga{\gamma}
\def\Ga{\Gamma}

\def\eps{\epsilon}
\def\si{\sigma}

\newcommand{\thirteengrid}{
\fmfstraight
\fmfleft{a1,a2,a3,a4,a5,a6,a7,a8,a9,a10,a11,a12,a13}
\fmfright{m1,m2,m3,m4,m5,m6,m7,m8,m9,m10,m11,m12,m13}
\fmf{phantom}{a1,b1,c1,d1,e1,f1,g1,h1,i1,j1,k1,l1,m1}
\fmf{phantom}{a2,b2,c2,d2,e2,f2,g2,h2,i2,j2,k2,l2,m2}
\fmf{phantom}{a3,b3,c3,d3,e3,f3,g3,h3,i3,j3,k3,l3,m3}
\fmf{phantom}{a4,b4,c4,d4,e4,f4,g4,h4,i4,j4,k4,l4,m4}
\fmf{phantom}{a5,b5,c5,d5,e5,f5,g5,h5,i5,j5,k5,l5,m5}
\fmf{phantom}{a6,b6,c6,d6,e6,f6,g6,h6,i6,j6,k6,l6,m6}
\fmf{phantom}{a7,b7,c7,d7,e7,f7,g7,h7,i7,j7,k7,l7,m7}
\fmf{phantom}{a8,b8,c8,d8,e8,f8,g8,h8,i8,j8,k8,l8,m8}
\fmf{phantom}{a9,b9,c9,d9,e9,f9,g9,h9,i9,j9,k9,l9,m9}
\fmf{phantom}{a10,b10,c10,d10,e10,f10,g10,h10,i10,j10,k10,l10,m10}
\fmf{phantom}{a11,b11,c11,d11,e11,f11,g11,h11,i11,j11,k11,l11,m11}
\fmf{phantom}{a12,b12,c12,d12,e12,f12,g12,h12,i12,j12,k12,l12,m12}
\fmf{phantom}{a13,b13,c13,d13,e13,f13,g13,h13,i13,j13,k13,l13,m13}
}

\usepackage{tikz}
\usetikzlibrary{calc}
\newcommand\softstaple[1][1]
{
 \begin{tikzpicture}[scale=#1]
  \coordinate (dx) at (0.15, 0);
  \coordinate (dy) at (0   , 0.1);
  \coordinate (bT) at (0.  , 0.05);
  \draw [line width=0.15mm] (0,0) -- ($(dx) + (dy)$) -- ($2*(dy)$) -- ($2*(dy)+(bT)$) -- ($(dx) + (dy) + (bT)$) -- ($(bT)$) -- (0,0);
  \draw [line width=0.12mm] (0,0) -- (0, -0.0055);
 \end{tikzpicture}
}

\title{Towards Double Parton Distributions from First Principles using Large Momentum Effective Theory}

\author[a,b]{Max Jaarsma,}
\author[a,b,c]{Rudi Rahn,}
\author[a,b]{Wouter J.~Waalewijn}

\affiliation[a]{Nikhef, Theory Group,
	Science Park 105, 1098 XG, Amsterdam, The Netherlands}
\affiliation[b]{Institute for Theoretical Physics Amsterdam and Delta Institute for Theoretical Physics, University of Amsterdam, Science Park 904, 1098 XH Amsterdam, The Netherlands}
\affiliation[c]{Department of Physics and Astronomy, University of Manchester, Manchester, M13 9PL,United
Kingdom}

\emailAdd{m.jaarsma@uva.nl}
\emailAdd{rudi.rahn@manchester.ac.uk}
\emailAdd{w.j.waalewijn@uva.nl}

\abstract{In double parton scattering (DPS), two partonic collisions take place between one pair of colliding hadrons. The effect of DPS can be significant for precision measurements due to the additional radiation from secondary partonic collisions, and especially for specific processes such as same-sign $WW$ production. Its effect is usually included through Monte Carlo parton showers. In a factorization approach to DPS, the initial state is described by double parton distributions (DPDs). These are currently poorly constrained by experiment, but provide a view on interesting correlations between partons in the hadron. 
Here we show that the Large Momentum Effective Theory approach can be applied to DPDs.
Specifically, we present a general matching relation between DPDs and lattice-calculable quasi-DPDs for general flavor, spin and color structures. We furthermore calculate the one-loop matching coefficients for the quark-quark DPDs, verifying that the infrared logarithms and divergences cancel in the matching. While we restrict to the flavor-non-singlet case, we do take color and spin correlations into account.
Interestingly, quasi-DPDs combines nontrivial features from both the collinear and transverse momentum dependent quasi-parton distribution functions.
This represents a first step in extending the quasi-PDF approach to DPDs, opening up a new way to constrain these distributions using lattice QCD.}

\begin{document} 
\maketitle
\flushbottom

\section{Introduction}
\label{sec:intro}
Experimental studies at hadron colliders largely focus on hard scattering processes, in which heavy particles (such as Higgs bosons or top quarks) or jets with large transverse momenta are produced. In the theoretical description of these processes, one usually considers a single partonic scattering between a pair of colliding hadrons. Monte Carlo parton showers account for the underlying event through multiple parton interactions~\cite{Sjostrand:1987su,Butterworth:1996zw,Sjostrand:2004ef,Bellm:2019icn} to describe the data. These additional partonic collisions are much less energetic, but the resulting radiation may e.g.~affect jet measurements such as the jet mass~\cite{Dasgupta:2012hg,Stewart:2014nna}. 

In a factorization approach to double parton scattering (DPS), the initial state is described by double parton distributions (DPDs). DPS was already considered in the early days of the parton model~\cite{Landshoff:1978fq,Goebel:1979mi,Takagi:1979wn,Mekhfi:1985dv,Paver:1982yp}, and since then there has been substantial progress in formalizing the theoretical framework~\cite{Paver:1982yp,Blok:2010ge,Diehl:2011tt,Diehl:2011yj,Ryskin:2011kk,Manohar:2012jr,Blok:2013bpa,Diehl:2015bca,Diehl:2017kgu,Buffing:2017mqm}, see ref.~\cite{Bartalini:2018qje} for a comprehensive review. DPDs describe the possibility of extracting two partons out of a hadron, in direct analogy to the parton distribution functions (PDFs) that describe the extraction of a single parton. DPDs depend on the transverse separation of the partons, as well as their flavor, spin and color states, opening up the exciting possibility of studying correlations between the partons in the hadron~\cite{Mekhfi:1985dv,Diehl:2011tt,Diehl:2011yj,Manohar:2012jr}.

While there is clear experimental evidence for double parton scattering, the result of such measurements is often expressed in terms of a single number: the effective cross section $\sigma_{\rm eff}$. This assumes that both scatterings are independent of each other, as summarized in the ``pocket formula'' for the DPS cross section: $\sigma_{\rm DPS} = \sigma_1 \sigma_2/(S \sigma_{\rm eff})$, where $\sigma_{1,2}$ are the cross sections of the individual partonic scatterings and $S$ is a symmetry factor~\cite{Paver:1982yp}. Though there are many measurements of $\sigma_{\rm eff}$ from different processes~\cite{AxialFieldSpectrometer:1986dfj,CDF:1993sbj,CDF:1997lmq,D0:2014owy,LHCb:2012aiv,ATLAS:2013aph,CMS:2013huw,D0:2014vql,LHCb:2015wvu,ATLAS:2016rnd,LHCb:2016wuo,CMS:2017han,CMS:2021wfx,CMS:2021lxi,CMS:2022pio}, the field has not progressed to the point that an extraction of DPDs is within reach. In the meantime, sum rules~\cite{Gaunt:2009re,Diehl:2020xyg} and positivity bounds have been investigated~\cite{Diehl:2013mla,Diehl:2021wvd}, DPDs have been studied using various models of the proton~\cite{Chang:2012nw,Rinaldi:2013vpa,Broniowski:2013xba,Rinaldi:2014ddl,Rinaldi:2015cya,Kasemets:2016nio,Broniowski:2016trx}, and moments of the DPDs have been extracted from lattice data~\cite{Bali:2018nde,Courtoy:2020tkd,Bali:2020mij} (similar to the lattice extraction of moments of parton distributions).

In recent years, a new method of obtaining PDFs form lattice QCD has been proposed~\cite{Ji:2013dva,Ji:2014gla,Ma:2014jla,Radyushkin:2017cyf,Orginos:2017kos,Ji:2020ect,Zhang:2018diq,Izubuchi:2018srq,Xiong:2013bka},  which makes use of Large Momentum Effective Theory (LaMET) and has been referred to the quasi-PDF approach (alternatively, there is the pseudo-PDF approach based on short-distance factorization). In this method one defines an analogue of the PDF in which the fields are now space-like separated, known as the quasi-PDF, which corresponds to the PDF under an infinite Lorentz boost. The quasi-PDF is defined such that it can be calculated using lattice methods. It agrees with the PDF in the infrared (nonperturbative) limit, and the difference in the ultraviolet limit can be encoded by a perturbative matching correction, in principle providing access to the entire momentum fraction dependence of the PDF. More recently, this method has been extended to the case of transverse momentum dependent parton distributions (TMDs)~\cite{Ji:2018hvs,Ebert:2018gzl,Ebert:2019okf,Ji:2019sxk,Ji:2019ewn,Ebert:2022fmh,Schindler:2022eva,Vladimirov:2020ofp,delRio:2023pse,LPC:2022zci}. This extension was rather non-trivial because the definition of the physical TMDs, the ones that enter the factorization formulae, contains a soft function. The soft function involves two opposite light-like directions, which presents a difficulty to implement on a Euclidean lattice. Fortunately, it has been shown that the soft function can be split up into a rapidity independent part and a part that only involves the Collins-Soper kernel and that each individual part can be calculated on the lattice ~\cite{Ji:2019sxk,LatticeParton:2020uhz,Shanahan:2020zxr,Ebert:2018gzl,Ebert:2020no,Schlemmer:2021aij}.

In this paper, we extend the quasi-PDF approach to the case of DPDs. We define lattice-calculable quasi-DPDs and construct a matching formula that relates them to their physical counterparts. The matching relation we present is general: it holds for DPDs of all flavor combinations, spin structures and color structures. We further present a one-loop calculation of the matching kernels that are relevant to the quark flavor non-singlet case, and we include results for spin and color correlations. This calculation verifies, at least to one-loop order, that the quasi-DPDs and the physical DPDs share the same infrared behaviour, which is a necessary condition for the matching to apply. With this matching relation, DPD and the nonperturbative correlations of partons in a hadron they encode, can be accessed through lattice QCD.

The outline of this paper is as follows: In \sec{lamet} we provide a brief introduction to LaMET and its application to quasi-PDFs and TMDs. Similarly, in \sec{dpds} we provide a brief introduction on DPDs, including their field-theoretic definition. (These sections can be skipped by those familiar with these topics.) The matching between quasi-DPDs and physical DPDs is discussed in \sec{quasidpd}, and an explicit one-loop calculation for the quark flavor non-singlet case is carried out in \sec{oneloop}, with expressions for individual diagrams relegated to \app{diagrams}. We conclude in \sec{conclusions}. Notation regarding plus distributions are summarized in \app{conventions}.

\section{Quasi-PDFs, TMDs, and LaMET}
\label{sec:lamet}
After establishing our notation and conventions in \sec{convention}, we start in \sec{light-cone} with a recap of the field-theoretic definition of parton distribution functions (PDFs) and transverse-momentum-dependent parton distribution functions (TMDs). We review the current approach to extract PDFs (\sec{Quasi-PDFs}) and TMDs (\sec{Quasi-TMDs}) from lattice calculations using Large Momentum Effective Theory (LaMET). 

We refer the reader to the original literature on
LaMET~\cite{Ji:2013dva,Ji:2014gla,Ji:2020ect}, as well as a paper~\cite{Ebert:2022fmh} proving the possibility of
determining TMDs using lattice calculations, on which much of our understanding is based. We generally follow the notation established in ref.~\cite{Ebert:2022fmh}.

%------------------------------------------------------------------------------------------------------------------
\subsection{Notation and conventions}
\label{sec:convention}
%------------------------------------------------------------------------------------------------------------------

In defining parton distribution functions, it is useful to work in lightcone coordinates. We denote the components of a vector in lightcone coordinates by $(v^+,v^-,\mathbf{v}_\perp)$, where
%%%
\begin{align}
    v^\pm = \frac{v^0 \pm v^z}{\sqrt{2}}\,,
    \quad
    \mathbf{v}_\perp = (v^x, v^y)
\,.\end{align}
%%%
The dot product takes the following form
%%%
\begin{align}
    v\cdot w
    &=
    v^+ w^- + v^- w^+ - \mathbf{v}_\perp \cdot \mathbf{w}_\perp \,,
\end{align}
%%%
and the factors of $\sqrt{2}$ ensure that the Jacobian of the coordinate transformation is unity, $\df^4 x = \df x^+ \df x^- \df^2 \mathbf{x}_\perp$. Corresponding to a transverse vector $\mathbf{b}_\perp$, we write 
%%%
\begin{align}
b_\perp = (0,0,\mathbf{b}_\perp)
\,,\end{align}
%%%
and we will use $\mathbf{b}_\perp$ and $b_\perp$ interchangeably when there is no potential for confusion.
We will use $n_a = (1^+,0^-,\mathbf{0}_\perp)$, $n_b = (0^+,1^-,\mathbf{0}_\perp)$ to denote the light-like basis vectors.

The definition of parton distributions involve Wilson lines, which are path ordered exponentials of gauge field operators. For a general path $\gamma$, the Wilson line is defined as
%%%
\begin{align}\label{eq:W_def}
    W[\gamma]=P\exp\biggl(\img g\int_\gamma\df x^\mu \mathcal{A}_\mu^a(x) t^a\biggr)\,.
\end{align}
%%%
In general the $t^a$ depends on the $SU(3)$ representation of the partons, but in this work we restrict ourselves to quarks and therefore only need the fundamental representation. Hence there is no need for a label to indicate the representation.

In the TMDs we will encounter Wilson lines that follow a staple-shaped path, for which we introduce the following notation 
%%%
\begin{align}\label{eq:W_staple_def}
    W_\sqsupset(b,\eta v,\delta)
    &=
    W\biggl[
    b\leftarrow b+\eta v-\frac{\delta}{2}\leftarrow\eta v+\frac{\delta}{2}\leftarrow0
    \biggr]\,.
\end{align}
%%%
This describes a Wilson line consisting of straight-line segments going from the origin to $b$ along a staple-shaped path, where the sides of the staple have length $\eta$ and lie in the direction of $v$. Note that our convention for the direction of the arrows is opposite that of \cite{Ebert:2022fmh}, i.e.~our ``$\leftarrow$" corresponds to their ``$\rightarrow$". The argument $\delta$ concerns the shape of the transverse segment of the Wilson line and is chosen such that the transverse segment is perpendicular to the longitudinal pieces (to avoid angle-dependence in the renormalization of the soft function). Besides the staple-shaped Wilson line, the definition of TMDs also involves a Wilson loop that is obtained by gluing together two staple-shaped Wilson lines at their end-points, which we denote by 
%%%
\begin{align}\label{eq:S_softstaple_def}
    S_{\softstaple}(b,\eta v,\bar\eta\bar v)
    &=
    \tr\bigl[W[b\leftarrow b+\bar \eta \bar v
    \leftarrow \bar \eta \bar v\leftarrow0\leftarrow \eta v\leftarrow b+ \eta v
    \leftarrow b]\bigr]\,.
\end{align}
%%%

%------------------------------------------------------------------------------------------------------------------
\subsection{Lightcone PDFs and TMDs}
\label{sec:light-cone}
%------------------------------------------------------------------------------------------------------------------

The (bare) parton distribution functions are defined as hadronic matrix elements of fields separated along the lightcone. For quarks, 
%%%
\begin{align}\label{eq:collpdf}
    f_q(x,\eps)
    &=
    \int\frac{\df b^-}{4\pi} \,e^{-\img x P^+ b^-}
    \bra{P} \bar{\psi}(b^-)\gamma^+ W[b^-\leftarrow0]\psi(0) \ket{P} \,,
\end{align}
%%%
where $(0^+,b^-,\mathbf{0}_\perp)$ is abbreviated to $b^-$ and $\eps$ regulates the UV divergences. Here, $W[b^-\leftarrow0]$ is a straight Wilson line from $0$ to $b^-$ that ensures the gauge invariance of the parton distribution, and is defined in \eq{W_def}. The finite length Wilson line in eq.~\eqref{eq:collpdf} can be regarded as the
remnant of two Wilson lines extending to infinity with opposite orientation, 
%%%
\begin{align} \label{eq:Wsplit}
W[b^-\leftarrow0] = W[b^-\leftarrow - \infty n_b]\, W[-\infty n_b \leftarrow0]
\,,\end{align}
%%%
describing the remaining color-charged objects in the process and accounting for
the interactions of the extracted parton with their color-potential. The bare lightcone PDFs have ultraviolet divergences and require renormalization, leading to a dependence on a renormalization scale $\mu$ in the renormalized PDFs.

Next, we consider the field theoretic definition of lightcone TMDs, for which we will only consider the quark case and hence suppress flavor labels.
Since TMDs also encode the dependence on the transverse momentum of the parton, they naively correspond to PDFs in which the fields also have a separation along the transverse directions. This transverse gap prevents the cancellation of the anti-parallel Wilson lines in \eq{Wsplit}. As a consequence, we encounter the rapidity divergences typically associated with infinite-length light-like Wilson lines.
Many different regulators have been introduced to handle these rapidity divergences, see e.g.~refs.~\cite{Chiu:2009yx,Becher:2010tm,Collins:2011zzd,Chiu:2011qc,Becher:2011dz,Echevarria:2015byo,Li:2016axz}. In this work we consider two of these: the off-lightcone regulator used in the Collins scheme~\cite{Collins:2011zzd} and the $\delta$-regulator~\cite{Chiu:2009yx,Echevarria:2015byo}. The off-lightcone regulator takes all light-like Wilson slightly off the lightcone,
%%%
\begin{align} \label{eq:offlightcone_reg}
    &W[a-\infty n_a\leftarrow a]\rightarrow W[a-\infty n_A(y_A)\leftarrow a]\,,
    \nonumber\\
    &W[a-\infty n_b\leftarrow a]\rightarrow W[a-\infty n_B(y_B)\leftarrow a]\,,
\end{align}
%%%
where $n_A(y_A)$ and $n_B(y_B)$ are space-like vectors with rapidities of respectively $y_A$ and $y_B$,
%%%
\begin{align} \label{eq:nAnB}
    &n_A(y_A)=(1,-e^{-2y_A},\mathbf{0}_\perp)\,,
    \nonumber\\
    &n_B(y_B)=(-e^{2y_B},1,\mathbf{0}_\perp)\,.
\end{align}
%%%
The delta regulator is implemented by modifying the definition of infinite-length Wilson lines as follows, 
%%%
\begin{align} \label{eq:delta_reg}
    W[a-\infty n_a\leftarrow a]
    &\rightarrow
    P \exp\biggl(\img g \int_0^{-\infty} \df s\, e^{s\delta^-}\,A^{c,-}(a+s n_a) t^c\biggr)
    \nn
    \\
    W[a-\infty n_b\leftarrow a]
    &\rightarrow
    P \exp\biggl(\img g \int_0^{-\infty} \df s\, e^{s\delta^+}\,A^{c,+}(a+s n_b) t^c\biggr)\,.
\end{align}
%%%
The effect of this regulator is that the $\pm \img 0$ in eikonal propagators get replaced by finite imaginary numbers.

To construct a lightcone TMD that is free of rapidity divergences, we begin by
defining a \emph{beam function}, also known as the unsubtracted TMD. In the Collins scheme the (bare) beam function is defined as 
%%%
\begin{align}\label{eq:beamfunctiondefinition}
    B(x,b_\perp,\epsilon,y_B,P^+)
    &=
    \int\frac{\df b^-}{4\pi}\, e^{-\img xP^+ b^-}
    \bra{P} \bar{\psi}(b) 
    \gamma^+ \, W_\sqsupset\bigl(b,-\infty n_B(y_B),b^- n_b\bigr) \, 
    \psi(0)\ket{P},
\end{align}
%%%
where $b=(0,b^-, \mathbf{b}_\perp)$ and $\mathcal{W}_\sqsupset$ is a staple-shaped Wilson line defined in \eq{W_staple_def}. It extends from one quark field along the lightcone to minus infinity, bridges the transverse gap, and returns to connect to the second quark field in the correlator. The third argument in this Wilson line is chosen such that the transverse gap is perpendicular to the longitudinal segments of the Wilson line. In the above, $y_B$ is associated with the rapidity of
$n_B$, as given in \eq{nAnB}, and acts as a rapidity regulator by taking the Wilson line slightly off the lightcone.

The dependence on a rapidity regulator is indicative of a missing piece in our
calculation, given here by a \emph{soft function} encoding the dependence on
soft emissions radiated by the energetic colour-charged particles in the process.
In the Collins scheme, the (bare) soft function is defined as
%%%
\begin{align}\label{eq:softfunctiondefinition}
    S(b_\perp,\epsilon,y_A,y_B)
    &=
    \frac{1}{N_c}\bra{0} S_{\softstaple}\bigl(b_\perp,-\infty n_A(y_A),-\infty n_B(y_B)\bigr) \ket{0}
\,,\end{align}
%%%
with the Wilson loop $S_{\softstaple}$ defined in \eq{S_softstaple_def}.
The soft function encodes information about the full process, as soft
emissions --- isotropic and long-ranged --- can mediate interactions between the
different collinear sectors. For the definition of one single TMD, this means
that only one Wilson staple's direction is fixed to match the beam function
($\sqsupset$), the other ($\sqsupset'$) must be matched up with the TMD describing the other incoming parton (or outgoing spray of hadrons, for semi-inclusive deep-inelastic scattering),
such that the full cross-section is well-defined. As a result, the geometry
encoded by the soft function resembles that of an open book, with the spine
along the transverse separation between the partons (and with infinitely
wide pages).
It carries two rapidity regulators, $y_A$ and $y_B$, due to the fact that there are
two staple-shaped Wilson lines in its definition (the two open
pages of the book).

The physical TMD is defined as the ratio of the beam function and the square root of the soft function\footnote{The reason we divide, rather than multiply, with by the square root of the soft function is due the zero-bin~\cite{Manohar:2006nz}, that accounts for the overlap between the beam function and soft function. Here this overlap is equal to the inverse of the soft function~\cite{Ji:2004wu,Idilbi:2007ff,Mantry:2009qz}.}. In the Collins scheme, 
%%%
\begin{align}\label{eq:tmddefinition}
    f(x,b_\perp,\mu,\zeta)
    &=
    \lim_{\epsilon\to0} Z_\text{uv}(\epsilon,\mu,\zeta)
    \lim_{y_B\to-\infty}
    \frac{B(x,b_\perp,\epsilon,y_B,P^+)}{\sqrt{S(b_\perp,\epsilon,y_A,y_B)}}\,,
\end{align}
%%%
which is free of rapidity divergences regularized
by $y_B$, but depends on an auxiliary variable 
%%%
\begin{align}\label{eq:yn}
y_n=\tfrac12(y_A+y_B)
\,.\end{align}
%%%
In \eq{tmddefinition} it is implied that one takes $y_B\to-\infty$ and $y_A\to+\infty$ while keeping $y_n$ fixed. The dependence on $y_n$ cancels in
the full cross-section, through a dependence on a rapidity scale
%%%
\begin{equation} \label{eq:zeta}
  \zeta=2(xP^+)^2e^{-2y_n}
\end{equation}
as a remnant of the rapidity regulator cancellation.
After the rapidity divergences are cancelled, the UV divergences
are regularized and renormalized through $Z_\text{uv}$, to arrive at a
TMD that can be used in calculations.

%------------------------------------------------------------------------------------------------------------------
\subsection{Quasi-PDFs and matching with lightcone-PDFs}
\label{sec:Quasi-PDFs}
%------------------------------------------------------------------------------------------------------------------

The PDFs and TMDs as defined above are, unfortunately, not compatible with calculations on the lattice. This is mainly due to the sign problem. Lattice QCD circumvents the sign problem by making use of a Euclidean lattice, prohibiting the calculation of matrix elements where operators are separated in time. Since the parton distributions in \eqs{collpdf}{beamfunctiondefinition} involve fields that are separated along the lightcone, and hence also separated in time, they cannot be directly accesed on the lattice.

A solution arises from the insight that the ultra-relativistic limit of a
space-like trajectory ``looks'' light-like, and so we may expect to be able to relate a highly-boosted off-lightcone parton distribution to a PDF (or TMD). This is the fundamental insight behind the quasi-PDF approach, and has its root in the view of the parton picture as envisioned by Large Momentum Effective Theory (LaMET)~\cite{Ji:2013dva,Ji:2014gla,Ji:2020ect}. LaMET posits that in the large-momentum limit the structure of a proton (typically chosen to travel along the $z$-direction) is independent of the exact value for $P^z$ --- a Large Momentum symmetry --- and so we should expect results for PDFs (or TMDs) defined with separation along the $z$-direction to agree with those separated along the lightcone up to corrections of $\mathcal{O}(\Lambda_{\rm QCD}^2/P_z^2)$\footnote{The power corrections also depend on $x, b_\perp$ and the mass of the hadron.}. As a consequence, we can define a quasi-PDF, with exactly this type of separation.

Quasi-PDFs have the same definition as their lightcone counterparts, but with the lightcone correlators replaced by correlators where the fields are only separated along the $z$-axis. For quarks, the (bare) quasi-PDF is defined as 
%%%
\begin{align}
    \tilde{f}_q(x,\eps, P^z)
    &=
    \int\frac{\df z}{4\pi}\, e^{\img xP^z z} 
    \bra{P} \bar{\psi}(z) \gamma^z W[z\leftarrow0]\psi(0)\ket{P}\,.
\end{align}
%%%
where we use a tilde to distinguish it from the light-cone PDF in \eq{collpdf}.
The time-independence of this matrix element makes the quasi-PDF well suited for lattice calculations. In contrast to the boost-invariant lightcone-PDF, the quasi-PDF is boost dependent, which is captured by its $P^z$ dependence. Applying an infinite boost to the quasi-PDF is therefore identical to considering the $P^z\rightarrow\infty$ limit.

As $P^z\rightarrow\infty$, one naively expects the quasi-PDF to approach the lightcone-PDF. However, they are inherently different due the order of limits concerning the ultraviolet regulator and $P^z\rightarrow\infty$: For the lightcone-PDF that enters in factorization formulae, a UV regulator is introduced after already having taken the limit of infinite hadron momentum. For the quasi-PDF, the infinite boost is only performed after UV divergences have been regulated. Instead of equality, one can however derive a matching relation relating the lightcone- and quasi-PDFs in the limit that $P^z$ is much larger than $\Lambda_\text{QCD}$ and the mass of the hadron $M$~\cite{Ji:2013dva,Ji:2014gla,Izubuchi:2018srq}: 
%%%
\begin{align}\label{eq:quasipdfmatching}
    \tilde{f}_a(x,\mu,P^z)
    &=
    \sum_{a'}\int_{-1}^1\frac{\df x'}{|x'|}\, 
    \mathcal{C}_{a a'}\Bigl(\frac{x}{x'},\frac{\mu}{|x'|P^z}\Bigr) f_{a'}(x',\mu) +\mathcal{O}\biggl(\frac{M^2}{P_z^2},\frac{\Lambda_\text{QCD}^2}{x^2 P_z^2},\frac{\Lambda_\text{QCD}^2}{(1-x)^2 P_z^2}\biggr)\,.
\end{align}
%%%
Here $\mathcal{C}$ is a perturbative matching kernel and the sum over $a'$ accounts for mixing between parton species. This matching relation also holds for polarized PDFs ($a=\Delta q, \delta q$). It should be noted that the above equation relates the renormalized lightcone- and quasi-PDFs, so the matching kernel depends on the renormalization schemes for the lightcone- and quasi-PDFs. Though scale-invariance requires that the matching coefficient depends on the value of $\mu$ and the partonic momentum $|x'|P^z$, any perturbative expansion will involve $\alpha_s(\mu)$ as well.

The matching kernel for the quark flavor non-singlet case ($\mathcal{C}_{qq}$) was first calculated to one-loop order for all polarizations in ref.~\cite{Xiong:2013bka} and is now known up to two-loop order~\cite{Chen:2020ody,Li:2020xml}. The complete matching for all parton species and polarizations has been performed to one-loop order~\cite{Wang:2019tgg}. 

Much progress has been made in extracting parton distributions from lattice data. The unpolarized quark flavor non-singlet distribution was recently extracted from lattice data using the two-loop matching kernel~\cite{Gao:2022uhg} and the gluon PDF has been calculated using the one-loop matching coefficient~\cite{Zhang:2018diq,Fan:2022kcb,HadStruc:2021wmh,HadStruc:2022yaw}. Additionally, the above matching relation has been extended to the case of generalized parton distributions (GPDs)~\cite{Cichy:2023dgk}.

%------------------------------------------------------------------------------------------------------------------
\subsection{Quasi-TMDs and matching with physical TMDs}
\label{sec:Quasi-TMDs}
%------------------------------------------------------------------------------------------------------------------

In this section we review the recent progress for quasi-TMDs. We start by providing their field-theoretic definition, present the matching relation with the physical TMDs, and then sketch the proof for this matching that was given in ref.~\cite{Ebert:2022fmh}.

The (bare) quasi-beam function corresponding to the beam function in \eq{beamfunctiondefinition} is defined as, 
%%%
\begin{align}\label{eq:quasibeamfunctiondefinition}
    \tilde{B}(x,b_\perp,a,\tilde \eta,x\tilde{P}^z)
    &=
    \int\frac{\text{d}z}{4\pi}\, e^{\img xP^z z}
    \bra{\tilde P} \bar{\psi}(b) 
    \gamma^z \, W_\sqsupset (b,\tilde{\eta}\hat{z},b^z \hat{z}) \,
    \psi(0)\ket{\tilde P} \,,
\end{align}
%%%
where $b=(0,\mathbf{b}_\perp,z)$,  $a$ denotes a UV regulator (e.g.~$\eps$ or the lattice spacing), and the Wilson staple is now along the $z$-axis. The finite length of the Wilson line $\tilde \eta$ renders this object calculable on the lattice. The limit of large $\tilde \eta$ is divergent, so $\tilde \eta$ has
to be chosen large but finite for calculations on the lattice. 
The external proton state and choice of rapidity regulator influence each other as will be discussed below (see \eq{yptilde}), which is the reason we write $\ket{\tilde P}$ and $\tilde P^z$ in this section (which was not needed in \sec{Quasi-PDFs}).

As for the physical TMD in \eq{tmddefinition}, the quasi-TMD requires a soft function. Defining a quasi-soft function that is related to the soft function by a Lorentz boost and is lattice calculable is challenging: the soft function knows about the color flow in the full process, as indicated by the two Wilson staples extending in \emph{different} light-like directions. 
Here we contend ourselves to provide a definition of a soft function that leads to a consistent matching, deferring a discussion of how to calculate it on the lattice to \sec{lattice_soft}. The quasi-soft function can be defined by taking the Collins soft function, where the two staples are slightly off the lightcone, and boosting it such that one of its staples lies along the $z$-axis. Since the soft function is boost invariant, one does not have to make this boost explicit. One does, however, need to take into account that the quasi-soft function is to be calculated on the lattice, and therefore one has to impose a restriction on the length of the Wilson lines. This leads to the following definition
%%%
\begin{align}\label{eq:quasisoftfunctiondefinition}
    \tilde{S}(b_\perp,a,\tilde\eta,y_A,y_B)
    &=
    \frac{1}{N_c}\bra{0} S_{\softstaple}\Bigl(b_\perp,-\tilde\eta\frac{n_A(y_A)}{|n_A(y_A)|},-\tilde\eta\frac{n_B(y_B)}{|n_B(y_B)|}\Bigr) \ket{0}\,,
\end{align}
%%%
Note that, because of the finite length $\tilde{\eta}$ of the Wilson lines, the quasi-soft function is rapidity finite. Rapidity divergences appear as $\tilde\eta\to\infty$.

The quasi-TMD is then defined as the ratio of the quasi-beam function in \eq{quasibeamfunctiondefinition} and the square root of the quasi-soft function in \eq{quasisoftfunctiondefinition},
%%%
\begin{align}\label{eq:qTMD_def}
    \tilde{f}(x,b_\perp,\mu,\tilde{\zeta},x\tilde{P}^z)
    &=
    \lim_{a \to 0}
    \tilde Z_{\rm uv}(a,\mu, y_A - y_B)\,
    \lim_{\tilde{\eta}\to\infty}
    \frac
    {\tilde{B}(x,b_\perp,a,\tilde \eta,x\tilde{P}^z)}
    {\sqrt{\tilde{S}(b_\perp,a,\tilde\eta,y_A,y_B)}}\,.
\end{align}
%%%
Here, the limit $\tilde\eta\to\infty$ is taken at fixed but arbitrarily large $y_A$ and $y_B$, and the divergences that appear as $\tilde{\eta}\to\infty$ cancel between the beam and the soft function. Since the quasi-beam and soft function can both be renormalized multiplicatively, we included one renormalization factor $Z_{\rm uv}$ for both. This renormalization factor only depends on the cusp of the Wilson lines, which can be expressed in terms of $y_A-y_B$. More notably, the renormalization does not depend on $\tilde{\eta}$ as this parameter only contains long-distance physics, and it is also insensitive to the parton momentum as is always the case for quasi parton distributions. Finally, note that the dependence on $y_A$ and $y_B$ on the right-hand side is hidden in the
%%%
\begin{align} \label{eq:tildezeta}
  \tilde \zeta = (2 x \tilde P^z e^{y_B - y_n})^2
\end{align}
%%%
on the left-hand side. This would seem to diverge as $y_A \to \infty$ and $y_B \to -\infty$ for fixed $y_n$, but this divergence can be shown to be spurious by expressing $\tilde P^z$ in terms of the momentum $P^+$ for the physical TMD using \eq{yptilde} given below. 

The quasi-TMD in \eq{qTMD_def} can be matched perturbatively onto the physical (Collins scheme) TMD,
%%%
\begin{align}\label{eq:tmd_matching}
    \tilde{f}(x,b_\perp,\mu,\tilde\zeta,x \tilde{P}^z)
    &=
    C(x\tilde{P}^z,\mu) \exp\biggl[\frac{1}{2}\gamma_\zeta(b_\perp,\mu)\ln\biggl(\frac{\tilde\zeta}{\zeta}\biggr)\biggr]
    f(x,b_\perp,\mu,\zeta)\,,
\end{align}
%%%
where $C(x\tilde{P}^z,\mu)$ is a perturbative matching factor. 
This matching was proven in ref.~\cite{Ebert:2022fmh} in two steps: First the quasi-TMD was related to a so-called large rapidity (LR) scheme TMD, which is then subsequently related to the physical (Collins scheme) TMD. The LR scheme can be viewed as an intermediate scheme between the quasi-TMD and the Collins-TMD, and is defined as the Collins scheme but with the order of the $y_B\to-\infty$ and $\epsilon\to0$ limits reversed: 
%%%
\begin{align}\label{eq:LRdefinition}
    f^\text{LR}(x,b_\perp,\mu,\zeta,y_P-y_B)
    &=
    \lim_{-y_B\gg1}
    \lim_{\epsilon\to0} 
    Z^\text{LR}_\text{uv}(\epsilon,\mu,y_n - y_B)
    \frac{B(x,b_\perp,\epsilon,y_P - y_B)}{\sqrt{S(b_\perp,\epsilon,y_A,y_B)}}\,,
\end{align}
%%%
where 
%%%
\begin{equation}
y_P = \tfrac12 \ln (P^+/P^-) = \ln [P^+/(\sqrt{2} m_h)]
\end{equation}
%%%
is the rapidity of the hadron momentum $P$. Here, both the left-hand-side and the right-hand-side depend on $y_A$ and $y_B$, through either $\zeta$ as given in~\eq{zeta} and~\eq{yn}, or through $y_n$ as given in~\eq{yn} directly. As $\zeta$ and $y_n$ are assumed to be fixed the value of $y_B$ therefore also determines the size of $y_A$. It was then shown, by an analysis of all Lorentz invariants that a TMD can depend on, that the quasi-TMD of \eq{qTMD_def} and the LR-scheme TMD in \eq{LRdefinition} are related by
%%%
\begin{align}\label{eq:quasi_LR_relation}
    \tilde{f}(x,b_\perp,\mu,\tilde{\zeta},x\tilde{P}^z)
    &=
    f^\text{LR}(x,b_\perp,\mu,\tilde{\zeta},y_P-y_B)\,.
\end{align}
%%%
 Additionally, this analysis leads to the conclusion that 
 %%%
 \begin{equation}\label{eq:yptilde}
 y_{\tilde{P}}=y_P-y_B
\,,\end{equation}
%%%
where $y_{\tilde{P}}$ is the rapidity of $\tilde{P}$, which is the hadron momentum for the quasi-TMD. Consequently this equation illustrates why we take $y_B$ to be large but finite.

Finally, the authors of~\cite{Ebert:2022fmh} relate the LR scheme TMD to the Collins TMD, which differ only by the order of limits $y_B\to-\infty$ and $\epsilon\to0$. Using asymptotic freedom, they argue that the difference between the two schemes can be accounted for by a perturbative matching factor. Combining this with the relation between the quasi- and LR scheme TMD of eq.~\eqref{eq:quasi_LR_relation}, they arrive at the matching relation in eq.~\eqref{eq:tmd_matching}.

The first lattice calculation of the unpolarized flavor non-singlet quark TMD was carried out in~\cite{LPC:2022zci} and an application to spin dependent TMDs can be found in \cite{Ji:2020jeb}. Additionally, the matching relation in eq.~\eqref{eq:tmd_matching} has been used to extract the Collins-Soper kernel from lattice calculations~\cite{Schlemmer:2021aij,Shanahan:2020zxr}. The matching factor in eq.~\eqref{eq:tmd_matching} is currently known to two-loops~\cite{delRio:2023pse}.

%------------------------------------------------------------------------------------------------------------------
\subsection{Lattice calculability of the quasi-soft function}
\label{sec:lattice_soft}
%------------------------------------------------------------------------------------------------------------------

Unfortunately, the quasi-TMD in \eq{qTMD_def} that appears in the matching relation \eq{tmd_matching} is not directly calculable on the lattice. The problem lies with the quasi-soft function that enters the definition of the quasi-TMD. The quasi-soft function consists of two Wilson line staples that are slightly off the lightcone staples, only one of which can be boosted to lie along the $z$-axis. Consequently, the matrix element defining the quasi-soft function is time dependent and cannot be calculated on the lattice directly.

In the literature this issue has been addressed by introducing a naive quasi-soft function. The (bare) naive quasi soft function is defined in terms of a rectangular Wilson loop, whose longitudinal sides lie along the $z$-axis,
%%%
\begin{align}\label{eq:naive_soft}
    \tilde{S}_\text{naive}(b_\perp,a,\tilde\eta)
    &=
    \frac{1}{N_c}\bra{0} S_{\softstaple}(b_\perp,\tilde{\eta}\hat{z},-\tilde{\eta}\hat{z}) \ket{0}\,.
\end{align}
%%%
This soft factor can then be used to define a naive quasi-TMD that is directly calculable on the lattice,
%%%
\begin{align}
    \tilde{f}_\text{naive}(x,b_\perp,\mu,x\tilde{P}^z)
    &=
    \lim_{\tilde{\eta}\to\infty}
    \frac
    {\tilde{B}(x,b_\perp,\mu,\tilde{\eta},x\tilde{P}^z)}
    {\sqrt{\tilde{S}_\text{naive}(b_\perp,\mu,\tilde{\eta})}}\,,
\end{align}
%%%
and the structure of the renormalization is the same as in \eq{qTMD_def}.
However, this naive quasi-TMD cannot directly be matched to the physical TMD, because it has different IR behavior.

The final step consists of relating the quasi- and naive quasi-TMD, by considering the ratio of the two functions
%%%
\begin{align}
    \frac
    {\tilde{f}(x,b_\perp,\mu,\tilde{\zeta},x\tilde{P}^z)}
    {\tilde{f}_\text{naive}(x,b_\perp,\mu,x\tilde{P}^z)}
    &=
    \lim_{\tilde{\eta}\to\infty}
    \sqrt{
    \frac
    {\tilde{S}_\text{naive}(b_\perp,\mu,\tilde{\eta})}
    {\tilde{S}(b_\perp,\mu,y_A,y_B,\tilde{\eta})}
    }
    =
    \sqrt{
    \frac
    {\tilde{S}_\text{naive}(b_\perp,\mu)}
    {S(b_\perp,\mu,y_A,y_B)}
    }\,,
\end{align}
%%%
where the dependence on $y_A$ and $y_B$ on the right-hand-side is hidden in the $\tilde{\zeta}$ and $\tilde{P}^z$, see \eqs{tildezeta}{yptilde}. In the second equality the divergences as $\tilde{\eta}\to\infty$ cancel in the ratio. Crucially, the $S(b_\perp,\mu,y_A,y_B)$ in the final expression is the same soft function that enters the definition of the physical Collins scheme TMD. We can further reduce the above ratio by using the fact that for large $y_A$ and $y_B$ the Collins soft function behaves as 
%%%
\begin{align} \label{eq:S_I}
    S(b_\perp,\mu,y_A,y_B)
    &=
    S_I(b_\perp,\mu) e^{(y_A-y_B)\gamma_\zeta(b_\perp,\mu)}\,.
\end{align}
%%%
Here, $S_I(b_\perp,\mu)$ is the rapidity-independent part of the soft function. It is referred to in the literature as the intrinsic soft function, and has been related to a lattice-calculable meson form factor~\cite{Ji:2019sxk}. Furthermore, the Collins-Soper kernel can be calculated on the lattice by calculating ratios of quasi-beam functions at different hadron momenta~\cite{Ebert:2018gzl}. Summarizing, the quasi-TMD is related to the naive quasi-TMD by
%%%
\begin{align}
    \tilde{f}(x,b_\perp,\mu,\tilde{\zeta},x\tilde{P}^z)
    &=
    \tilde{f}_\text{naive}(x,b_\perp,\mu,x\tilde{P}^z)
    \sqrt{\frac{\tilde{S}_\text{naive}(b_\perp,\mu)}{S_I(b_\perp,\mu)}}
    \exp\biggl[-\frac{1}{2}\gamma_\zeta(b_\perp,\mu)\ln\biggl(\frac{(2x\tilde{P}^z)^2}{\zeta}\biggr)\biggr],
\end{align}
%%%
where all ingredients on the right-hand-side can be calculated on the lattice. Results for lattice calculations of the soft function can be found in~\cite{LatticeParton:2020uhz,Zhang:2020hy}.

\section{Double parton distribution functions}
\label{sec:dpds}
We start with a brief introduction to double parton scattering in \sec{intro_dps}, including a short overview of the theoretical framework. For a more comprehensive presentation, we refer to the book on this subject~\cite{Bartalini:2018qje}. The definition of the double parton distributions are discussed in \sec{dpd_defs}, which is the starting point for their extracting from lattice QCD using LaMET, and their renormalization is treated in \sec{dpd_ren}.

%------------------------------------------------------------------------------------------------------------------
\subsection{Introduction to double parton scattering} 
\label{sec:intro_dps}
%------------------------------------------------------------------------------------------------------------------

Double parton scattering (DPS) refers to two partonic scatterings between the same colliding hadrons. In contrast to pile-up, in which there are  collisions between different hadrons in the same bunch crossing, the two partonic collisions in DPS are \emph{not} independent of each other. Within the area of multi-parton interactions, there are different kinematic regions of interest: For many LHC measurements, there is a single energetic collision and additional partonic scatterings take place at lower energies. These additional partonic scatterings still produce radiation that affect measurements and are modeled in Monte Carlo parton showers. On the other hand, there is also an interest in two energetic collisions, for which a field-theoretic description in terms of factorization formulae is available. This is the case we focus on, and the one investigated in measurements of DPS that extract $\sigma_{\rm eff}$.

A given process, say double Drell-Yan for concreteness, can receive contributions from both DPS as well as single parton scattering (SPS). In the SPS contribution, only one parton is extracted from each of the colliding hadrons, and the two electroweak bosons are produced by a single hard scattering, rather than from two separate partonic scatterings.  The contribution of DPS to the total cross section is suppressed by $\Lambda_{\rm QCD}^2/Q^2$ compared to SPS, where $Q$ is the typical energy scale of the hard collisions (the invariant mass of an electroweak boson in double Drell-Yan). To enhance the contribution of DPS, one can consider processes where the energies of the hard collisions are fairly low, such as charm production~\cite{Luszczak:2011zp}. Alternatively, one can restrict to the region of phase-space where the total transverse momentum of each of the individual hard scatterings  is small (for double Drell-Yan this is the transverse momentum of each of the electroweak bosons). For single parton scattering the contribution is also power suppressed in this region as the two electroweak bosons are unlikely to each have a small transverse momentum, so the size of the SPS contribution will be of the same order as the DPS contribution~\cite{Diehl:2011tt}.

The factorization formula for the double Drell-Yan cross section takes the following form~\cite{Manohar:2012jr,Diehl:2017kgu}
%%%
\begin{align} \label{eq:si_DPS}
\frac{\df \si_{\text{DPS}}}{\df Q_1^2\, \df Y_1\, \df Q_2^2\, \df Y_2}
&= \frac{1}{S}  \! \sum_{a_1,a_2,a_3,a_4} \sum_{R_1,R_2,R_3,R_4} c_{R_1,R_2,R_3,R_4} \int\! \frac{\df x_1}{x_1} \frac{\df x_2}{x_2} \frac{\df x_3}{x_3} \frac{\df x_4}{x_4} \int\! \df^2 \mathbf{b}_\perp   \Phi^2(|\mathbf{b}_\perp| \nu)\, 
\nn \\&\qquad \times
\si_{a_1 a_3}(x_1, x_3, Q_1,Y_1, \mu)\, \si_{a_2 a_4}(x_2,x_4,Q_2,Y_2,\mu)\, 
\nn \\&\qquad \times
 {^{R_1 R_2}}\!F_{a_1 a_2}(x_1,x_2,\mathbf{b}_\perp,\mu,\zeta_p) \, {^{R_3 R_4}}\!F_{a_3 a_4}(x_3,x_4,\mathbf{b}_\perp,\mu,\bar \zeta_p)\,.
\end{align}
%%%
Here ${}^{R_1 R_2}\!F_{a_1a_2}(x_1,x_2,\mathbf{b}_\perp,\mu,\zeta)$ is the double parton distribution (DPD) describing the probability of extracting partons $a_1$ and $a_2$ from a proton with moment fraction $x_1$ and $x_2$ and transverse separation $\mathbf{b}_\perp$. The superscripts $R_i$ label the color representation of the partons, and in principle there are also interference effects in fermion number (that we will not consider). Because color singlets are produced, the dependence on the color representation can simply be encoded in an overall coefficient $c_{R_1,R_2,R_3,R_4}$. The regulator $\Phi^2(|\mathbf{b}_\perp| \nu)$ will be discussed below.

The partonic cross section $\si_{a_1 a_3}(x_1,x_3,Q_1,Y_1,\mu)$ in \eq{si_DPS} describes the scattering process in which the partons $a_1$ and $a_3$ collide to form an electroweak boson with invariant mass $Q_1$ and rapidity $Y_1$. Beyond leading order in perturbation theory, additional partons are produced in the final state and infrared poles in the partonic cross sections need to be subtracted.
The coefficient $S$ describes a symmetry factor.
 The parton flavors $a_i$ and representations  $R_i$ are summed over, and the momentum fractions $x_i$ and the transverse separation $\mathbf{b}_\perp$ between the two collisions are integrated over. 

Even for unpolarized protons, the cross section in \eq{si_DPS} receives contributions from DPDs such as $F_{\Delta q \Delta q}$, that describes extracting two longitudinally polarized quarks. While a non-vanishing distribution for a single longitudinal quark requires a polarized proton, $F_{\Delta q \Delta q}$ describes spin correlations of the two partons in an unpolarized proton. The sum over polarizations is included in the sum over flavors, e.g.~$a_1 = \Delta q$ is part of the sum over $a_1$. 

As an example of color correlations encoded in the superscripts $R_1, R_2, R_3, R_4$, two pairs of color-correlated (anti-)quarks can produce two color singlet electroweak bosons, corresponding to a term ${}^{88}\!F_{qq} {}^{88}\!F_{\bar q \bar q}$ in the DPS cross section.\footnote{Since the proton is colorless, there is no corresponding distribution for single parton scattering. However, the proton is not an electroweak singlet, so there is a corresponding (perturbative) electroweak effect~\cite{Manohar:2018kfx}.} The tree-level cross section  for double Drell-Yan, including spin and color correlations and interference effects, is shown in eq.~(43) of ref.~\cite{Manohar:2012jr}. For color correlations, the currents in the DPDs at position 0 and $\mathbf{b}_\perp$ are not color singlets. (The complete operator in the DPD of course is.) Consequently, soft radiation  resolving the large distance $\mathbf{b}_\perp\sim 1/ \Lambda_{\rm QCD}$ between the currents must be included. The corresponding soft functions are shown explicitly in ref.~\cite{Manohar:2012jr} but have been absorbed in the DPDs in \eq{si_DPS}. As for the case of TMDs in \sec{light-cone}, there is an associated rapidity resummation, for which a corresponding argument $\zeta_p$ and $\bar \zeta_p$ is included in the DPDs (this is absent for the color-summed DPDs).

DPDs can receive contributions from PDFs in the $\mathbf{b}_\perp \to 0$ limit~\cite{Kirschner:1979im,Shelest:1982dg}, known as the double parton splitting singularity~\cite{Nagy:2006xy}. For example, $F_{q \bar q}$ receives a contribution from $f_g$ where the gluon splits, $g \to q \bar q$. In particular, one has to make sure to avoid double counting between DPS and SPS. While several proposals for how to address this problem were put forward~\cite{Ryskin:2011kk,Blok:2011bu,Ryskin:2012qx,Blok:2017alw,Manohar:2012pe}, a first complete solution was presented in ref.~\cite{Diehl:2017kgu}. This leads to the inclusion of the regulator $\Phi^2(|\mathbf{b}_\perp| \nu)$ in \eq{si_DPS}, which goes to zero for $\mathbf{b}_\perp \to 0$ and becomes one for large $|\mathbf{b}_\perp|$. This does not play a role in our calculations because we only consider the flavor-non-singlet DPDs and restrict to the one-loop matching. One crucial ingredient in proving factorization is the cancellation of Glauber gluons, which was established for double Drell-Yan in ref.~\cite{Diehl:2015bca}.

We will conclude this section by explaining how the ``pocket formula" with $\sigma_{\rm eff}$ arises from \eq{si_DPS}. We assume all spin and color correlations and interferences effects can be ignored, such that we only need to consider ${}^{11}\! F_{a_1a_2}$ and can drop the argument $\zeta$. 
The following ansatz~\cite{Paver:1982yp}
%%%
\begin{align}
{}^{11}\! F_{a_1a_2}(x_1,x_2,\mathbf{b}_\perp,\mu) = f_a(x_1,\mu) f_b(x_2,\mu)  G(\mathbf{b}_\perp,\mu)
\,,\end{align}
%%%
is then made, where $f_{a,b}$ are the collinear PDFs. Due to the constraint $x_1 + x_2 < 1$, this can only hold if $x_i$ is small (which we assume).
This implies 
%%%
\begin{align}
\frac{\df \si_{\text{DPS}}}{\df Q_1^2\, \df Y_1\, \df Q_2^2\, \df Y_2}
 &= \frac{1}{S}\,\frac{\df \si}{\df Q_1^2\, \df Y_1}\, \frac{\df \si}{\df Q_2^2\, \df Y_2}  
 \underbrace{\int\! \df \mathbf{b}_\perp  \Phi^2(|\mathbf{b}_\perp| \nu) G(\mathbf{b}_\perp,\mu)}_{1/\sigma_{\rm eff}}
\,,\end{align}
%%%
so the two partonic scattering can be treated independently. The effective cross section can then be interpreted as a measure of the transverse size of the proton. It will be interesting to confront these assumptions with lattice data, using the LaMET approach.

%------------------------------------------------------------------------------------------------------------------
\subsection{Field-theoretical definition of DPDs} \label{sec:dpd_defs}
%------------------------------------------------------------------------------------------------------------------

Definitions of DPDs as proton matrix elements of operators in quantum field theory, have e.g.~been given in refs.~\cite{Diehl:2011yj,Manohar:2012jr}. For convenience, we repeat the definition for the quark-quark DPD here.

As for TMDs, the definition of DPDs involve infinite-length light-like Wilson lines and therefore contain rapidity divergences. These divergences cancel in the cross section when the DPDs are combined with their associated soft functions. The DPDs that appear in \eq{si_DPS} can be defined as the ratio of an unsubtracted DPD and the square root of the corresponding soft function. 

Let us now define the unsubtracted quark-quark DPDs, for general color and spin structures: 
%%%
\begin{align}
    \label{eq:dpddefinition}
    &F^\text{unsub}_{ijk\ell}(x_1,x_2,\mathbf{b}_\perp,\epsilon,y_B,P^+)
    \\
    &\quad=
    -\pi P^+ 
    \int \frac{\df b_1^-}{2\pi} \frac{\df b_2^-}{2\pi} \frac{\df b_3^-}{2\pi}\,
    e^{-\img x_1 P^+ b_1^-} e^{-\img x_2 P^+ b_2^-} e^{\img x_1 P^+ b_3^-}
    \nonumber
    \\
    &\quad\qquad\times
    \bra{P} \bar{T}
    \Big\{ 
    \Big[\bar{\psi}(b_1)W\big[b_1 \leftarrow b_\perp-\infty n_B(y_B) \big]\Big]_i  
    \Big[\bar{\psi}(b_2)W\big[b_2\leftarrow -\infty n_B(y_B)\big]\Big]_j 
    \Big\}
    \nonumber
    \\
    &\qquad\qquad\ \ \times
    T
    \Big\{ 
    \Big[W\big[b_\perp-\infty n_B(y_B) \leftarrow b_3 \big]\psi(b_3)\Big]_k
    \Big[W\big[-\infty n_B(y_B) \leftarrow 0 \big]\psi(0)\Big]_\ell
    \Big\} 
    \ket{P}.
    \nonumber
\end{align}
%%%
Here $i,j,k,\ell$ denote the color \emph{and} spin indices, which are uncontracted,
$T$ ($\bar T$) denotes (anti-)time ordering. The coordinates $b_i$ are
%%%
\begin{equation}
    b_1 = (0^+,b_1^-,\mathbf{b}_\perp) \, ,
    \quad
    b_2= (0^+,b_2^-,\mathbf{0}_\perp) \, ,
    \quad
    b_3 = (0^+,b_3^-,\mathbf{b}_\perp) \, ,
    \quad
    b_\perp = (0^+,0^-,\mathbf{b}_\perp) \, ,
\end{equation}
%%%
The Wilson line $W$ was defined in \eq{W_def}, and for definiteness we use the off-lightcone rapidity regulator $y_B$ in \eq{nAnB}. 

The quark-quark DPD as defined above can be decomposed into several color and spin structures. Following the conventions of ref.~\cite{Buffing:2017mqm},
%%% 
\begin{align}
    \label{eq:csdpddefinition}
    {^{11}}\!F_{q_1q_2} 
    &= (\Gamma_{q_1})_{ik} (\Gamma_{q_2})_{j\ell} F_{ijk\ell} \,,
    \\
    \label{eq:ccdpddefinition}
    {^{88}}\!F_{q_1 q_2}
    &=   \frac{2N_c}{\sqrt{N_c^2-1}}\, (\Gamma_{q_1} t^c)_{ik} (\Gamma_{q_2} t^c)_{j\ell}\, F_{ijk\ell} 
\,,\end{align}
%%%
where $N_c=3$ is the number of colors, $t^c$ is the SU(3) generator in the fundamental representation and the Dirac structures $\Gamma_{q_1},\Gamma_{q_2} \in\{\gamma^+,\gamma^+\gamma^5,\gamma^+\gamma_\perp^\mu\gamma^5\}$ are labelled $q$, $\Delta q$ and $\delta q$, respectively. The free Lorentz index in the $\delta q$ can be contracted with $\mathbf{b}_\perp$ or the index of another $\delta q$.

For the color correlated DPD, the currents at $\mathbf{b}_\perp$ and $\mathbf{0}_\perp$ are not separately gauge invariant and a transverse Wilson line at infinity is required. While this transverse Wilson line can be eliminated in covariant gauges by setting the gauge field at infinity to zero, they will be important when we introduce a lattice calculable quasi-DPD in \sec{quasidpddefinition}. In that case the Wilson lines are located at a finite distance $\tilde \eta$ along the $z$ direction, and can no longer be set to unity.

Next we discuss the soft functions needed to obtain the physical DPD from the unsubtracted one in \eq{dpddefinition}.
For the case of the quark-quark DPDs there are only two soft functions corresponding to the two different color structures: ${}^1\!S$ and ${}^8\!S$ (in principle there are more when one considers interference contributions). The soft function corresponding to the color-summed DPD is trivial, ${}^1\!S=1$, so in this case no soft subtraction is needed. 
The (bare) soft function for the color-correlated DPD can be written as 
%%%
\begin{align}\label{eq:ccsoft}
   {}^8\!S(b_\perp,\epsilon,y_A,y_B)
    =
    -\frac{1}{2N_c C_F} 
    +\frac{1}{2N_c C_F}\bra{0}\, 
    &S_{\softstaple}^\dagger\bigl(b_\perp,-\infty n_B(y_B),-\infty n_A(y_A)\bigr)
    \nn \\
    \times&
    S_{\softstaple}\bigl(b_\perp,-\infty n_B(y_B),-\infty n_A(y_A)\bigr) \ket{0}\,,
\end{align}
%%%
where $y_A$ and $y_B$ are two off-lightcone rapidity regulators and $S_{\softstaple}$ is the Wilson loop defined in \eq{S_softstaple_def}.

For \emph{general} flavor, spin and color we now define the (bare) DPDs by performing the soft subtraction: 
%%%
\begin{align}\label{eq:DPD_def_subtracted_bare}
    {^{R_1R_2}}\!F^\text{bare}_{a_1 a_2}(x_1,x_2,b_\perp,\epsilon,\zeta_p)
    &=
    \lim_{y_B\rightarrow-\infty}
    \frac
    {{^{R_1R_2}}\!F^\text{unsub}_{a_1 a_2}(x_1,x_2,b_\perp,\epsilon,y_B)}
    {\sqrt{{^{R_{1/2}}}\!S(b_\perp,\epsilon,y_A,y_B)}}\,.
\end{align}
%%%
This ratio is finite as $y_B\to-\infty$, but a dependence on a rapidity scale $\zeta_p$ remains. This rapidity scale dependence is described by a Collins-Soper evolution~\cite{Buffing:2017mqm},
%%%
\begin{align}
    \frac{\partial}{\partial\ln\zeta_p}
    {^{R_1R_2}}\!F^\text{bare}_{a_1 a_2}(x_1,x_2,b_\perp,\epsilon,\zeta_p)
    &=
    \frac{1}{2} {^{R_{1/2}}}\!J^\text{bare}(b_\perp,\eps)\,
    {^{R_1R_2}}\!F^\text{bare}_{a_1 a_2}(x_1,x_2,b_\perp,\epsilon,\zeta_p)\,,
\end{align}
%%%
with a similar equation being satisfied by the renormalized DPDs. Here ${^R}\!J$ is the rapidity anomalous dimension or rapidity evolution kernel. This kernel depends only on the dimension of the representation $R$ with $|R_1|=|R_2|$ being implied. The rapidity evolution kernels are related to the familiar Collins-Soper kernel by a color factor. Note that in contrast to the TMD case in \eq{zeta}, the rapidity scale associated to DPDs contains no momentum fractions,
%%%
\begin{align} \label{eq:zeta_p}
\zeta_p = 2(P^+)^2e^{-2y_n}\,.
\end{align}
%%%

This general form will be used in \sec{quasidpd}, but in our actual calculations in \sec{oneloop} we will only consider the color-summed ($R_1 = R_2 = 1$) and color-correlated ($R_1 = R_2 = 8$) quark-quark DPDs.

%------------------------------------------------------------------------------------------------------------------
\subsection{Renormalization of DPDs} \label{sec:dpd_ren}
%------------------------------------------------------------------------------------------------------------------

The bare DPDs defined in \eq{DPD_def_subtracted_bare} contain UV divergences and need to be renormalized. The reason for discussing renormalization in some detail here is that the structure of the renormalization group equations will play a key role in constructing the matching relation between the lightcone- and quasi-DPDs. This renormalization has been discussed extensively in refs.~\cite{Buffing:2017mqm,Diehl:2019rdh,Diehl:2021wpp,Diehl:2022rxb}, and includes results at order $\alpha_s^2$.

The renormalization of DPDs can be performed either in position or momentum space. These \emph{bare} DPDs are related by a Fourier transform 
%%%
\begin{align}\label{eq:DPD_momentumspace}
    {^{R_1 R_2}}\!F^\text{bare}_{a_1 a_2}(x_1,x_2,\Delta_\perp,\epsilon,\zeta)
    &=
    \int\! \df^{d-2} \mathbf{b}_{\perp}\,
    e^{\img \mathbf{b}_\perp\cdot\mathbf{\Delta}_\perp}\,
    {^{R_1 R_2}}\!F^\text{bare}_{a_1 a_2}(x_1,x_2,b_\perp,\epsilon,\zeta)\,.
\end{align}
%%%
This same formula does not hold for the renormalized DPDs, because of the double parton splitting singularity which occurs in the $\mathbf{b}_\perp\to0$ limit (see the discussion of the function $\Phi$ in \eq{si_DPS}).

In position space, the currents that enter the definition of the bare DPDs are separated by a distance $\mathbf{b}_\perp$. Because of this transverse separation, the two operators renormalize separately\footnote{In principle one could consider two separate renormalization scales for the two operators as they renormalize independently.}, 
%%%
\begin{align}
    F
    &=
    Z\otimes_1
    Z\otimes_2
    F^\text{bare}
\,.\end{align}
%%%
Explicitly, this reads
%%%
\begin{align} \label{eq:DPD_renormalization}
    {^{R_1 R_2}}\!F_{a_1 a_2}(x_1,x_2,b_\perp,\mu,\zeta_p)    
    &= \sum_{a_1',a_2'} \sum_{R_1', R_2'}
    \int\frac{\df x_1'}{x_1'}\frac{\df x_2'}{x_2'}\,
    {^{R_1 \bar{R_1'}}}\!Z_{a_1 a_1'}\Bigl(\frac{x_1}{x_1'},\mu,x_1^2\zeta_p\Bigr)
    \\ & \qquad \times
    {^{R_2 \bar{R_2'}}}\!Z_{a_2 a_2'}\Bigl(\frac{x_2}{x_2'},\mu,x_2^2\zeta_p\Bigr)
    {^{R_1' R_2'}}\!F^\text{bare}_{a_1' a_2'}(x_1',x_2',b_\perp,\mu,\zeta_p).\nn
\end{align}
%%%
Here it is clear why $\zeta_p$ is defined without momentum fractions, since they differ in the various terms. For color-summed DPDs the rapidity scale dependence drops out. 

In momentum space, additional singularities are generated due to the $1\to2$ splitting mechanism. This leads to mixing with single PDFs, 
%%%
\begin{align}
    F
    &=
    Z\otimes_1
    Z\otimes_2
    F^\text{bare}
    +
    Z_s\otimes f^\text{bare}\,.
\end{align}
%%%
Explicitly,
%%%
\begin{align}
    {^{R_1 R_2}}\!F_{a_1 a_2}(x_1,x_2,\Delta_\perp,\mu,\zeta_p)    
    &=
     \sum_{a_1',a_2'} \sum_{R_1', R_2'}
    \int\frac{\df x_1'}{x_1'}\frac{\df x_2'}{x_2'}\,
    {^{R_1 \bar{R_1'}}}\!Z_{a_1 a_1'}\Bigl(\frac{x_1}{x_1'},\mu,x_1^2\zeta_p\Bigr)
    \\ & \qquad \times
    {^{R_2 \bar{R_2'}}}\!Z_{a_2 a_2'}\Bigl(\frac{x_2}{x_2'},\mu,x_2^2\zeta_p\Bigr)
    {^{R_1' R_2'}}\!F^\text{bare}_{a_1' a_2'}(x_1',x_2',\Delta_\perp,\mu,\zeta_p)
    \nonumber\\
    &\quad
    +
     \sum_{a'}
    \int\frac{\df x'}{(x')^2}\,
    {^{R_1 R_2}}\!Z_{a_1 a_2,a'}\Bigl(\frac{x_1}{x'},\frac{x_2}{x'},\mu,x_1 x_2 \zeta_p\Bigr)
    f^\text{bare}_{a'}(x',\mu,\epsilon)\,.
    \nonumber
\end{align}
%%%

Because the renormalization of DPDs is different in position space and momentum space, the matching with quasi-DPDs will also be different in position space and momentum space. Note that this only affects the mixing with single PDFs, which is absent from our explicit one-loop calculations that are limited to the flavor-non-singlet case.

\section{Quasi-double parton distributions}
\label{sec:quasidpd}
In this section we take the first steps in extending the quasi-PDF approach to the case of double parton distributions. We define quasi-DPDs in \sec{quasidpddefinition}, which are related to the physical lightcone-DPD by an infinite Lorentz boost. The lattice-calculability of the DPD soft function is discussed in \sec{dpd_soft} (paralleling the discussion in \sec{lattice_soft} for the TMD soft function). In \sec{quasidpdfactorization}, we construct a matching relation between the physical and quasi-DPD. Although we do not prove the matching relation in this work, we verify its consistency to one-loop order in \sec{oneloop} for the flavor non-singlet quark-quark DPD.

%------------------------------------------------------------------------------------------------------------------
\subsection{Defining quasi-DPDs} \label{sec:quasidpddefinition}
%------------------------------------------------------------------------------------------------------------------

First we define (bare and unsubtracted) quasi-DPDs. Their definition can be obtained straightforwardly from their lightcone counterparts by essentially replacing all appearances of $n_a$ and $n_b$ by $\hat{z}$ and $-\hat{z}$.
For the quark-quark DPD with spin and color indices $i,j,k,\ell$ uncontracted, we obtain 
%%%
\begin{align}\label{eq:generalquasidpd}
    \tilde{F}^\text{unsub}_{ijk\ell}(x_1,x_2,\mathbf{b}_\perp,a,\tilde{\eta},\tilde P^z)
    &=
    -\pi \tilde P^z 
    \int\frac{\df b_1^z}{2\pi} \frac{\df b_2^z}{2\pi} \frac{\df b_3^z}{2\pi} \,
    e^{\img x_1 \tilde P^z b_1^z} e^{\img x_2 \tilde P^z b_2^z} e^{-\img x_1 \tilde P^z b_3^z}
    \\
    &\qquad\times
    \bra{\tilde P}  
    \Big[\bar{\psi}(b_1) W\big[b_1\leftarrow b_\perp+\tilde{\eta}\hat{z}\big]\Big]_i
    \Big[\bar{\psi}(b_2) W\big[b_2 \leftarrow \tilde{\eta}\hat{z}\big]\Big]_j
    \nonumber
    \\
    &\qquad\qquad\times
    \Big[W\big[b_\perp+\tilde{\eta}\hat{z}\leftarrow b_3\big]\psi(b_3)\Big]_k
    \Big[W\big[\tilde{\eta}\hat{z}\leftarrow0\big]\psi(0)\Big]_\ell 
    \ket{\tilde P},
    \nonumber
\end{align}
%%%
as the quasi-analogue of \eq{dpddefinition}. The coordinates $b_i$ are now
%%%
\begin{align}
    b_1 = (0,\mathbf{b}_\perp,b_1^z) \, ,
    \quad
    b_2 = (0,\mathbf{0}_\perp,b_2^z) \, ,
    \quad
    b_3 = (0,\mathbf{b}_\perp,b_3^z) \,,
    \quad
    b_\perp = (0,\mathbf{b}_\perp,0) \,,
\end{align}
%%%
so (anti-)time ordering is no longer relevant. The Wilson line $W$ is defined in \eq{W_def}, and we replaced the off-lightcone rapidity regulator in  \eq{dpddefinition} by a finite length $\tilde \eta$ of the Wilson lines, i.e.~$-\infty n_B(y_B) \to \tilde \eta \hat z$. This is necessary as the quasi-DPD is calculated on a lattice of finite size.

The color and spin decomposition of quasi-DPDs is essentially the same as for lightcone-DPDs. For definiteness, the decomposition of the quark-quark quasi-DPDs is given by 
%%%
\begin{align}
    &{^{11}}\!\tilde{F}^\text{unsub}_{q_1 q_2}(x_1,x_2,b_\perp,a,\tilde{\eta},\tilde P^z)
    \label{eq:quasi_DPD_unsub_cs}
    \\
    &\quad =
    -\pi \tilde P^z 
    \int\frac{\df b_1^z}{2\pi} \frac{\df b_2^z}{2\pi} \frac{\df b_3^z}{2\pi} \,
    e^{ix_1 \tilde P^z b_1^z} e^{ix_2 \tilde P^z b_2^z} e^{-ix_1 \tilde P^z b_3^z}
    \nonumber
    \\
    &\qquad\times
    \bra{\tilde P}  
    \bar{\psi}(b_1)\tilde{\Gamma}_{q_1}
        W[b_1\leftarrow b_3]\psi(b_3) \
    \bar{\psi}(b_2)\tilde{\Gamma}_{q_2}
        W[b_2\leftarrow0]\psi(0)
    \ket{\tilde P},
    \nonumber\\[1ex]
    &{^{88}}\!\tilde{F}^\text{unsub}_{q_1 q_2}(x_1,x_2,b_\perp,a,\tilde{\eta},\tilde P^z)
    \label{eq:quasi_DPD_unsub_cc}
    \\
    &\quad =
    -\pi \tilde P^z 
    \int\frac{\df b_1^z}{2\pi} \frac{\df b_2^z}{2\pi} \frac{\df b_3^z}{2\pi} \,
    e^{\img x_1 \tilde P^z b_1^z} e^{\img x_2 \tilde P^z b_2^z} e^{-\img x_1 \tilde P^z b_3^z}
    \nonumber
    \\
    &\qquad\times
    \bra{\tilde P}  
    \bar{\psi}(b_1) \tilde{\Gamma}_{q_1}
    W[b_1\leftarrow b_\perp+\tilde{\eta}\hat{z}]\,t^c\,
    W[b_\perp+\tilde{\eta}\hat{z}\leftarrow b_3]\psi(b_3)
    \nonumber
    \\
    &\qquad\quad \ \ \times
    \bar{\psi}(b_2) \tilde{\Gamma}_{q_2}
    W[b_2 \leftarrow \tilde{\eta}\hat{z}]\,t^c\,
    W[\tilde{\eta}\hat{z}\leftarrow0]\psi(0)
    \ket{\tilde P}.
    \nonumber
\end{align}
%%%
Note that $q_i$ denotes both the flavor of the quark field $\psi$ as well as the Dirac structure $\tilde \Gamma_{q_i}$.
These Dirac structures are related to those of the lightcone-DPD by
%%%
\begin{align} \label{eq:quasigamma}
    \frac{1}{p^z} \bar{u}(p) \tilde{\Gamma} u(p)
    =
    \frac{1}{p^+} \bar{u}(p) \Gamma u(p) \, ,
\end{align}
%%%
with $u(p)$ the spinor for a massless quark with momentum in the $z$ direction. Note that in this equation, the overall magnitude of $p$ is irrelevant because it cancels in the ratio, so $p$ can be replaced by $\tilde p$ if necessary.
In principle there exists a universality class of valid choices of $\tilde{\Gamma}$, obtained by replacing $\gamma^+$ in the lightcone-DPD definition by a linear combination of $\gamma^0$ and $\gamma^z$ (except for $\gamma^0 - \gamma^z$). For definiteness, we take $\tilde{\Gamma}=\gamma^z, \,\gamma^z \gamma^5, \,\gamma^z \gamma_\perp \gamma^5$ for unpolarized $(q)$, helicity $(\Delta q)$ and transversity $(\delta q)$, respectively.

We note that the color non-singlet distribution ${^{88}}\!\tilde{F}$ as written in \eq{quasi_DPD_unsub_cs} is not automatically gauge invariant due to the appearance of SU(3) generators $t^c$ in between the Wilson lines. While this issue could be ignored for the DPD in \sec{dpd_defs} (at least as long as covariant gauges are used), here we need to be more careful. We address this by first applying completeness relations for the generators which reconnects the color indices of the end-points of the Wilson lines, and then introducing transverse gauge links at spatial infinity where needed. As an example, in the quark-quark case we first use
%%%
\begin{align}
    t^a_{ij} t^a_{k\ell}=T_F\Bigl(\delta_{i\ell}\delta_{kj}
        -\frac{1}{N_c}\delta_{ij}\delta_{k\ell}\Bigr),
\end{align}
%%%
and then add the missing transverse Wilson lines between $\tilde{\eta}\hat{z}$ and $b_\perp+\tilde{\eta}\hat{z}$. This leads to the following gauge invariant definition of the color-correlated quark-quark quasi-DPD 
%%%
\begin{align} 
    &{^{88}}\!\tilde{F}^\text{unsub}_{q_1 q_2}(x_1,x_2,b_\perp,a,\tilde{\eta},\tilde P^z)
    \label{eq:quasi_DPD_unsub_cc_gi}
    \nn \\
    &\quad =
    -\frac{T_F}{N} 
    {^{11}}\!\tilde{F}^\text{unsub}_{q_1 q_2}(x_1,x_2,b_\perp,a,\tilde{\eta},\tilde P^z)
    -\pi \tilde P^z\int\frac{\df b_1^z}{2\pi}\frac{\df b_2^z}{2\pi}\frac{\df b_3^z}{2\pi}\,
    e^{ix_1 \tilde P^z b_1^z} e^{ix_2 \tilde P^z b_2^z} e^{-ix_1 \tilde P^z b_3^z}
    \nonumber
    \\
    &\qquad\times
    \bra{\tilde P}  
    \Bigl[\bigl(\bar{\psi}(b_1) \tilde{\Gamma}_{q_1}\bigr)_\alpha
        W[b_1\leftarrow b_\perp+\tilde{\eta}\hat{z}
        \leftarrow\tilde{\eta}\hat{z}\leftarrow0]\psi_\beta(0)\Bigr]
    \nonumber
    \\
    &\qquad\quad \ \ \times
    \Bigl[\bigl(\bar{\psi}(b_2) \tilde{\Gamma}_{q_2}\bigr)_\beta 
        W[b_2\leftarrow\tilde{\eta}\hat{z}\leftarrow b_\perp+\tilde{\eta}\hat{z}
        \leftarrow b_3]\psi_\alpha(b_3)\Bigr]
    \ket{\tilde P},
\end{align}
%%%
where $\al, \bt$ denote the spin indices. (Note that we already used this approach to obtain a gauge-invariant soft function in \eq{ccsoft}.)

Next we define a quasi-DPS soft function. As for the TMD soft function, it is not possible to define a matrix element that is both time-independent and is related to the lightcone soft function by a boost. To make sure that the quasi- and lightcone distributions possess the same infrared behaviour, we follow the same approach as in \sec{Quasi-TMDs} for TMDs, defining the quasi-soft function as the off-lightcone regularized soft function with finite-length Wilson lines. Explicitly, the DPS quasi-soft function can be obtained by taking the off-lightcone regularized soft function and making the replacement 
%%%
\begin{align}
    \infty n_A(y_A)\rightarrow\tilde{\eta}\frac{n_A(y_A)}{|n_A(y_A)|}\,,
    \qquad
    \infty n_B(y_B)\rightarrow\tilde{\eta}\frac{n_B(y_B)}{|n_B(y_B)|}\,.
\end{align}
%%%
For the quark-quark case, the color-summed quasi-soft function is ${^1}\!\tilde{S}=1$ and the color-correlated quasi-soft function is defined as 
%%%
\begin{align} \label{eq:quasi_dpd_soft}
    {}^{8}\!\tilde{S}(b_\perp,a,\tilde{\eta},y_A,y_B)
    &=
    -\frac{1}{2N_c C_F}
    +\frac{1}{2N_c C_F}\bra{0} 
    S_{\softstaple}^\dagger\biggl(b_\perp, 
        -\tilde{\eta}\frac{n_B(y_B)}{|n_B(y_B)|},
        -\tilde{\eta}\frac{n_A(y_A)}{|n_A(y_A)|}\biggr)
    \\
    &\qquad \times
    S_{\softstaple}\biggl(b_\perp, 
        -\tilde{\eta}\frac{n_B(y_B)}{|n_B(y_B)|},
        -\tilde{\eta}\frac{n_A(y_A)}{|n_A(y_A)|}\biggr)
    \ket{0}
    \,,
    \nonumber
\end{align}
%%%
with $S_{\softstaple}$ defined in \eq{S_softstaple_def}.

We now define quasi-DPDs, which we will match onto the physical lightcone-DPDs in the next section. First we use the quasi-soft functions to subtract the singularities for $\tilde{\eta}\to\infty$ (known as pinch-pole singularities) from the unsubtracted quasi-DPDs, and then we perform the renormalization, 
%%%
\begin{align} \label{eq:quasi-DPD_renorm}
    {^{R_1 R_2}}\!\tilde{F}_{a_1 a_2}
    (x_1,x_2,b_\perp,\mu,\tilde{\zeta}_p,\tilde{P}^z)
    &=
    \lim\limits_{\substack{a\rightarrow0\\\tilde{\eta}\rightarrow\infty}}
    {^{R_1 R_2}}\!\tilde Z_{a_1,a_2}(a,\mu,y_A,y_B)\,
    \frac
    {{^{R_1 R_2}}\!\tilde{F}_{a_1 a_2}(x_1,x_2,b_\perp,a,\tilde{\eta},\tilde P^z)}
    {\sqrt{{^R}\!\tilde{S}(b_\perp,a,\tilde{\eta},y_A,y_B)}}\,,
\end{align}
%%%
Note that, as in the case of quasi-PDFs and quasi-TMDs, quasi-DPDs can be renormalized multiplicatively. This is in contrast to the lightcone-DPDs, which are renormalized by a convolution in the two momentum fractions, see \eq{DPD_renormalization}.

%------------------------------------------------------------------------------------------------------------------
\subsection{Lattice calculability of the DPD quasi-soft function}
\label{sec:dpd_soft}
%------------------------------------------------------------------------------------------------------------------

The quasi-DPD as defined above cannot be calculated directly on the lattice because the matrix element defining the quasi-soft function is time-dependent. Therefore, analogous to the TMD case in \sec{lattice_soft}, we define naive quasi-DPDs, which can be calculated on the lattice directly and can be related to the proper quasi-DPDs via an intrinsic soft function. For the quark-quark DPD, we simply have ${^1}\!S_\text{naive}=1$ in the color-summed case, while for the color-correlated case we define the (bare) naive soft function as 
%%%
\begin{align}
    {^8}\!S_\text{naive}(b_\perp,a,\tilde{\eta})
    &=
    -\frac{1}{2N_c C_F}
    +\frac{1}{2N_c C_F}
    \bra{0}
    S_{\softstaple}^\dagger\big[\mathbf{b}_\perp,-\tilde \eta \hat{z},\tilde \eta \hat{z}]
    S_{\softstaple}\big[\mathbf{b}_\perp,-\tilde \eta \hat{z},\tilde \eta \hat{z}\big]
    \ket{0}\,.
\end{align}
%%%
We then define the naive quasi-DPDs as 
%%%
\begin{align}\label{eq:DPD_naive}
    {^{R_1 R_2}}\!\tilde{F}^\text{naive}_{a_1 a_2}
    (x_1,x_2,b_\perp,\mu,\tilde P^z)
    &=
    \lim\limits_{\substack{a\rightarrow0\\\tilde{\eta}\rightarrow\infty}}
    {^{R_1 R_2}}\!\tilde Z_{a_1 a_2}(a,\mu,y_A,y_B)
    \frac
    {{^{R_1 R_2}}\!\tilde{F}_{a_1 a_2}(x_1,x_2,\mathbf{b}_\perp,a,\tilde{\eta},\tilde P^z)}
    {\sqrt{{^R}\!\tilde{S}_\text{naive}(b_\perp,a,\tilde{\eta})}} \,,
\end{align}
%%%
The naive quasi-DPD does not posses the correct IR behaviour to be used to in the matching relation, as it is not related to the lightcone-DPD by a boost. However, it can be used in lattice calculations to subtract the pinch-pole singularities of the unsubtracted quasi-DPD, that appear as divergences in the limit $\tilde{\eta}\rightarrow\infty$.

To relate the naive quasi-DPD to the quasi-DPD that is used in the matching to physical DPDs, we need to define an analogue of the intrinsic soft function of \eqref{eq:S_I} for the double parton case. To define an intrinsic soft function it is necessary that the rapidity divergences exponentiate, which was proven to be the case in~\cite{Vladimirov:2017ksc}. This allows us to define an intrinsic soft function for the DPD case by\footnote{Note that we assume a basis where mixing between different soft functions under rapidity evolution is absent.}
%%%
\begin{align}\label{eq:DPD_S_I}
    {^R}\!S(b_\perp,\mu,y_A,y_B)
    &=
    {^R}\!S_I(b_\perp,\mu)
    e^{(y_A-y_B)\,{^R}\!J(b_\perp,\mu)}\,.
\end{align}
%%%
Following the same arguments as for the TMD case, we then find the following relation between the quasi- and the naive quasi-DPDs,
%%%
\begin{align}\label{eq:DPD_quasi_vs_naive}
    \tilde{F}(x_1,x_2,b_\perp,\tilde{P}^z,\mu,\tilde{\zeta})
    &=
    \tilde{F}_\text{naive}(x_1,x_2,b_\perp,\tilde{P}^z,\mu,\tilde{\zeta})
    \sqrt{\frac{{^R}\!\tilde{S}_\text{naive}(b_\perp,\mu)}{{^R}\!S_I(b_\perp,\mu)}}
    \\
    &\quad\times
    \exp\biggl[-\frac{1}{4} {^R}\!J(b_\perp,\mu) 
            \ln\biggl(\frac{(2x_1\tilde{P}^z)^2}{x_1^2 \zeta_p}\biggr)
        -\frac{1}{4} {^R}\!J(b_\perp,\mu) 
            \ln\biggl(\frac{(2x_2\tilde{P}^z)^2}{x_2^2 \zeta_p}\biggr)\biggr].
    \nonumber
\end{align}
%%%

%------------------------------------------------------------------------------------------------------------------
\subsection{Factorization for quasi-DPDs} \label{sec:quasidpdfactorization}
%------------------------------------------------------------------------------------------------------------------

We will now present a matching relation between quasi-DPDs and the physical DPDs for \emph{general} flavor, color and spin. In position space the matching relation reads
%%%
\begin{align} \label{eq:match_nonsinglet}
    &{^{R_1 R_2}}\!\tilde{F}_{a_1 a_2}
        (x_1,x_2,b_\perp,\mu,\tilde{\zeta}_p,\tilde{P}^z)
    \\
    &\quad =
    \sum_{R_1',R_2'} \sum_{a_1',a_2'}
    \int_0^1 \frac{\df x_1'}{x_1'} \frac{\df x_2'}{x_2'}\,
    {^{R_1 R_1'}}\!C_{a_1 a_1'}
        \Bigl(\frac{x_1}{x_1'},x_1' \tilde{P}^z,\mu\Bigr)\,
    {^{R_2 R_2'}}\!C_{a_2 a_2'}
        \Bigl(\frac{x_2}{x_2'},x_2' \tilde{P}^z,\mu\Bigr)
    \nonumber
    \\
    &\qquad\qquad\times
    \exp\biggl[\frac{1}{2} {^{R_{1/2}'}}\!J(b_\perp,\mu) 
        \ln\biggl(\frac{\tilde{\zeta}_p}{\zeta_p}\biggr)\biggr]
    {^{R_1' R_2'}}\!F_{a_1' a_2'}(x_1',x_2',b_\perp,\mu,\zeta_p)\,,
    \nonumber
\end{align}
%%%
while in momentum space the matching relation reads
%%%
\begin{align}\label{eq:match_nonsinglet_momspace}
    &{^{R_1 R_2}}\!\tilde{F}_{a_1 a_2}
    (x_1,x_2,\Delta_\perp,\mu,\tilde{\zeta}_p,\tilde{P}^z)
    \\
    &\quad =
    \sum_{R_1',R_2'} \sum_{a_1',a_2'}
    \int_0^1 \frac{\df x_1'}{x_1'} \frac{\df x_2'}{x_2'}\,
    {^{R_1 R_1'}}\!C_{a_1 a_1'}
        \Bigl(\frac{x_1}{x_1'},x_1' \tilde{P}^z,\mu\Bigr)\,
    {^{R_2 R_2'}}\!C_{a_2 a_2'}
        \Bigl(\frac{x_2}{x_2'},x_2' \tilde{P}^z,\mu\Bigr)
    \nonumber
    \\
    &\qquad\qquad\times
    {^{R_1' R_2'}}\!F_{a_1' a_2'}(x_1',x_2',\Delta_\perp,\mu,\tilde{\zeta}_p)
    \nonumber
    \\
    &\qquad
    +\sum_{a'}
    \int_0^1 \frac{\df x'}{x'^2}\,
    {^{R_1 R_2}}\!C_{a_1 a_2,a'}
    \Bigl(\frac{x_1}{x'},\frac{x_2}{x'},x' \tilde P^z,\mu\Bigr)
    f_{a'}(x',\mu)\,.
    \nonumber
\end{align}
%%%
Here, ${^{R R'}}\!C_{a a'}$ and ${^{R R'}}\!C_{a_1 a_2,a'}$ are perturbative matching kernels. The rapidity evolution kernel ${^{R}}\!J$ only depends on the dimensionality of the representation, and $|R_1|=|R_2|$ is necessary to have a non-vanishing DPD. In the momentum-space matching the lightcone- and quasi-DPD are evaluated at the same rapidity scale, to avoid the more complicated momentum-space rapidity evolution. Furthermore we wish to stress that the position space and momentum space matching relations \eq{match_nonsinglet} and \eq{match_nonsinglet_momspace} are not equivalent due to the single-PDF contribution that arises at vanishing transverse separation.

As we will now show, the above matching relations are consistent with the renormalization group evolution and the rapidity evolution of the lightcone- and quasi-DPDs. The rapidity evolution is satisfied trivially, as both the lightcone- and the quasi-DPD satisfy a Collins-Soper evolution. Note that this relies on the fact there is no matching between quasi-DPDs onto lightcone-DPDs with a different rapidity anomalous dimension and that the matching kernels are rapidity-scale independent.

To show that the matching relation is also consistent with renormalization group evolution, we start by presenting the structure of evolution equations for the quasi-DPD. Their evolution can be inferred from \eq{DPD_quasi_vs_naive} by using the fact that the unsubtracted quasi-DPDs have a multiplicative renormalization and the quasi-soft soft function satisfies \eq{DPD_S_I}. This leads to the following evolution equation for the quasi-DPDs,
%%%
\begin{align}
    &\frac{\df}{\df\ln\mu^2}\,
    {^{R_1 R_2}}\!\tilde{F}_{a_1 a_2}
        (x_1,x_2,b_\perp,\mu,\tilde{\zeta}_p,\tilde{P}^z)
    \\
    &\quad =
    \biggl[{^{R_1}}\!\gamma_{a_1}
        +\frac{1}{4}{^{R_1}}\gamma_J
            \ln\biggl(\frac{(2x_1\tilde{P}^z)^2}{x_1^2 \zeta_p}\biggr)
        +(1\leftrightarrow2)\biggr]
    {^{R_1 R_2}}\!\tilde{F}_{a_1 a_2}
        (x_1,x_2,b_\perp,\mu,\tilde{\zeta}_p,\tilde{P}^z)\,.
    \nonumber
\end{align}
%%%
Here ${^{R}}\!\gamma_{a}$ corresponds to the anomalous dimension of one of the current operators in the unsubtracted quasi-DPD and ${^{R}}\!\gamma_J$ is the anomalous dimension of the rapidity evolution kernel,
%%%
\begin{align}
    \frac{\df}{\df\ln\mu^2}\,
    {^{R}}\!J(b_\perp,\mu)
    &=
    -{^{R}}\!\gamma_{J}(\mu)\,.
\end{align}
%%%

On the other hand, for the physical DPDs in position space, the renormalization scale dependence is given by~\cite{Buffing:2017mqm,Diehl:2021wpp}
%%%
\begin{align}
    &\frac{\df}{\df\ln\mu^2}\,
    {^{R_1 R_2}}\!F_{a_1 a_2}(x_1,x_2,b_\perp,\mu,\zeta_p)
    \\
    &\quad =
    \sum_{R_1' R_2'} \sum_{a_1',a_2'}
    \int\frac{\df x_1'}{x_1'}\frac{\df x_2'}{x_2'}\,
    \biggl[
    {^{R_1 \bar{R}_1'}}\!P_{a_1 a_1'}
        \Bigl(\frac{x_1}{x_1'},\mu,x_1^2\zeta_p\Bigr)\,
    \delta_{R_2' R_2} \delta_{a_2' a_2}\, \delta\Bigl(1-\frac{x_2}{x_2'}\Bigr)
    +(1\leftrightarrow2)
    \biggr]
    \nonumber
    \\
    &\qquad\qquad\times
    {^{R_1' R_2'}}\!F_{a_1' a_2'}(x_1',x_2',b_\perp,\mu,\zeta_p)\,,
    \nonumber
\end{align}
%%%
where the ${^{R\bar{R}'}}\!P_{a a'}$ are referred to as the color-dependent DGLAP splitting kernels. These kernels can be split up into a rapidity-independent part $\hat{P}$ and rapidity-dependent part, 
%%%
\begin{align}
    {^{R\bar{R}'}}\!P_{a a'}(x,\mu,\zeta)
    &=
    {^{R\bar{R}'}}\!\hat{P}_{a a'}(x,\mu)
    -\frac{1}{4}\delta_{R R'} \delta_{a a'}\, \delta(1-x)\,
        {^{R}}\!\gamma_J\,
        \ln\biggl(\frac{\zeta}{\mu^2}\biggr)\,.
\end{align}
%%%

Using these evolution equations, one can see that the matching relation is consistent with the UV and large rapidity behaviour of the lightcone- and quasi-DPDs. Furthermore, one can conclude that the matching kernels satisfy,
%%%
\begin{align}
    \frac{\df}{\df\ln\mu^2}\, {^{R R'}}\!C_{a a'}
        \Bigl(\frac{x}{x'},x'\tilde{P}_z,\mu\Bigr)
    &=
    \biggl[{^{R}}\!\gamma_{a} + \frac{1}{4}{^{R}}\gamma_J
        \ln\biggl(\frac{(2x'\tilde{P}^z)^2}{\mu^2}\biggr)\biggr]
    {^{R R'}}\!C_{a a'}\Bigl(\frac{x}{x'},x'\tilde{P}^z\mu\Bigr)
    \\
    &\quad
    -\sum_{R''} \sum_{a''}\int_0^1\frac{\df x''}{|x''|}\,
    {^{R'' \bar{R}'}}\!\hat{P}_{a'' a'}\Bigl(\frac{x''}{x'},\mu\Bigr)\,
    {^{R R''}}\!C_{a a''}
        \Bigl(\frac{x}{x''},x''\tilde{P}^z,\mu\Bigr).\nn
\end{align}
%%%

In this paper we do not provide a proof for the matching relation. However, for the color-summed case, the proof for the single-PDF case~\cite{Izubuchi:2018srq}, which makes use of the operator product expansion (OPE), directly carries over for the position-space matching in \eq{match_nonsinglet}. This is because the operator that defines the color-summed DPDs consists of two copies of the single-PDF operators separated by a finite distance $\mathbf{b}_\perp$. The finite transverse separation ensures that no additional terms can arise in the OPE and so one can apply the OPE to each current operator separately. This leads to the conclusion that the matching kernels for the color-summed DPDs will be identical to the matching kernels of the single parton case,
%%%
\begin{align} \label{eq:C_rel_cs}
    {^{11}}\!C_{a a'}
        \Bigl(\frac{x}{x'},x' \tilde{P}^z,\mu\Bigr)
    &=
    \mathcal{C}_{a a'}\Bigl(\frac{x}{x'},\frac{\mu}{|x'|\tilde P^z}\Bigr)\,,
\end{align}
%%%
where the $\mathcal{C}_{aa'}$ are the matching kernels that appear in \eqref{eq:quasipdfmatching}. Note that this proof does not directly carry over to the momentum-space matching relation of \eq{match_nonsinglet_momspace}, as the OPE relies on the position-space formulation where $\mathbf{b}_\perp$ is taken to be finite, whereas the mixing term involving single PDFs arises from the region where $\mathbf{b}_\perp\to0$.

For DPDs with a non-trivial color structure the proof would be more complicated. In principle, one could extend the proof of the TMD matching relation in ref.~\cite{Ebert:2022fmh} to the case of color-correlated DPDs. This proof was based on the decomposition of the hadronic correlator in terms of all available Lorentz structures, which was used to show that the correlators describing the quasi- and LR scheme TMDs agree in the infinite boost limit, up to higher twist terms. An extension of this proof to the case of DPDs would require an analysis of the Lorentz decomposition of the hadronic correlator defining DPDs. Furthermore, this proof cannot be carried over directly to our case because DPDs have a substantially different UV behaviour.

\section{One-loop matching for the flavor-non-singlet case}
\label{sec:oneloop}
In this section we will calculate the quark-quark DPD matching kernel ${^{R R'}}\!C_{q q'}$ for all color and spin structures to one-loop order. We consider the flavor non-singlet DPDs, which can be defined as 
%%%
\begin{align} \label{eq:flavor_nons}
    {^{R_1 R_2}}F^\text{NS}_{q_1q_2}
    &=
    {^{R_1 R_2}}F_{u_1 u_2}
    -{^{R_1 R_2}}F_{u_1 d_2}
    -{^{R_1 R_2}}F_{d_1 u_2}
    +{^{R_1 R_2}}F_{d_1 d_2}\,.
\end{align}
%%%
Here $q_i$ denote the spin structure $q,\Delta q, \delta q$ and \emph{not} the quark flavor, while $u_i$ and $d_i$ denote different quark flavors and have the same spin structure as the corresponding $q_i$.  By considering the flavor non-singlet case, we do not need to consider mixing with different flavors, but mixing with color and spin is still present. In this case the matching relation of \eq{match_nonsinglet} simplifies to
%%%
\begin{align} \label{eq:match_sec5}
    &{^{R_1 R_2}}\!\tilde{F}^{\text{NS}}_{q_1 q_2}
        (x_1,x_2,b_\perp,\mu,\tilde{\zeta}_p,\tilde{P}^z)
    \\
    &=
    \sum_{R_1',R_2'} \sum_{q_1',q_2'}
    \int_0^1 \frac{\df x_1'}{x_1'} \frac{\df x_2'}{x_2'}\,
    {^{R_1 R_1'}}\!C_{q_1 q_1'}
        \Bigl(\frac{x_1}{x_1'},x_1'\tilde{P}^z,\mu\Bigr)
    {^{R_2 R_2'}}\!C_{q_2 q_2'}
        \Bigl(\frac{x_2}{x_2'},x_2'\tilde{P}^z,\mu\Bigr)
    \nonumber
    \\
    &\qquad\qquad\times
    \exp\biggl[\frac{1}{2} {^{R_{1/2}}}\!J(b_\perp,\mu) 
        \ln\biggl(\frac{\tilde{\zeta}_p}{\zeta_p}\biggr)\biggr]
    {^{R_1' R_2'}}\!F^{\text{NS}}_{q_1' q_2'}(x_1',x_2',b_\perp,\mu,\zeta_p)\,.
    \nonumber
\end{align}
%%%
where the sum on $q_i'$ runs only over the spin structures, not quark flavors. The only nonvanishing color structures are $R_1 = R_2$ and $R_1' = R_2'$ which can be either 1 or 8.

The one-loop matching kernels can be extracted from the one-loop corrections to the lightcone- and quasi-DPDs with the proton replaced by partonic states. As we discuss in \sec{partonicstates}, the use of partonic states leads to ill-defined expressions. We resolve this by using different in- and out-states, only taking them equal at the end of the calculation. The ``master formula" for extracting the one-loop matching kernel from the various ingredients is derived in \sec{oneloopconstruction}. Next, \sec{oneloopdpds} provides an example calculation for one of the diagrams contributing to the one-loop lightcone- and quasi-DPD, with our conventions for plus distributions given in \app{conventions}. The results for all one-loop diagrams are presented in \app{diagrams}. Finally, in \sec{result} we obtain the one-loop matching coefficients and verify their perturbative nature by showing that they are free of infrared logarithms. We also provide the one-loop expressions for the remaining ingredients that enter the matching relation: the lightcone- and quasi-soft functions, the rapidity evolution kernel and the relevant renormalization factors.

%------------------------------------------------------------------------------------------------------------------
\subsection{Handling divergences in calculating DPDs for partonic states} \label{sec:partonicstates}
%------------------------------------------------------------------------------------------------------------------

The matching kernel in \eq{match_sec5} is independent of the external state $\ket{P}$. We can therefore replace the proton by a suitable partonic state to calculate the matching kernel in perturbation theory. However, the DPDs as defined in the previous section are not well-suited for partonic states. This is because the definition of DPDs leads to the square of a delta function when applied to identical partonic in- and out-states. Here we demonstrate how this issue arises, and how it can be resolved by using different in- and out states in intermediate steps of the calculation.

The simplest state that gives a non-trivial result for the DPDs is a di-quark state where the quarks have definite (on-shell) momenta $p_1$ and $p_2$. Neglecting the transverse momentum of the two quarks, we take
%%%
\begin{align} \label{eq:diquark}
    p_1 = \omega_1 p \ , \quad 
    p_2 = \omega_2 p \ , \quad
    \text{with}\quad
    \omega_1 + \omega_2 = 1\,,
\end{align}
%%%
where $p=(p^+,0^-,\mathbf{0})=(p^z,\mathbf{0},p^z)$, and we will use lowercase $p$ for partonic momenta throughout the calculation.

Since we calculate the matching kernels for all color and spin structures, we do not average over the color and spin of the external partons. For notation convenience, we denote products of spinors as
%%%
\begin{align} \label{eq:spin_color}
    \bar{u}(p_1)\Gamma_{q_1} T_{R_1} u(p_1) \ \bar{u}(p_2) \Gamma_{q_2} T_{R_2} u(p_2)
    &\equiv
    (T_{R_1}\otimes T_{R_2}) \ \Gamma_{q_1}\otimes\Gamma_{q_2} \ ,
\end{align}
%%%
where $T_{R_i}$ are the generators in the representation $R_i$ and $\Gamma_{q_i}$ are Dirac structures.

When calculated on di-parton states, the DPDs defined in \eq{dpddefinition} diverge. Explicitly, for the color-summed DPD at tree-level, using the di-quark state in \eq{diquark}
%%%
\begin{align}
    {^{11}}\!F_{qq}^{(0)} (x_1,x_2)
    &=
    -\pi p^+ \int 
    \frac{\df b_1^-}{2\pi} \frac{\df b_2^-}{2\pi} \frac{\df b_3^-}{2\pi}\,
    e^{-\img x_1 p^+ b_1^-} e^{-\img x_2 p^+ b_2^-} e^{\img x_1 p^+ b_3^-}
    \nonumber\\
    &\qquad\times
    \bar{u}(p_1) e^{\img\omega_1 p^+ b_1^-}\gamma^+ e^{-\img \omega_1 p^+ b_3^-}u(p_1)\,
    \bar{u}(p_2) e^{\img\omega_2 p^+ b_1^-}\gamma^+ u(p_2)\,.
\end{align}
%%%
The Fourier transforms over $b_1^-$ and $b_2^-$ result in delta functions of the two momentum fractions, 
%%%
\begin{align}\label{eq:dpddivergence}
    {^{11}}\!F_{qq}^{(0)} (x_1,x_2)
    &=
    -\pi (1\otimes1) \,
    \frac{\gamma^+\otimes\gamma^+}{p_1^+ p_2^+} \, 
    \delta\big(1 - \tfrac{x_1}{\omega_1}\big)
    \delta\big(1 - \tfrac{x_2}{\omega_2}\big)
    \int \frac{\df(p^+ b_3^-)}{2\pi} \ 
    e^{\img (x_1 - \omega_1) p^+ b_3^-} .
\end{align}
%%%
The combination of Dirac matrices and $p_i^+$ is chosen to match the form in \eq{quasigamma}. 
The remaining $b_3^-$ integral gives a delta function involving $x_1$, which has already been fixed to $x_1=\omega_1$ by the Fourier transform over $b_1^-$, therefore resulting in a square of a delta function.

To avoid this extra delta function, we introduce slightly different in- and out states, which we use in intermediate steps of the calculation of the matching coefficients. Specifically, we temporarily change the in-state to
%%%
\begin{align}\label{eq:instate}
    \ket{p_1 p_2}
    \rightarrow
    \int \df \omega_3\,\Psi(\omega_3)\ket{p_3 p_4}\,,
\end{align}
with $p_3=\omega_3 p$ and $p_4=\omega_4 p=(1-\omega_3)p$, while keeping the out-state as $\bra{p_1 p_2}$. The above in-state replaces the unwanted extra delta function by $\Psi(\omega_1)$: 
%%%
\begin{align}\label{eq:instatetreelevel}
    {^{11}}\!F_{qq}^{(0)}(x_1,x_2)
    &=
    -\pi (1\otimes1) \,
    \frac{\gamma^+\otimes\gamma^+}{p_1^+ p_2^+} \, 
    \Psi(\omega_1) \,
    \delta\big(1 - \tfrac{x_1}{\omega_1}\big) 
    \delta\big(1 - \tfrac{x_2}{\omega_2}\big) \ .
\end{align}
%%%
In the limit that the in- and out-states are identical, i.e.~when $\Psi(\omega_3)$ is narrowly peaked around $\omega_3=\omega_1$, the above factor $\Psi(\omega_1)$ can be treated as an infinite normalization factor. Because the tree-level lightcone- and the quasi-DPD share this normalization factor, it directly drops out of the matching coefficient at this order, see \eq{C_0} below. 
Beyond tree-level care should be taken in treating $\Psi(\omega_3)$ as narrowly peaked, as we will see in our one-loop calculation in \sec{oneloopdpds}. In practice, we will set $\omega_3$ equal to $\omega_1$ whenever that is possible without generating a divergence.

%------------------------------------------------------------------------------------------------------------------
\subsection{Constructing the one-loop matching kernel} \label{sec:oneloopconstruction}
%------------------------------------------------------------------------------------------------------------------

Here we lay out the details that enter the calculation of the one-loop matching kernel in \eq{match_sec5}. We denote the perturbative expansion of the matching kernels by 
%%%
\begin{align} \label{eq:pert_exp}
    C
    &=
    C^{(0)}
    +\frac{\alpha_s}{4\pi} C^{(1)}
    +\mathcal{O}(\alpha_s^2)\,,
\end{align}
%%%
and use a similar notation for the perturbative expansion of the other objects. At tree-level, the matching kernel can be constructed from the tree-level lightcone- and quasi-DPDs, which respectively read
%%%
\begin{align} \label{eq:dpd_tree}
    {^{R_1 R_2}}\!F_{q_1 q_2}^{(0)} (x_1,x_2)
    &=
    -\pi \Psi(\omega_1) 
    \delta\bigl(1-\tfrac{x_1}{\omega_1}\bigr)
    \delta\bigl(1-\tfrac{x_2}{\omega_2}\bigr) \ 
    (T_{R_1}\otimes T_{R_2}) \ 
    \frac{\Gamma_{q_1}\otimes\Gamma_{q_2}}{p_1^+ p_2^+}\,,
    \\
    {^{R_1 R_2}}\!\tilde{F}_{q_1 q_2}^{(0)} (x_1,x_2)
    &=
    -\pi \Psi(\omega_1) 
    \delta\bigl(1-\tfrac{x_1}{\omega_1}\bigr)
    \delta\bigl(1-\tfrac{x_2}{\omega_2}\bigr) \ 
    (T_{R_1}\otimes T_{R_2}) \ 
    \frac{\tilde{\Gamma}_{q_1}\otimes\tilde{\Gamma}_{q_2}}{p_1^z \ p_2^z}\,.
\end{align}
%%%
The tree-level matching kernel is then 
%%%
\begin{align} \label{eq:C_0}
    {^{R R'}}\!C^{(0)}_{q q'}
    \Bigl(\frac{x}{\omega}, \omega\tilde{p}^z,\mu\Bigr)
    &=
    \delta_{R R'} \delta_{q q'}\, \delta\bigl(1-\tfrac{x}{\omega}\bigr)\,.
\end{align}
%%%

Using these tree-level results in the expansion of \eq{match_sec5} to one-loop order, leads to
%%%
\begin{align}
    &{^{R_1 R_2}}\!\tilde{F}^{(1)}_{q_1 q_2}
    (x_1,x_2,b_\perp,\mu,\tilde{\zeta}_p,\tilde{p}^z)
    \\
    &\quad =
    {^{R_1 R_2}}\!F^{(1)}_{q_1 q_2}(x_1,x_2,b_\perp,\mu,\zeta_p)
    \nonumber
    \\
    &\qquad
    -
    \sum_{R_1'} \sum_{q_1'}
    \pi \Psi(\omega_1)\,
    \delta\bigl(1-\tfrac{x_2}{\omega_2}\bigr)\, 
    (T_{R_1'}\otimes T_{R_2})\, 
    \frac{\Gamma_{q_1'}\otimes\Gamma_{q_2}}{p_1^+ p_2^+}\,
    {^{R_1 R_1'}}\!C^{(1)}_{q_1 q_1'}
    \Bigl(\frac{x_1}{\omega_1},\omega_1 p^z,\mu\Bigr)
    \nonumber
    \\
    &\qquad
    -
    \sum_{R_2'} \sum_{q_2'}
    \pi \Psi(\omega_1)\,
    \delta\bigl(1-\tfrac{x_1}{\omega_1}\bigr)\,
    (T_{R_1}\otimes T_{R_2'})\, 
    \frac{\Gamma_{q_1}\otimes\Gamma_{q_2'}}{p_1^+ p_2^+}\,
    {^{R_2 R_2'}}\!C^{(1)}_{q_2 q_2'}
    \Bigl(\frac{x_2}{\omega_2},\omega_2 p^z,\mu\Bigr)
    \nonumber
    \\
    &\qquad
    -\pi \Psi(\omega_1)\,
    \delta\bigl(1-\tfrac{x_1}{\omega_1}\bigr)\,
    \delta\bigl(1-\tfrac{x_2}{\omega_2}\bigr)\,
    (T_{R_1}\otimes T_{R_2})\, 
    \frac{\Gamma_{q_1}\otimes\Gamma_{q_2}}{p_1^+ p_2^+}\,
    \frac{1}{2} {^R}\!J^{(1)}(b_\perp,\mu)
        \ln\biggl(\frac{\tilde{\zeta}_p}{\zeta_p}\biggr)\,.
\nn\end{align}
%%%
The one-loop corrections to the DPDs that appear in this expression are the renormalized subtracted DPDs. Following eqs.~\eqref{eq:DPD_def_subtracted_bare}, \eqref{eq:DPD_renormalization} and \eqref{eq:quasi-DPD_renorm}, these one-loop corrections can be written in terms of the \emph{bare unsubtracted} DPDs ${^{R_1 R_2}}\!F^{\text{unsub}}_{q_1 q_2}$ and ${^{R_1 R_2}}\!\tilde{F}^{\text{unsub}}_{q_1 q_2}$ as
%%%
\begin{align}
    &{^{R_1 R_2}}\!F_{q_1 q_2}^{(1)}(x_1,x_2,b_\perp,\mu,\zeta_p)
    =
    \lim_{\epsilon\to0}
    \lim_{\delta^+\to0}
    \biggl\{
    {^{R_1 R_2}}\!F^{\text{unsub}(1)}_{q_1 q_2}
        (x_1,x_2,b_\perp,\epsilon,\delta^+,p^+)
    \\
    &\qquad\quad
    +\pi \Psi(\omega_1)\,
    \delta\bigl(1-\tfrac{x_1}{\omega_1}\bigr)\,
    \delta\bigl(1-\tfrac{x_2}{\omega_2}\bigr)\,
    (T_{R_1}\otimes T_{R_2})\,
    \frac{\Gamma_{q_1}\otimes\Gamma_{q_2}}{p_1^+ p_2^+}\,
    \frac{1}{2} {^R}\!S^{(1)}(b_\perp,\epsilon,\delta^+,e^{2y_n}\delta^+)
    \nonumber
    \\
    &\qquad\quad
    -\pi \Psi(\omega_1)\,
    \delta\bigl(1-\tfrac{x_2}{\omega_2}\bigr)\,
    (T_{R_1'}\otimes T_{R_2})\,
    \frac{\Gamma_{q_1'}\otimes\Gamma_{q_2}}{p_1^+ p_2^+}\,
    {^{R_1 \bar{R_1'}}}\!Z_{q_1 q_1'}^{(1)}
        \Bigl(\frac{x_1}{\omega_1},\mu,x_1^2\zeta_p\Bigr)
    \nonumber
    \\
    &\qquad\quad
    -\pi \Psi(\omega_1)\,
    \delta\bigl(1-\tfrac{x_1}{\omega_1}\bigr)\,
    (T_{R_1}\otimes T_{R_2'})\,
    \frac{\Gamma_{q_1}\otimes\Gamma_{q_2'}}{p_1^+ p_2^+}\,
    {^{R_2 \bar{R_2'}}}\!Z_{q_2 q_2'}^{(1)}
        \Bigl(\frac{x_2}{\omega_2},\mu,x_2^2\zeta_p\Bigr)
    \biggr\}\,,
    \nonumber
    \\[2ex]
    &{^{R_1 R_2}}\!\tilde{F}_{q_1 q_2}^{(1)}
    (x_1,x_2,b_\perp,\mu,\tilde{\zeta}_p,\tilde{p}^z)
    =
    \lim_{\tilde{\eta}\to\infty}\lim_{\epsilon\to0}
    \biggl\{
    {^{R_1 R_2}}\!\tilde{F}^{\text{unsub}(1)}_{q_1 q_2}
        (x_1,x_2,b_\perp,\epsilon,\tilde{\eta},\tilde{p}^z)
    \\
    &\qquad\quad
    +\pi \Psi(\omega_1)\,
    \delta\bigl(1-\tfrac{x_1}{\omega_1}\bigr)\,
    \delta\bigl(1-\tfrac{x_2}{\omega_2}\bigr)\, 
    (T_{R_1}\otimes T_{R_2})\,
    \frac{\tilde{\Gamma}_{q_1}\otimes\tilde{\Gamma}_{q_2}}
        {\tilde{p}_1^z \ \tilde{p}_2^z}\,
    \frac{1}{2}{^R}\!\tilde{S}^{(1)}(b_\perp,\epsilon,\tilde{\eta},y_A,y_B)
    \nonumber
    \\
    &\qquad\quad
    -\pi \Psi(\omega_1)\,
    \delta\bigl(1-\tfrac{x_1}{\omega_1}\bigr)\,
    \delta\bigl(1-\tfrac{x_2}{\omega_2}\bigr)\, 
    (T_{R_1}\otimes T_{R_2})\,
    \frac{\tilde{\Gamma}_{q_1}\otimes\tilde{\Gamma}_{q_2}}
        {\tilde{p}_1^z \ \tilde{p}_2^z}\,
    {^{R_1 R_2}}\!\tilde Z_{q_1 q_2}^{(1)}(\epsilon,\mu,y_A,y_B)
    \biggr\}\,.
    \nonumber
\end{align}
%%%
Here we switched from the off-lightcone regulator in \eq{offlightcone_reg} to the delta regulator in \eq{delta_reg}.

\begin{figure}[t!]
\begin{fmffile}{oneloopcorrections1}
\begin{align*}
&
\begin{gathered}
\begin{fmfgraph*}(110,65)
%Setting up the grid
\thirteengrid
%drawing the actual diagram
\fmf{fermion,tension=0}{a3,e3}
\fmf{fermion,tension=0}{a11,d11}
\fmf{plain,tension=0}{d11,e11}
\fmf{fermion,tension=0}{i3,m3}
\fmf{fermion,tension=0}{j11,m11}
\fmf{plain,tension=0}{i11,j11}
\fmf{gluon,tension=0,right}{d11,j11}
\fmfdot{e3}\fmfdot{e11}\fmfdot{i3}\fmfdot{i11}
\fmflabel{$0^-,\mathbf{0}_\perp$}{e3}
\fmflabel{$b_3^-,\mathbf{b}_\perp$}{e11}
\fmflabel{$b_2^-,\mathbf{0}_\perp$}{i3}
\fmflabel{$b_1^-,\mathbf{b}_\perp$}{i11}
\fmffreeze
\fmfcmd{style_def karrow expr p = drawarrow subpath (45/100, 60/100) of p shifted 6 up
        withpen pencircle scaled 0.4; label.lrt(btex $k$ etex, point 0.48 of p
        shifted 4 up); enddef;}
\fmfcmd{style_def parrow expr p = drawarrow subpath (6/10, 9/10) of p shifted 6 down
        withpen pencircle scaled 0.4; label.bot(btex $p_1$ etex, point 0.8 of p
        shifted 6 down); enddef;}
\fmfcmd{style_def qarrow expr p = drawarrow subpath (6/10, 9/10) of p shifted 6 up
        withpen pencircle scaled 0.4; label.top(btex $p_2$ etex, point 0.8 of p
        shifted 6 up); enddef;}
\fmf{karrow}{a3,m3}
\fmf{parrow}{i11,m11}
\fmf{qarrow}{i3,m3}
\fmflabel{$(A)$}{g1}
\end{fmfgraph*}
\end{gathered}
\qquad\qquad
\begin{gathered}
\begin{fmfgraph*}(110,65)
%Setting up the grid
\thirteengrid
%drawing the actual diagram
\fmf{fermion,tension=0}{a3,e3}
\fmf{fermion,tension=0}{a11,e11}
\fmf{fermion,tension=0}{i3,m3}
\fmf{fermion,tension=0}{k11,m11}
\fmf{plain,tension=0}{i11,k11}
\fmf{double,tension=0}{e3,g3}
\fmf{gluon,tension=0,right}{k3,f3}
\fmfdot{e3}\fmfdot{e11}\fmfdot{i3}\fmfdot{i11}
\fmflabel{$(B)$}{g1}
\end{fmfgraph*}
\end{gathered}
\qquad\qquad
\begin{gathered}
\begin{fmfgraph*}(110,65)
%Setting up the grid
\thirteengrid
%drawing the actual diagram
\fmf{fermion,tension=0}{a3,c3}
\fmf{fermion,tension=0}{a11,c11}
\fmf{fermion,tension=0}{k3,m3}
\fmf{fermion,tension=0}{k11,m11}
\fmf{double,tension=0}{c3,e3}
\fmf{double,tension=0}{i3,k3}
\fmf{gluon,tension=0,right}{i3,e3}
\fmfdot{c3}\fmfdot{c11}\fmfdot{k3}\fmfdot{k11}
\fmflabel{$(C)$}{g1}
\end{fmfgraph*}
\end{gathered}
\\
&
\\
&
\begin{gathered}
\begin{fmfgraph*}(110,65)
%Setting up the grid
\thirteengrid
%drawing the actual diagram
\fmf{fermion,tension=0}{a3,c3}
\fmf{fermion,tension=0}{a11,c11}
\fmf{fermion,tension=0}{g3,m3}
\fmf{fermion,tension=0}{g11,m11}
\fmf{gluon,tension=0,right}{h11,l11}
\fmfdot{c3}\fmfdot{c11}\fmfdot{g3}\fmfdot{g11}
\fmflabel{$(D)$}{g1}
\end{fmfgraph*}
\end{gathered}
\qquad\qquad
\begin{gathered}
\begin{fmfgraph*}(110,65)
%Setting up the grid
\thirteengrid
%drawing the actual diagram
\fmf{fermion,tension=0}{a3,e3}
\fmf{fermion,tension=0}{a11,e11}
\fmf{fermion,tension=0}{i3,m3}
\fmf{fermion,tension=0}{k11,m11}
\fmf{plain,tension=0}{i11,k11}
\fmf{double,tension=0}{g11,i11}
\fmf{gluon,tension=0,right}{h11,k11}
\fmfdot{e3}\fmfdot{e11}\fmfdot{i3}\fmfdot{i11}
\fmflabel{$(E)$}{g1}
\end{fmfgraph*}
\end{gathered}
\qquad\qquad
\begin{gathered}
\begin{fmfgraph*}(110,65)
%Setting up the grid
\thirteengrid
%drawing the actual diagram
\fmf{fermion,tension=0}{a3,c3}
\fmf{fermion,tension=0}{a11,c11}
\fmf{fermion,tension=0}{k3,m3}
\fmf{fermion,tension=0}{k11,m11}
\fmf{double,tension=0}{g11,k11}
\fmf{gluon,tension=0,right}{g11,j11}
\fmfdot{c3}\fmfdot{c11}\fmfdot{k3}\fmfdot{k11}
\fmflabel{$(F)$}{g1}
\end{fmfgraph*}
\end{gathered}
\end{align*}
\end{fmffile}
\caption{The one-loop corrections to the quark-quark quasi-DPD that involve only a single quark line. For the corresponding lightcone diagrams, a cut should be inserted vertically in the middle of the diagram. The top and bottom row have different color factors for the color-correlated DPD, and consequently require a rapidity regulator.}
\label{fig:quasidpddiagrams1}
\end{figure}
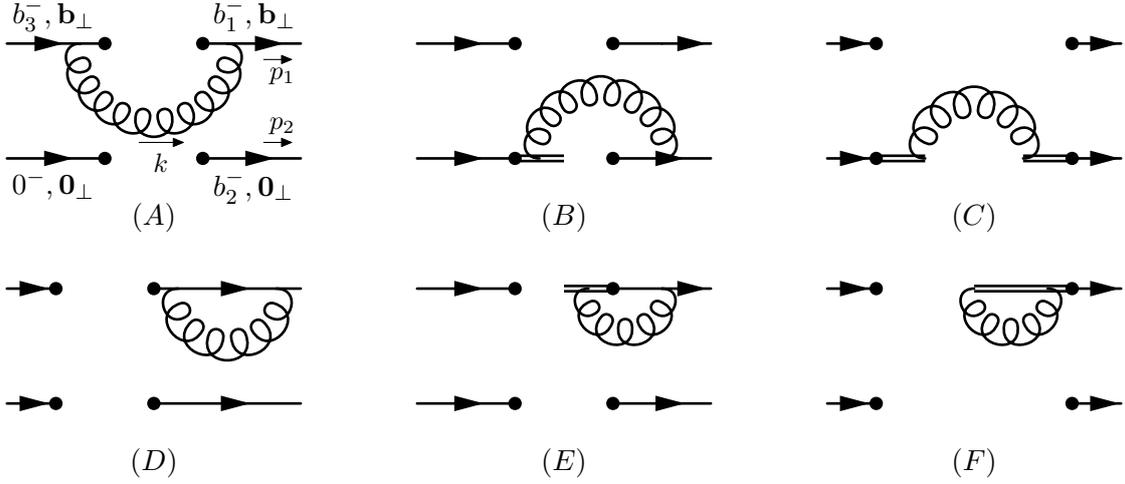

\begin{figure}
\begin{fmffile}{oneloopcorrections2}
\begin{align*}
&
\begin{gathered}
\begin{fmfgraph*}(110,65)
%Setting up the grid
\thirteengrid
%drawing the actual diagram
\fmf{fermion,tension=0}{a3,c3,e3}
\fmf{fermion,tension=0}{a11,e11}
\fmf{fermion,tension=0}{i3,m3}
\fmf{fermion,tension=0}{i11,k11,m11}
\fmf{gluon,tension=0}{c3,k11}
\fmfdot{e3}\fmfdot{e11}\fmfdot{i3}\fmfdot{i11}
\fmflabel{$0^-,\mathbf{0}_\perp$}{e3}
\fmflabel{$b_3^-,\mathbf{b}_\perp$}{e11}
\fmflabel{$b_2^-,\mathbf{0}_\perp$}{i3}
\fmflabel{$b_1^-,\mathbf{b}_\perp$}{i11}
\fmffreeze
\fmfcmd{style_def karrow expr p = drawarrow subpath (5/10, 7/10) of p shifted 12 down
        withpen pencircle scaled 0.4; label.lrt(btex $k$ etex, point 0.6 of p
        shifted 12 down); enddef;}
\fmfcmd{style_def parrow expr p = drawarrow subpath (6/10, 9/10) of p shifted 6 down
        withpen pencircle scaled 0.4; label.bot(btex $p_1$ etex, point 0.8 of p
        shifted 6 down); enddef;}
\fmfcmd{style_def qarrow expr p = drawarrow subpath (6/10, 9/10) of p shifted 6 up
        withpen pencircle scaled 0.4; label.top(btex $p_2$ etex, point 0.8 of p
        shifted 6 up); enddef;}
\fmf{karrow}{e5,k11}
\fmf{parrow}{i11,m11}
\fmf{qarrow}{i3,m3}
\fmflabel{$(G)$}{g1}
\end{fmfgraph*}
\end{gathered}
\qquad\qquad
\begin{gathered}
\begin{fmfgraph*}(110,65)
%Setting up the grid
\thirteengrid
%drawing the actual diagram
\fmf{fermion,tension=0}{a11,e11}
\fmf{fermion,tension=0}{a3,e3}
\fmf{fermion,tension=0}{k11,m11}
\fmf{fermion,tension=0}{i3,m3}
\fmf{plain,tension=0}{i11,k11}
\fmf{double,tension=0}{e3,g3}
\fmf{gluon,tension=0}{f3,k11}
\fmfdot{e3}\fmfdot{e11}\fmfdot{i3}\fmfdot{i11}
\fmflabel{$(H)$}{g1}
\end{fmfgraph*}
\end{gathered}
\qquad\qquad
\begin{gathered}
\begin{fmfgraph*}(110,65)
%Setting up the grid
\thirteengrid
%drawing the actual diagram
\fmf{fermion,tension=0}{a3,c3}
\fmf{fermion,tension=0}{a11,c11}
\fmf{fermion,tension=0}{k3,m3}
\fmf{fermion,tension=0}{k11,m11}
\fmf{double,tension=0}{c3,e3}
\fmf{double,tension=0}{i11,k11}
\fmf{gluon,tension=0}{e3,i11}
\fmfdot{c3}\fmfdot{c11}\fmfdot{k3}\fmfdot{k11}
\fmflabel{$(I)$}{g1}
\end{fmfgraph*}
\end{gathered}
\\
&
\\
&
\begin{gathered}
\begin{fmfgraph*}(110,65)
%Setting up the grid
\thirteengrid
%drawing the actual diagram
\fmf{fermion,tension=0}{a3,e3}
\fmf{fermion,tension=0}{a11,e11}
\fmf{fermion,tension=0}{i3,k3,m3}
\fmf{fermion,tension=0}{i11,k11,m11}
\fmf{gluon,tension=0}{k3,k11}
\fmfdot{e3}\fmfdot{e11}\fmfdot{i3}\fmfdot{i11}
\fmflabel{$(J)$}{g1}
\end{fmfgraph*}
\end{gathered}
\qquad\qquad
\begin{gathered}
\begin{fmfgraph*}(110,65)
%Setting up the grid
\thirteengrid
%drawing the actual diagram
\fmf{fermion,tension=0}{a11,e11}
\fmf{fermion,tension=0}{a3,e3}
\fmf{fermion,tension=0}{k11,m11}
\fmf{fermion,tension=0}{i3,m3}
\fmf{plain,tension=0}{i11,k11}
\fmf{double,tension=0}{g3,i3}
\fmf{gluon,tension=0}{h3,k11}
\fmfdot{e3}\fmfdot{e11}\fmfdot{i3}\fmfdot{i11}
\fmflabel{$(K)$}{g1}
\end{fmfgraph*}
\end{gathered}
\qquad\qquad
\begin{gathered}
\begin{fmfgraph*}(110,65)
%Setting up the grid
\thirteengrid
%drawing the actual diagram
\fmf{fermion,tension=0}{a3,c3}
\fmf{fermion,tension=0}{a11,c11}
\fmf{fermion,tension=0}{k3,m3}
\fmf{fermion,tension=0}{k11,m11}
\fmf{double,tension=0}{i3,k3}
\fmf{double,tension=0}{i11,k11}
\fmf{gluon,tension=0}{i11,i3}
\fmfdot{c3}\fmfdot{c11}\fmfdot{k3}\fmfdot{k11}
\fmflabel{$(L)$}{g1}
\end{fmfgraph*}
\end{gathered}
\end{align*}
\end{fmffile}\noindent
\caption{The one-loop corrections to the quark-quark quasi-DPD that involve a gluon exchange between the two quark lines. For the corresponding lightcone diagrams, a cut should be inserted vertically in the middle of the diagram.}
\label{fig:quasidpddiagrams2}
\end{figure}
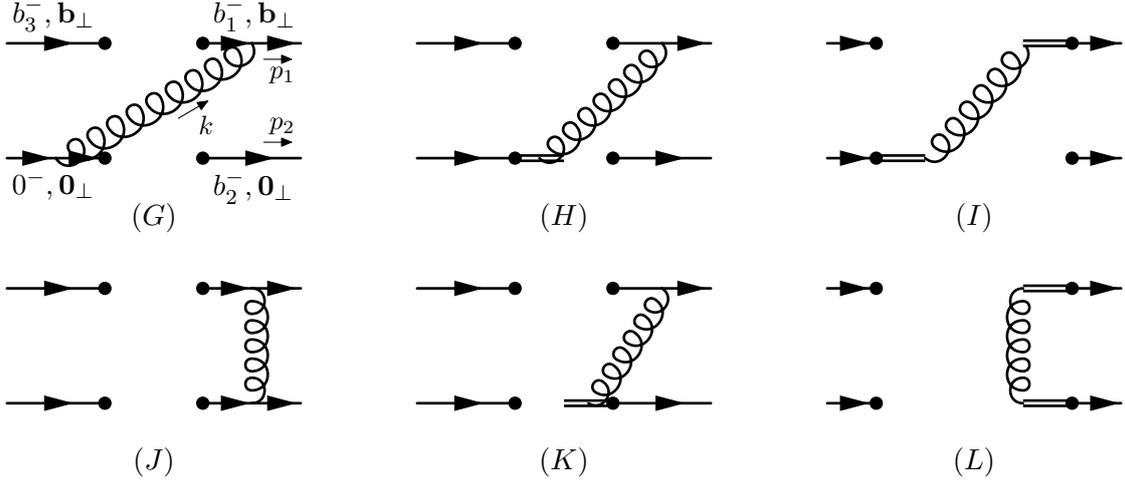

The diagrams for the one-loop corrections to the bare unsubtracted DPDs are shown in figures \ref{fig:quasidpddiagrams1} and \ref{fig:quasidpddiagrams2}. For simplicity, we only show the diagrams belonging to the quasi-DPD, as those for the lightcone-DPD look identical up to a vertical cut through the middle of the diagram due to the time-ordering prescription in \eq{dpddefinition}. For identical quark flavors one could also consider ``crossing" quark lines, but this turns out to be irrelevant.
The one-loop corrections can be classified as diagrams where only a single quark line is involved (\fig{quasidpddiagrams2}) and diagrams where a gluon connects the two quark lines (\fig{quasidpddiagrams1}). This classification is useful, as the diagrams in the first category are identical to the one-loop corrections to regular PDFs, up to an overall normalization and the presence of the other quark line. Note however that these diagrams must be calculated with a rapidity regulator, as the diagrams in the top and bottom row receive different color factors for the color-correlated DPD, preventing the cancellation of rapidity divergences. The diagrams shown do not include the contributions of the transverse Wilson that are necessary to ensure gauge invariance of the color-correlated quasi-DPD (see \eq{quasi_DPD_unsub_cc} vs.~\eq{quasi_DPD_unsub_cc_gi}). However, we have explicitly verified that at one-loop order the contribution of the transverse Wilson line cancels between the unsubtracted quasi-DPD and the quasi-soft function.

It is useful to separate the color structure from the rest of the diagram such that the resulting expressions can be used to calculate both the color-summed and color-correlated DPDs:
%%%
\begin{align} \label{eq:oneloop_decomp}
    {^{R_1 R_2}}\!F^{\text{unsub}\,(1)}_{q_1 q_2}
    &=
    \sum_{i\in\text{diagrams}} \sum_{R_1',R_2'} \sum_{q_1',q_2'}
    {^{R_1 R_2, R_1' R_2'}}\!c^i
    \bigl[F_{q_1 q_2}^i\bigr]_{q_1'q_2'} \ 
    (T_{R_1'}\otimes T_{R_2'})
    \frac{\Gamma_{q_1'} \otimes \Gamma_{q_2'}}{p_1^+ p_2^+}\,.
\end{align}
%%%
Here $F^i_{q_1 q_2}$ is the expression for diagram $i$ contributing to ${^{R_1 R_2}}\!F_{q_1 q_2}^{(1)}$ with color factors absorbed into ${^{R_1 R_2, R_1' R_2'}}\!c^i$. Conveniently, the diagrams in figures \ref{fig:quasidpddiagrams1} and \ref{fig:quasidpddiagrams2} are ordered such that diagrams in each row have an identical color structure. We furthermore use $\bigl[F_{q_1 q_2}^i\bigr]_{q_1'q_2'}$ to denote its contribution to the spin structure $\Gamma_{q_1'} \otimes \Gamma_{q_2'}$. A similar expression holds for the quasi-DPDs $\tilde F$, with the appropriate replacement of $\Ga$ and $p_i^+$, see \eq{quasigamma}.

Before writing down an expression for the one-loop matching kernel, let us discuss how the divergent factor $\Psi(\omega_1)$ drops out of the matching kernel. The individual expressions for the one-loop lightcone- and quasi-DPDs contain non-trivial integrals involving $\Psi(\omega_3)$. However, in constructing the matching kernel, these terms cancel between the lightcone- and the quasi-DPDs. Even at the level of individual diagrams, the difference between the one-loop corrections to the lightcone- and quasi-DPDs factors as
%%%
\begin{align} \label{eq:Delta_def}
    &\lim_{p_3,p_4\to p_1,p_2}
    \bigl[\tilde{F}_{q_1 q_2}^i\bigr]_{q_1' q_2'}
        (x_1,x_2,b_\perp,\epsilon,\tilde{\eta},\tilde{p}^z)
    -\bigl[F_{q_1 q_2}^i\bigr]_{q_1' q_2'}
        (x_1,x_2,b_\perp,\epsilon,\delta^+,p^+)
    \\
    &\quad\qquad=
    -\pi \Psi(\omega_1) \bigl[\Delta_{q_1 q_2}^i\bigr]_{q_1' q_2'}
    (x_1,x_2,b_\perp,\epsilon,\{\tilde{\eta},\tilde{p}^z\},\{\delta^+,p^+\})\,,
    \nonumber
\end{align}
%%%
such that the divergent factor $\Psi(\omega_1)$ drops out of the matching kernel.

Combining all the perturbative expansions and organizing by color and spin structures, the one-loop matching kernels for the flavor non-singlet case can be expressed as 
%%%
\begin{align} \label{eq:oneloop_final}
    &\delta_{R_2 R_2'} \delta_{q_2 q_2'}
    \delta\bigl(1-\tfrac{x_2}{\omega_2}\bigr)
    {^{R_1 R_1'}}\!C^{(1)}_{q_1 q_1'}
    \Bigl(\frac{x_1}{\omega_1},\omega_1\tilde{p}^z,\mu\Bigr)\,
    +(1\leftrightarrow2)
    \\
    &\quad
    =
    \lim_{\tilde{\eta}\to\infty}\lim_{\epsilon\to0}\lim_{\delta^+\to0}
    \biggl\{\sum_i 
    {^{R_1 R_2, R_1' R_2'}}\!c^i
    \bigl[\Delta_{q_1 q_2}^i\bigr]_{q_1'q_2'}
        (x_1,x_2,b_\perp,\epsilon,\{\tilde{\eta},\tilde{p}^z\},\{\delta^+,p^+\})
    \nonumber
    \\
    &\qquad\quad
    +\frac{1}{2}\delta_{R_1 R_1'} \delta_{R_2 R_2'}
    \delta_{q_1 q_1'} \delta_{q_2 q_2'}
    \delta\bigl(1-\tfrac{x_1}{\omega_1}\bigr)
    \delta\bigl(1-\tfrac{x_2}{\omega_2}\bigr)
    \nonumber
    \\
    &\qquad\qquad\times
    \biggl[{^R}\!S^{(1)}(b_\perp,\epsilon,\delta^+,e^{2y_n}\delta^+)
        -{^R}\!\tilde{S}^{(1)}(b_\perp,\epsilon,\tilde{\eta},y_A,y_B)
        -{^R}\!J^{(1)}(b_\perp,\mu)\ln\biggl(\frac{\tilde{\zeta}_p}{\zeta_p}\biggr)
        \biggr]
    \nonumber
    \\
    &\qquad\quad
    +\delta_{R_1 R_1'} \delta_{R_2 R_2'}
    \delta_{q_1 q_1'} \delta_{q_2 q_2'}
    \delta\bigl(1-\tfrac{x_1}{\omega_1}\bigr)
    \delta\bigl(1-\tfrac{x_2}{\omega_2}\bigr)
    {^{R_1 R_2}}\!\tilde Z_{q_1 q_2}^{(1)}(\epsilon,\mu,y_A,y_B)
    \nonumber
    \\
    &\qquad\quad
    -\delta_{R_2 R_2'} \delta_{q_2 q_2'}
    {^{R_1 \bar{R_1'}}}\!Z_{q_1 q_1'}^{(1)}
        \Bigl(\frac{x_1}{\omega_1},\mu,x_1^2\zeta_p\Bigr)\,
    \delta\bigl(1-\tfrac{x_2}{\omega_2}\bigr)
    \nonumber
    \\
    &\qquad\quad
    -\delta_{R_1 R_1'} \delta_{q_1 q_1'}
    {^{R_2 \bar{R_2'}}}\!Z_{q_2 q_2'}^{(1)}
        \Bigl(\frac{x_2}{\omega_2},\mu,x_2^2\zeta_p\Bigr)\,
    \delta\bigl(1-\tfrac{x_1}{\omega_1}\bigr)\biggr\}\,.
    \nonumber
\end{align}
%%%
The order of the $\tilde \eta$ and $\eps$ limits does not matter~\cite{Ebert:2022fmh}.

%------------------------------------------------------------------------------------------------------------------
\subsection{Lightcone- and quasi-DPDs at one-loop} \label{sec:oneloopdpds}
%------------------------------------------------------------------------------------------------------------------

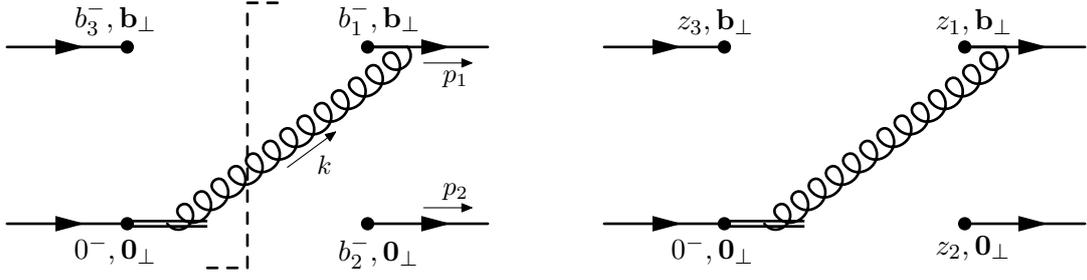
\begin{figure}[t!]
\begin{fmffile}{dpddiagrame}
\begin{align*}
\begin{gathered}
\begin{fmfgraph*}(180,100)
%Setting up the grid
\thirteengrid
%drawing the actual diagram
\fmf{fermion,tension=0}{a3,d3}
\fmf{fermion,tension=0}{a11,d11}
\fmf{fermion,tension=0}{j3,m3}
\fmf{fermion,tension=0}{j11,m11}
\fmf{double,tension=0}{d3,f3}
\fmf{plain,tension=0}{j11,k11}
\fmf{gluon,tension=0}{e3,k11}
\fmfdot{d3}\fmfdot{d11}\fmfdot{j3}\fmfdot{j11}
\fmflabel{$0^-,\mathbf{0}_\perp$}{e3}
\fmflabel{$b_3^-,\mathbf{b}_\perp$}{e11}
\fmflabel{$b_2^-,\mathbf{0}_\perp$}{i3}
\fmflabel{$b_1^-,\mathbf{b}_\perp$}{i11}
\fmf{dashes,tension=0}{f1,g1,g13,h13}
\fmffreeze
\fmfcmd{style_def karrow expr p = drawarrow subpath (5/10, 7/10) of p shifted 12 down
        withpen pencircle scaled 0.4; label.lrt(btex $k$ etex, point 0.6 of p
        shifted 12 down); enddef;}
\fmfcmd{style_def parrow expr p = drawarrow subpath (6/10, 9/10) of p shifted 6 down
        withpen pencircle scaled 0.4; label.bot(btex $p_1$ etex, point 0.8 of p
        shifted 6 down); enddef;}
\fmfcmd{style_def qarrow expr p = drawarrow subpath (6/10, 9/10) of p shifted 6 up
        withpen pencircle scaled 0.4; label.top(btex $p_2$ etex, point 0.8 of p
        shifted 6 up); enddef;}
\fmf{karrow}{e3,k11}
\fmf{parrow}{i11,m11}
\fmf{qarrow}{i3,m3}
\end{fmfgraph*}
\end{gathered}
\qquad\qquad
\begin{gathered}
\begin{fmfgraph*}(180,100)
%Setting up the grid
\thirteengrid
%drawing the actual diagram
\fmf{fermion,tension=0}{a3,d3}
\fmf{fermion,tension=0}{a11,d11}
\fmf{fermion,tension=0}{j3,m3}
\fmf{fermion,tension=0}{j11,m11}
\fmf{double,tension=0}{d3,f3}
\fmf{plain,tension=0}{j11,k11}
\fmf{gluon,tension=0}{e3,k11}
\fmfdot{d3}\fmfdot{d11}\fmfdot{j3}\fmfdot{j11}
\fmflabel{$0^-,\mathbf{0}_\perp$}{e3}
\fmflabel{$z_3,\mathbf{b}_\perp$}{e11}
\fmflabel{$z_2,\mathbf{0}_\perp$}{i3}
\fmflabel{$z_1,\mathbf{b}_\perp$}{i11}
\end{fmfgraph*}
\end{gathered}
\end{align*}
\end{fmffile}\noindent
\centering
\caption{Diagram $H$ for the lightcone- (left) and quasi-DPD (right).}
\label{fig:dpddiagramb}
\end{figure}

Here we will provide an example calculation of one of the diagrams that contributes to the one-loop lightcone- and quasi-DPDs. A complete overview of the results of all diagrams is given in \app{diagrams}. The diagram we will calculate here is diagram $H$, also shown in \fig{dpddiagramb}. This diagram exhibits most features that distinguish DPDs from PDFs and TMDs, as it involves both quark lines.

First we calculate the lightcone diagram, shown on the left in \fig{dpddiagramb}. To keep the notation compact, we will use the shorthand for the graph labelled with $n$ 
%%%
\begin{align} \label{eq:Fn_in_calc}
   F^n=
    \frac{\alpha_s}{4\pi}
    \sum_{q_1',q_2'}
    [F^n_{q_1 q_2}]_{q_1'q_2'} \frac{\Gamma_{q_1'} \otimes \Gamma_{q_2'}}{p_1^+ p_2^+}\,,
\end{align}
%%%
and $\tilde F^n$ for the corresponding quasi-DPDs.  From the Feynman rules one can derive the following expression,
%%%
\begin{align} \label{eq:diagram_H_explicit}
    F^H
    &=
    -\pi p^+ \mu_0^{2\epsilon}
    \int\!\df\omega_3\, \Psi(\omega_3)
    \int\!\frac{\df^d k}{(2\pi)^d}\,
    e^{\img \mathbf{k}_\perp \cdot \mathbf{b}_\perp}\,
    \biggl[(\img g \gamma^\mu) \frac{\img}{\fsl{p}_1 - \fsl{k}} \Gamma_{q_1}
    \otimes
    \Gamma_{q_2} (\img g n_a^\nu) \frac{\img}{k^+ + \img \delta^+}\biggr]
    \nonumber\\
    &\qquad\times
    (-g_{\mu\nu})\,2\pi\delta(k^2)\theta(k^0)\,
    \delta\bigl[x_1 p^+ - (p_1^+ - k^+)\bigr]\,
    \delta\bigl[x_2 p^+ - p_2^+\bigr]\,
    \delta\bigl[x_1 p^+ - p_3^+\bigr]
    \nonumber\\
    &=
    -\alpha_s \Psi(x_1)\,
    \delta\bigl(1-\tfrac{x_2}{\omega_2}\bigr)\,
    \frac{\gamma^+ \gamma_\mu \Gamma_{q_1}\otimes\Gamma_{q_2}}{p_1^+ p_2^+}\,
    \frac{1}{(\omega_1-x_1)p^+ + \img \delta^+}
    \nonumber\\
    &\qquad\times
    4\pi^3 \mu_0^{2\epsilon}
    \int\! \frac{\df^d k}{(2\pi)^d}\,
    e^{\img \mathbf{k}_\perp \cdot \mathbf{b}_\perp}\,
    \frac{(p_1-k)^\mu}{k^-}\,
    \delta(k^2) \theta(k^0)\,
    \delta\bigl[k^+ - (\omega_1-x_1)p^+\bigr]\,,
\end{align}
%%%
where $\mu_0$ is the scale associated with dimensional regularization. The remaining momentum integral in \eq{diagram_H_explicit} can be simplified by considering the contribution of each component of $(p_1-k)^\mu$ separately: First, since $(\gamma^+)^2=0$, the contribution of $(p_1-k)^-$ vanishes. Second, since $\gamma_\perp^\mu$ can be anti-commuted through $\gamma^+$, and because $\gamma^+ \Gamma=0$ for all $\Gamma\in\{\gamma^+,\gamma^+\gamma^5,\gamma^+ \gamma^\perp \gamma^5\}$ the contribution of $(p_1-k)_\perp^\mu$ vanishes as well. This leaves us only with the combination $\gamma^- (p_1-k)^+ = x_1 p^+ \gamma^-$, which gives
%%%
\begin{align}
    F^H
    &=
    -\frac{\al_s}{4\pi}\,\Psi(x_1)\,
    \delta\big(1-\tfrac{x_2}{\omega_2}\big)\,
    \frac{\Gamma_{q_1}\otimes\Gamma_{q_2}}{p_1^+ p_2^+}\,
    \frac{2x_1}{(\omega_1-x_1) + \img \delta^+/p^+}
    \\
    &\qquad\times
    (4\pi^2 \mu_0^2)^\epsilon
    \int\df^d k\, 
    e^{\img \mathbf{k}_\perp \cdot \mathbf{b}_\perp}\,
    \frac{1}{k^-}\,
    \delta(k^2) \theta(k^0)\,
    \delta\big[k^+ - (\omega_1-x_1)p^+\big]
    \,.\nn
\end{align}
%%%
The remaining integral is given in \eq{int_for_fig_e_lc}, resulting in
%%%
\begin{align}
    F^H
    &=
    -\frac{\al_s}{4\pi}\,\pi \Psi(x_1)\,
    \delta\big(1-\tfrac{x_2}{\omega_2}\big)\,
    \frac{\Gamma_{q_1}\otimes\Gamma_{q_2}}{p_1^+ p_2^+}\,
    \frac{2x_1}{(\omega_1-x_1) + \img\delta^+/p^+}
    \frac{2\sqrt{\pi} \Gamma(-2\epsilon_\text{ir})}
        {\Gamma\big(\tfrac{1}{2} - \epsilon\big)}
    (4\pi \mu_0^2 \mathbf{b}_\perp^2)^\epsilon\,,
\end{align}
where $d=4-2\epsilon_\text{ir/uv}$ and we only sometimes add the subscript ``ir'' or ``uv'' to indicate the origin of the poles. Finally, introducing a plus distribution using the convention in \app{conventions}, and expanding in $\epsilon$ and dropping terms of $\mathcal{O}(\epsilon)$
%%%
\begin{align}
    F^H
    &=
    \frac{\al_s}{4\pi}\,
    2\pi\Psi(x_1)\delta\big(1-\tfrac{x_2}{\omega_2}\big) \ 
    \frac{\Gamma_{q_1}\otimes\Gamma_{q_2}}{p_1^+ p_2^+}
    \bigg[\frac{1}{\epsilon_\text{ir}} 
        +\ln\bigg(\frac{\mu^2 \mathbf{b}_\perp^2}{b_0^2}\bigg)\bigg] \ 
    \\
    &\qquad\times
    \bigg\{
    \biggl[\frac{\frac{x_1}{\omega_1}}{1-\frac{x_1}{\omega_1}}\biggr]^{[0,1]}_+
    -
    \delta\big(1-\tfrac{x_1}{\omega_1}\big)
    \bigg[1\!+\!\ln\bigg(\frac{\delta^+}{p_1^+}\bigg)\!+\!\frac{\img \pi}{2}\bigg]
    \bigg\},
    \nonumber
\,.\end{align}
%%%
Here the ratio $x_1/\omega_1$ is the variable of the plus distribution, i.e.~integrating it over the interval $[0,1]$ will give zero. The $\mu_0$ is related to the $\overline{\text{MS}}$-scale $\mu$ and the $b_0$ is given by
%%%
\begin{align} \label{eq:mu0_b0}
\mu_0^2 = \mu^2\, \frac{e^{\ga_E}}{4\pi}
\,,\qquad
b_0 = 2 e^{-\ga_E}
\,.\end{align}
%%%

Next we calculate diagram $H$ for the quasi-DPD. This matrix element can be treated as time-ordered and therefore does not include a cut gluon propagator. From the Feynman rules we derive 
%%%
\begin{align}
    \tilde{F}^H
    &=
    -\pi p^z \mu_0^{2\epsilon}
    \int\!\df\omega_3\,\Psi(\omega_3)
    \int\!\frac{\df^d k}{(2\pi)^d}\,
    e^{\img\mathbf{k}_\perp \cdot \mathbf{b}_\perp}\,
    \biggl[(\img g \gamma^\mu) \frac{\img}{\fsl{p}_1 - \fsl{k}} \tilde{\Gamma}_{q_1}
    \otimes\tilde{\Gamma}_{q_2} (\img g \hat z^\nu) \frac{\img}{k^z}\biggr]
    \nonumber\\
    &\qquad\times
    \frac{-\img g_{\mu\nu}}{k^2}
    \bigl(e^{-\img k^z \tilde \eta} - 1\bigr)\,
    \delta\bigl[x_1 p^z - (p_1^z - k^z)\bigr]\,
    \delta\bigl[x_2 p^z - p_2^z\bigr]\,
    \delta\bigl[x_1 p^z - p_3^z\bigr]
    \nonumber\\
    &=
    \alpha_s \Psi(x_1)\,
    \delta\big(1-\tfrac{x_2}{\omega_2}\big)\,
    \frac{\gamma^z\gamma_\mu\tilde{\Gamma}_{q_1}\otimes\tilde{\Gamma}_{q_2}}
        {p_1^z p_2^z}
    \\
    &\qquad\times
    4\img \pi^2 p_1^z \mu_0^{2\epsilon}
    \int\!\frac{\df^d k}{(2\pi)^d}\,
    e^{\img \mathbf{k}_\perp \cdot \mathbf{b}_\perp}\,
    \frac{(p_1-k)^\mu}{k^2 (p_1-k)^2}\,
    \frac{1-e^{-\img k^z \tilde \eta}}{k^z}\,
    \delta\big[k^z - (\omega_1-x_1)p^z\big]\,.
    \nonumber
\end{align}
%%%
Note that  $\hat z \cdot p = - p^z$ due to the signature of our metric.
To simplify the above expression, we use that for an on-shell quark spinor $u(p)$ with $p=(p^z,0,0,p^z)$ we have $\gamma^0 u(p)=\gamma^z u(p)$. Additionally, we can ignore the term $(p_1-k)_\perp^\mu$ as its contribution is power-suppressed by $1/(|\mathbf{b}_\perp| p^z)$. This leads to
%%%
\begin{align}
    \tilde{F}^H
    &=
    \frac{\al_s}{4\pi}\,
    16\img \pi^3 \Psi(x_1)\,
    \delta\big(1-\tfrac{x_2}{\omega_2}\big) \ 
    \frac{\tilde{\Gamma}_{q_1}\otimes\tilde{\Gamma}_{q_2}}{p_2^z}
    \\
    &\qquad\times
    \mu_0^{2\epsilon}
    \int\! \frac{\df^d k}{(2\pi)^d} \ 
    e^{\img \mathbf{k}_\perp \cdot \mathbf{b}_\perp} \, 
    \frac{(p_1-k)^0 + (p_1-k)^z}{k^2 (p_1-k)^2}\, 
    \frac{1-e^{-\img k^z \tilde \eta}}{k^z}\,
    \delta\big[k^z - (\omega_1-x_1)p^z\big] \ .
    \nonumber
\end{align}
%%%

We find it convenient to already introduce a plus distribution at this stage of the calculation. That way we may omit the regulator $\tilde{\eta}$ inside the plus distribution, as the behavior of the function at $x_1=\omega_1$ is contained in the delta-function term. To calculate the resulting momentum integrals we combine denominators by introducing a Feynman parameter $v$ and use \eq{int_for_fig_e_q},
%%%
\begin{align}
    \tilde{F}^H
    &=
    \frac{\al_s}{4\pi}\biggl\{
    - \Psi(x_1)\,
    \delta\big(1-\tfrac{x_2}{\omega_2}\big) \ 
    \frac{\tilde{\Gamma}_{q_1}\otimes\tilde{\Gamma}_{q_2}}{p_1^z \ p_2^z}
    \\
    &\qquad\times
    \bigg(\frac{\mu_0^2}{p_{1z}^2}\bigg)^\epsilon
    \bigg[
    \frac{1}{1-\frac{x_1}{\omega_1}}
    \int_0^1\! \df v \ \big(v+\tfrac{x_1}{\omega_1}\big)
    \bigg(\frac{2\pi|\mathbf{b}_\perp| p_1^z}{|v-\frac{x_1}{\omega_1}|}\bigg)^{\frac{1}{2}+\epsilon} 
    K_{\frac{1}{2}+\epsilon}\big(|v-\tfrac{x_1}{\omega_1}||\mathbf{b}_\perp| p_1^z\big)
    \bigg]_+
    \nonumber\\
    &\quad
    +
    \pi \Psi(\omega_1)\,
    \delta\big(1-\tfrac{x_1}{\omega_1}\big)\,
    \delta\big(1-\tfrac{x_2}{\omega_2}\big)\, 
    \frac{\tilde{\Gamma}_{q_1}\otimes\tilde{\Gamma}_{q_2}}{p_1^z\,p_2^z}
    \nn \\
    &\qquad\times
    16\img \pi^2 p_1^z \mu_0^{2\epsilon}
    \int_0^1\! \df v
    \int_{-\infty}^{\infty}\! \df y\,
    \frac{1\!-\!e^{-\img y p_1^z \tilde \eta}}{y}
    \int\! \frac{\df^d\ell}{(2\pi)^d}\,
    e^{\img \boldsymbol{\ell}_\perp \cdot\, \mathbf{b}_\perp}
    \frac{2\!-\!y\!-\!v}{\ell^4}\,
    \delta\big[\ell^z - (y\!-\!v)p_1^z\big]\biggr\}\,.
    \nonumber
\end{align}
%%%
Finally, we perform the integral over the Feynman parameter $v$ and expand in $\epsilon$ to obtain
%%%
\begin{align}
    \tilde{F}^H
    &=
    \frac{\al_s}{4\pi} \biggl\{
    \pi \Psi(x_1)\,
    \delta\bigl(1-\tfrac{x_2}{\omega_2}\bigr)\,
    \frac{\tilde{\Gamma}_{q_1}\otimes\tilde{\Gamma}_{q_2}}{p_1^z\,p_2^z}\,
    \bigg[\frac{1}{\epsilon_\text{ir}} 
        +\ln\bigg(\frac{\mu^2 \mathbf{b}_\perp^2}{b_0^2}\bigg)\bigg]
    \bigg[\frac{2\frac{x_1}{\omega_1}}{1-\frac{x_1}{\omega_1}}\biggr]^{[0,1]}_+
    \\
    &\quad
    +\pi \Psi(\omega_1)\,
    \delta\big(1-\tfrac{x_1}{\omega_1}\big)\,
    \delta\big(1-\tfrac{x_2}{\omega_2}\big)\,
    \frac{\tilde{\Gamma}_{q_1}\otimes\tilde{\Gamma}_{q_2}}{p_1^z\,p_2^z}
    \nn \\
    &\qquad\times
    16\img \pi^2 p_1^z \mu_0^{2\epsilon}
    \int_{-\infty}^{\infty}\df y
    \frac{1\!-\!e^{-\img y p_1^z \tilde \eta}}{y}
    \int_0^1\df v
    \int\frac{\df^d \ell}{(2\pi)^d}\,
    e^{\img \boldsymbol{\ell}_\perp \cdot \mathbf{b}_\perp}\,
    \frac{2\!-\!y\!-\!v}{\ell^4}\,
    \delta\big[\ell^z - (y\!-\!v)p_1^z\big]\biggr\}\,.
    \nonumber
\end{align}
%%%
While we were able to obtain a closed-form expression for the remaining integral, we find that it substantially simplifies  after combining it with other diagrams, so we only present results for the sum. This is discussed at the end of \app{diagrams}.

Let us now write down the contribution of diagram $H$ to the matching kernel. First, note that as both the lightcone- and quasi diagrams are diagonal in spin, this diagram does not contribute to mixing between spin structures. Second, note that the difference of the two diagrams can be written as \eq{Delta_def}.
Including the contributions of the three sister topologies of this diagram, we find that
%%%
\begin{align}
    &\bigl[\Delta_{q_1 q_2}^H\bigr]_{q_1' q_2'}
    (x_1,x_2,b_\perp,\epsilon,\{\tilde{\eta},\tilde{P}^z\},\{\delta^+,p^+\})
    \\
    &=
    -4
    \delta_{q_1 q_1'} 
    \delta_{q_2 q_2'}
    \delta\bigl(1-\tfrac{x_1}{\omega_1}\bigr)
    \delta\bigl(1-\tfrac{x_2}{\omega_2}\bigr) \ 
    \bigg[\frac{1}{\epsilon_\text{ir}} 
        +\ln\bigg(\frac{\mu^2 \mathbf{b}_\perp^2}{b_0^2}\bigg)\bigg]
    \bigg[1 + \ln\biggl(\frac{\delta^+}{p_1^+}\biggr)\bigg]
    \nonumber\\
    &\quad
    -
    2
    \delta_{q_1 q_1'} 
    \delta_{q_2 q_2'}
    \delta\big(1-\tfrac{x_1}{\omega_1}\big)
    \delta\big(1-\tfrac{x_2}{\omega_2}\big)
    \nonumber
    \\
    &\qquad\times
    16\img \pi^2
    \tilde{p}_1^z \mu_0^{2\epsilon}
    \int_{-\infty}^{\infty}\! \df y\,
    \frac{1-\cos(y \tilde{p}_1^z \tilde \eta)}{y}
    \int_0^1\!\df v
    \int\!\frac{\df^d \ell}{(2\pi)^d}\,
    e^{\img \boldsymbol{\ell}_\perp \cdot \mathbf{b}_\perp}\,
    \frac{2\!-\!y\!-\!v}{\ell^4}\, 
    \delta\big[\ell^z - (y\!-\!v)\tilde{p}_1^z\big]
    \nonumber\\
    &\quad
    +(1\leftrightarrow2)
    \,.\nn
\end{align}
%%%
Inserting this into \eq{oneloop_final}, gives its contribution to the one-loop matching coefficient.

%------------------------------------------------------------------------------------------------------------------
\subsection{Result for one-loop matching kernel} \label{sec:result}
%------------------------------------------------------------------------------------------------------------------

Here we present the one-loop matching kernels for the flavor non-singlet case of \eq{flavor_nons} for all color- and spin structures, which can be calculated from the master formula in \eq{oneloop_final}.

Here we list all the one-loop ingredients that are needed to construct the matching kernels. The results for the one-loop diagrams for the lightcone- and quasi-DPDs can be found in \app{diagrams}. The one-loop soft functions for the lightcone-DPD can be obtained from the TMD case~\cite{Echevarria:2012js} by an appropriate modification of the color factor 
%%%
\begin{align}
    {^{88}}\!S^{(1)}(b_\perp,\epsilon,\delta^+,\delta^-)
    &=
    -4C_A\biggl\{\frac{1}{\epsilon_\text{uv}^2}
    \!+\!\biggl[\frac{1}{\epsilon_\text{uv}}
            \!+\!\ln\Bigl(\frac{\mu^2 \mathbf{b}_\perp^2}{b_0^2}\Bigr)\biggr]
        \ln\Bigl(\frac{\mu^2}{2\delta^+ \delta^-}\Bigr)
    \!-\!\frac{1}{2}\ln^2 \Bigl(\frac{\mu^2 \mathbf{b}_\perp^2}{b_0^2}\Bigr)
    \!-\!\frac{\pi^2}{12}\biggr\}\,,
\end{align}
%%%
We have calculated the corresponding quasi-soft function, obtaining
%%%
\begin{align}
    {^{88}}\!\tilde{S}^{(1)}(b_\perp,\epsilon,\tilde{\eta},y_A,y_B)
    &=
    4C_A
    \biggl\{\biggl[\frac{1}{\epsilon_\text{uv}}
            +\ln\Bigl(\frac{\mu^2\mathbf{b}_\perp^2}{b_0^2}\Bigr)\biggr]
        \Bigl[2-(y_A-y_B)\Bigr]
    +\frac{2\pi \tilde{\eta}}{|\mathbf{b}_\perp|}\biggr\}\,.
\end{align}
%%%
Note that the above expression for the quasi-soft function does not contain the contribution from the transverse Wilson line. Though non-zero, its contribution cancels between the quasi-soft function and the unsubtracted quasi-DPD. The one-loop rapidity evolution kernel is given by
%%%
\begin{align}
    {^{8}}\!J^{(1)}(b_\perp,\mu)
    &=
    -4C_A\ln\Bigl(\frac{\mu^2 \mathbf{b}_\perp^2}{b_0^2}\Bigr)\,.
\end{align}
%%%

For completeness we also present the renormalization kernels and factors that define the renormalized distributions. The renormalization kernels for the lightcone-DPD read~\cite{Diehl:2022rxb}
%%%
\begin{align}
    {^{11}}\!Z_{qq}^{(1)}(x,\mu,\zeta)
    &=
    -\frac{1}{\epsilon_\text{uv}} 2C_F \biggl[\frac{1+x^2}{1-x}\biggr]^{[0,1]}_+\,,
    \\
    {^{11}}\!Z_{\Delta q \Delta q}^{(1)}(x,\mu,\zeta)
    &=
    -\frac{1}{\epsilon_\text{uv}} 2C_F \biggl[\frac{1+x^2}{1-x}\biggr]^{[0,1]}_+\,,
    \nn \\
    {^{11}}\!Z_{\delta q \delta q}^{(1)}(x,\mu,\zeta)
    &=
    -\frac{1}{\epsilon_\text{uv}}C_F\biggl(
    4\biggl[\frac{x}{1-x}\biggr]^{[0,1]}_+
    +\delta(1-x)\biggr)\,,
\nn\end{align}
%%%
for the color-singlet case. Those for the color non-singlet case are related to their color-singlet counterpart by an modification of the color factor and an additional piece 
%%%
\begin{align} \label{eq:Z88}
    {^{88}}\!Z_{q q}^{(1)}(x,\mu,\zeta)
    &=
    \Bigl(1-\frac{C_A}{2C_F}\Bigr)
    {^{11}}\!Z_{q q}^{(1)}(x,\mu,\zeta)
    -
    C_A \delta(1-x)
    \biggl\{\frac{1}{\epsilon_\text{uv}^2}+\frac{1}{\epsilon_\text{uv}}
        \biggl[\ln\biggl(\frac{\mu^2}{\zeta}\biggr)+\frac{3}{2}\biggr]\biggr\}\,.
\end{align}
%%%
For the color-singlet quasi-DPDs, we find that only the transversity distribution needs renormalization,
%%%
\begin{align}
    {^{11}}\!\tilde Z_{q q}^{(1)}(\epsilon,\mu,y_A,y_B)
    &=
    0
    \,, \\
    {^{11}}\!\tilde Z_{\Delta q \Delta q}^{(1)}(\epsilon,\mu,y_A,y_B)
    &=
    0
    \,, \nn \\
    {^{11}}\!\tilde Z_{\delta q \delta q}^{(1)}(\epsilon,\mu,y_A,y_B)
    &=
    -\frac{1}{\epsilon_\text{uv}} 2C_F\,.
\nn\end{align}
%%%
In direct analogy to \eq{Z88}, the renormalization factors for the color-correlated quasi-DPDs are related to their color-singlet counterpart by
%%%
\begin{align}
    {^{88}}\!\tilde Z_{qq}^{(1)}(\epsilon,\mu,y_A,y_B)
    &=
    \Bigl(1-\frac{C_A}{2C_F}\Bigr)
    {^{11}}\!\tilde Z_{qq}^{(1)}(\epsilon,\mu,y_A,y_B)
    -
    \frac{1}{\epsilon_\text{uv}}
    C_A\Bigl[1+2(y_A-y_B)\Bigr]\,.
\end{align}
%%%
Note that all renormalization kernels are diagonal in color- and spin structures.

We find that at one-loop order there is no mixing between color structures. This can be understood from the fact that the matching coefficients are related to the difference between the order of the UV and large rapidity limits (as discussed at the end of \sec{Quasi-TMDs}). However, only diagrams where a gluon connects the two quark lines can lead to mixing between color structures, but these do not have UV divergences since the quark lines are separated by $b_\perp$. It should be noted that individual diagrams of this type can contribute to the matching kernel due to the different treatment of rapidity divergences for the lightcone- and quasi-DPDs, but their contribution should cancel once all diagrams are combined. This argument is expected to hold at higher orders in perturbation theory as well.

For the color-singlet case, we verify that the matching kernel is related to that of the ordinary PDF case, see \eq{C_rel_cs}. We have verified that this holds, and for completeness we present the matching kernels in the $\overline{\text{MS}}$ scheme,
%%%
\begin{align}
    {^{11}}\!C^{(1)}_{qq}(x,\tilde{p}^z,\mu)
    &=
    -2C_F
    \biggl\{
    \biggl[
    \frac{1+x^2}{1-x}
        \biggl[\ln\Bigl(\frac{\mu^2}{4(1-x)^2\tilde{p}_z^2}\Bigr)
            -1\biggr]
    +\frac{4x}{1-x}
    \biggr]^{[0,1]}_+
    \\
    &\qquad\qquad\quad
    -\biggl[\text{sgn}(x)
    \biggl(1+\frac{1+x^2}{1-x}\ln\Bigl|\frac{x}{1-x}\Bigr|\biggr)
    \biggr]^{(-\infty,+\infty)}_+
    \biggr\}\,,
    \nonumber
    \\[1.5ex]
    {^{11}}\!C^{(1)}_{\Delta q \Delta q}(x,\tilde{p}^z,\mu)
    &=
    -2C_F
    \biggl\{
    \biggl[
    \frac{1+x^2}{1-x}
        \biggl[\ln\Bigl(\frac{\mu^2}{4(1-x)^2\tilde{p}_z^2}\Bigr)
            +3\biggr]
    -\frac{4x}{1-x}
    \biggr]^{[0,1]}_+
    \\
    &\qquad\qquad\quad
    -\biggl[\text{sgn}(x) 
    \biggl(1+\frac{1+x^2}{1-x}\ln\Big|\frac{x}{1-x}\Big|\biggr)
    \biggr]^{(-\infty,+\infty)}_+
    \biggr\}\,.
    \nonumber
    \\[1.5ex]
    {^{11}}\!C^{(1)}_{\delta q \delta q}(x,\tilde{p}^z,\mu)
    &=
    -2C_F
    \biggl\{
    \biggl[
    \frac{2x}{1-x}
        \biggl[\ln\Bigl(\frac{\mu^2}{4(1-x)^2 \tilde{p}_z^2}\Bigr)
            +1\biggr]
    \biggr]^{[0,1]}_+
    \\
    &\qquad\qquad\quad
    -\biggl[\text{sgn}(x) \frac{x}{1-x} \ln\Bigl|\frac{x}{1-x}\Bigr|
    \biggr]^{(-\infty,+\infty)}_+
    \biggr\}\,.
    \nonumber
\end{align}
%%%
The matching kernels for the color-correlated case are identical to those for the color-summed case, up to a color factor and an additional $x\tilde P^z$-dependent piece,
%%%
\begin{align}\label{eq:match8}
    {^{88}}\!C^{(1)}_{qq}\Bigl(\frac{x}{x'},x'\tilde{P}^z,\mu\Bigr)
    &=
    \Bigl(1-\frac{C_A}{2C_F}\Bigr)\,
    {^{11}}\!C^{(1)}_{qq}\Bigl(\frac{x}{x'},x'\tilde{P}^z,\mu\Bigr)
    \\
    &\quad
    -\delta\Bigl(1-\frac{x}{x'}\Bigr)
    \frac{1}{2}C_A
    \biggl[
    \ln^2\biggl(\frac{\mu^2}{(2x'\tilde{P}^z)^2}\biggr)
    +2\ln\biggl(\frac{\mu^2}{(2x'\tilde{P}^z)^2}\biggr)
    +4-\frac{\pi^2}{6}
    \biggr]\,,
    \nonumber
\end{align}
%%%
where we switched back to expressing partonic momenta as fractions of hadronic momenta.
Note that all matching kernels which are off-diagonal in color or spin vanish (though this is not true at the level of individual diagrams). Note also that the form of~\eqref{eq:match8} arises because the contribution of all graphs in fig.~\ref{fig:quasidpddiagrams2} to the matching kernel is proportional to delta functions in both momentum fractions, although the contribution of some individual graphs to the DPD or quasi-DPD is not.

\section{Conclusions}
\label{sec:conclusions}
Double parton scattering can significantly affect  precision measurements due to the radiation from a secondary partonic collision. For certain processes, such as same-sign $WW$ production, its contribution can be on par with that of single parton scattering. Currently, the double parton distributions (DPDs) that enter in the factorization theorems for these cross sections are poorly constrained experimentally: Essentially only a single number, the effective cross section, has been measured for a range of different processes. At the same time, these DPDs provide a window on a range of interesting correlations of partons inside the proton.

Inspired by the substantial progress in the quasi-PDF approach to extract (single) PDFs from lattice QCD, we have taken the first steps in this paper to extend this approach to DPDs. We have put forward a general matching relation, whose form is constrained using the renormalization group equations. This shares similarities with both the quasi-PDF approach to parton distribution functions (convolutions in momentum fractions and flavor mixing) and transverse momentum distributions (rapidity divergences, requiring a soft function). We have obtained explicit results for the flavor non-singlet quark-quark DPD at one-loop order, showing that the matching coefficients do not involve the infrared scale $b_\perp$. For the color-summed case, the kernel can directly be expressed in terms of that for the single PDF case.

There are several open questions left that we wish to explore in future work: On the conceptual side, the method to obtain the double-parton scattering soft function from the lattice requires further investigation. On the calculational side, there is the obvious extension to other flavors, for which the mixing with single PDFs may need to be taken into account (in transverse momentum space), as well as the extension to interference DPDs. The following issues related to lattice calculations will also need to be addressed:  The nonperturbative renormalization, conversion to the RI/MOM scheme and the mixing of operators. We expect that in the coming years this effort will lead to a substantial improvement of our understanding of DPDs, that can be confronted with measurements of double parton scattering and unveil more of the fascinating structure of the proton.

\vspace{1.5ex}

\noindent \textbf{Note added:} While this manuscript was in preparation, ref.~\cite{Zhang:2023wea} appeared. It discusses the color-singlet quark-quark DPD, showing that the matching can be expressed in terms of that for single PDFs, due to the spatial separation between the currents. We reach the same conclusion, as discussed at the end of \sec{quasidpdfactorization}. However, we take a broader perspective, presenting a matching relation for general flavor, spin and color correlations. The latter particularly complicates things due to the presence of rapidity divergences and the need to subtract a soft factor. In our calculations we restrict to the non-singlet quark-quark DPD, but account for general spin and color correlations.

\acknowledgments

We thank Martha Constantinou, Markus Diehl, Florian Fabry, Iain Stewart and Alexey Vladimirov for discussions.
This work is supported by the Royal Society through grant URF\textbackslash R1\textbackslash 201500, the NWO projectruimte 680-91-122, and the D-ITP consortium, a program of NWO that is funded by the Dutch Ministry of Education, Culture and Science (OCW). 

\appendix

\section{Plus distributions}
\label{app:conventions}
In this paper we will encounter singularities requiring a plus prescription at $x=1$. Since quasi-parton distributions have a range outside the domain $[0,1]$, we have modified the plus prescription such that $[f(x)]_+^I$ only has support on the finite or infinite length interval $I$ (i.e.~the theta functions ensuring this are included in the definition of the plus distribution) and satisfies
%%%
\begin{align}\label{eq:plusprescription}
    \int_I\! \df x\, [f(x)]_+^I = 0
\,.\end{align}
%%%
In practice we decompose a function regulated by e.g.~$\eps$ in terms of plus distributions by writing
%%%
\begin{align}
    f(x,\eps) =
    \big[f(x,\eps)\big]_+^I + \delta(1-x)\int_I \df y \,f(y,\eps) \,,
\end{align}
%%%
and then expanding (with respect to the regulator) the expression \emph{in} the plus distribution and the \emph{result} of the integral.
The following identity still holds for these plus distributions 
%%%
\begin{align}\label{eq:plusintegration}
    \int_I\! \df x \,g(x) \big[f(x)\big]_+^I
    &=
    \int_I\! \df x \,\big[g(x) - g(1)\big] f(x) \,.
\end{align}
%%%

\section{One-loop diagrams}
\label{app:diagrams}
In this appendix we present the calculation of the one-loop lightcone- and quasi-DPDs, defined in \secs{dpd_defs}{quasidpddefinition}. 
The one-loop diagrams for the quasi-DPD are shown in \figs{quasidpddiagrams1}{quasidpddiagrams2}. The diagrams for the lightcone-DPD are identical up to a cut that goes vertically through the middle of the diagram. 
In each figure, the diagrams in the top row correspond to real-emission diagrams while those on the bottom are virtual corrections. 
These diagrams are decomposed according to \eq{oneloop_decomp}, and the color factors belonging to these diagram are shown in table \ref{tab:colorfactors}. 
\renewcommand{\arraystretch}{1.5}
\begin{table}
\centering
\begin{tabular}{ |c|c|c|c|c| }
 \hline
 diagram & $R_1 = 1, R_1' = 1$ & $R_1 = 1, R_1' = 8$ & $R_1 = 8, R_1'=1$ & $R_1 = 8, R_1'=8$ \\ 
 \hline\hline
 $A,B,C$ & $C_F$ & 0 & 0
    & $C_F - \tfrac{C_A}{2}$ \\
 \hline
 $D,E,F$ & $C_F$ & 0 & 0 & $C_F$ \\
 \hline
 $G,H,I$ & 0 & 1 &
    $\tfrac{C_F}{2N}$ & $2C_F - \tfrac{C_A}{2}$ \\ 
 \hline
 $J,K,L$ & 0 & 1 & 
    $\tfrac{C_F}{2N}$ & $2C_F - C_A$\\
 \hline
\end{tabular}
\caption{Color factors ${^{R_1 R_2, R_1' R_2'}}\!c^i$ for the $i$-th diagram, as defined in \eq{oneloop_decomp}. Note that only $R_1 = R_2$ and $R_1'=R_2'$ are allowed, which is why we just list $R_1$ and $R_1'$. }
\label{tab:colorfactors}
\end{table}
\renewcommand{\arraystretch}{1.0}

The following integrals are convenient for calculating the one-loop corrections to the lightcone-DPD
%%%
\begin{alignat}{2}
    &
    \int \df^d k\,
    \frac{\delta(k^+ - \xi)}{k^2 (k - p)^2}\, 
    e^{\img \mathbf{k}_\perp \cdot \mathbf{b}_\perp}
    &&=
    \frac{\img \pi^2}{p^+}\,
    \theta(\xi)\theta(p^+ - \xi)\,
    \frac{\pi^{\frac{1}{2}-\epsilon} \Gamma(-\epsilon_\text{ir})}
        {\Gamma\bigl(\tfrac{1}{2}-\epsilon\bigr)} \,
     (\mathbf{b}_\perp^2)^\epsilon \,,
    \\
    &
    \int \df^d k \, 
    \frac{\delta(k^+ - \xi)}{k^2 (k - p)^2}
    &&=
    \frac{\img \pi^2}{p^+}\,
    \theta(\xi)\theta(p^+ - \xi)
    \biggl(\frac{1}{\epsilon_\text{uv}} - \frac{1}{\epsilon_\text{ir}}\biggr) \, ,
    \\
    &
    \int \df^d k \, 
    \delta_+(k^2)\,
    \frac{\delta(k^+ - \xi)}{k^-}\, 
    e^{\img \mathbf{k}_\perp \cdot \mathbf{b}_\perp}
    &&=
    \pi \theta(\xi)\,
    \frac{2\pi^{\frac{1}{2}-\epsilon} \Gamma(-2\epsilon_\text{ir})}
    {\Gamma\bigl(\tfrac{1}{2} - \epsilon\bigr)}\,
    (\mathbf{b}_\perp^2)^\epsilon \,,
    \label{eq:int_for_fig_e_lc}
    \\
    &
    \int \df^d k \, 
    \delta_+(k^2)\,
    \frac{\delta(k^+ - \xi)}{k^-} 
    &&=
    \pi \theta(\xi)
    \biggl(\frac{1}{\epsilon_\text{uv}} - \frac{1}{\epsilon_\text{ir}}\biggr) \,.
\end{alignat}
%%%
For the calculation of the one-loop quasi-DPD the integrals below can be used,
%%%
\begin{alignat}{2}
    &     \label{eq:int_for_fig_e_q}
    \int\frac{\df^d k}{(2\pi)^d} \, 
    \frac{\delta(k^z - \xi)}{k^4}\,
    e^{\img \mathbf{k}_\perp \cdot \mathbf{b}_\perp}
    &&=
    \frac{\img}{16\pi^3} \,
    \biggl(\frac{2\pi |\mathbf{b}_\perp|}{|\xi|}\biggr)^{\frac{1}{2}+\epsilon} 
    K_{\frac{1}{2}+\epsilon}\bigl(|\xi||\mathbf{b}_\perp|\bigr) \, ,
    \\
    &\label{eq:masterint}
    \int\frac{\df^d k}{(2\pi)^d} \, 
    \frac{\delta(k^z - \xi)}{k^4}
    &&=
    \frac{\img}{16\pi^2} \, 
    \frac{(4\pi)^\epsilon \Gamma\bigl(\tfrac{1}{2}+\epsilon\bigr)}{\sqrt{\pi}}\,
    |\xi|^{-1-2\epsilon} \,,
    \\
    &
    \int\frac{\df^d k}{(2\pi)^d} \,
    \frac{\delta(k^z - \xi)}{k^2}\,
    e^{\img \mathbf{k}_\perp \cdot \mathbf{b}_\perp}
    &&=
    -\frac{\img}{4\pi^2}\,
    \biggl(\frac{2\pi |\mathbf{b}_\perp|}{|\xi|}\biggr)^{-\frac{1}{2}+\epsilon}
    K_{-\frac{1}{2}+\epsilon} \bigl(|\xi||\mathbf{b}_\perp|\bigr) \, ,
    \\
    &
    \int\frac{\df^{d}k}{(2\pi)^d} \, 
    \frac{\delta(k^z - \xi)}{k^2}
    &&=
    -\frac{\img}{8\pi^2} \ 
    \frac{(4\pi)^{\epsilon}\Gamma\bigl(\!-\!\tfrac{1}{2} + \epsilon\bigr)}{\sqrt{4\pi}}\,
    |\xi|^{1-2\epsilon} \, ,
\end{alignat}
%%%
where $K$ denotes the modified Bessel function of the second kind.

We will now discuss each diagram in turn. To keep the notation compact, we will use the shorthand introduced in \eq{Fn_in_calc}
%%%
\begin{align}
   F^I=
    \frac{\alpha_s}{4\pi}
    \sum_{q_1',q_2'}
    [F^I_{q_1 q_2}]_{q_1'q_2'} \frac{\Gamma_{q_1'} \otimes \Gamma_{q_2'}}{p_1^+ p_2^+}\,.
\end{align}
%%%
In the following we will assume that the arguments of the plus distributions are always the ratios $x_i/\omega_i$.

\subsubsection*{diagram A}
This diagram corresponds to the emission of a real gluon. For the lightcone diagram the gluon propagator is cut and the diagram yields
%%%
\begin{align}\label{B:LCa}
    F^A
     &=
    -\pi p^+ \mu_0^{2\epsilon}
    \int \df \omega_3 \, \Psi(\omega_3)
    \int\frac{\df^d k}{(2\pi)^d} \ 
    (\img g \gamma^\mu) \frac{\img}{\fsl{p}_1 - \fsl{k}} \Gamma_{q_1}
    \frac{\img}{\fsl{p}_3 - \fsl{k}} (\img g \gamma^\nu) \otimes \Gamma_{q_2}
    \\
    &\quad\times
    (-g_{\mu\nu})\, 2\pi \delta_+(k^2)\,
    \delta\bigl[x_1 p^+ - (p_1^+ - k^+)\bigr]\,
    \delta\bigl[x_2 p^+ - p_2^+\bigr]\,
    \delta\bigl[x_1 p^+ - (p_3^+ - k^+)\bigr]
    \nonumber
    \\[1.5ex]
    &=
    -\alpha_s
    \Psi(\omega_1)\, \delta\bigl(1 - \tfrac{x_2}{\omega_2}\bigr) \, 
    \frac{\Gamma_{q_1}\otimes\Gamma_{q_2}}{p_1^+ p_2^+}\, 
    \frac{\rho_{q_1}}{2} 
    \biggl(\frac{1}{\epsilon_\text{uv}}-\frac{1}{\epsilon_\text{ir}}\biggr)
    \bigl[1 - \tfrac{x_1}{\omega_1}\bigr]_+^{[0,1]}
    \nonumber\\
    &\quad
    -\alpha_s
    \Psi(\omega_1)\, 
    \delta\bigl(1 - \tfrac{x_1}{\omega_1}\bigr)\,
    \delta\bigl(1 - \tfrac{x_2}{\omega_2}\bigr) \, 
    \frac{\Gamma_{q_1}\otimes\Gamma_{q_2}}{p_1^+ p_2^+}\, 
    \frac{\rho_{q_1}}{4} 
    \biggl(\frac{1}{\epsilon_\text{uv}}-\frac{1}{\epsilon_\text{ir}}\biggr)\, .
    \nn
\end{align}
%%%
The factors $\rho_{q_1}$ are spin-dependent and given by
%%%
\begin{align}
    \rho_q = 1 \, , \qquad
    \rho_{\Delta q} = 1 \, , \qquad
    \rho_{\delta q} = 0 \, .
\end{align}
%%%
In principle these factors include terms of $\mathcal{O}(\epsilon)$ and beyond, but these are irrelevant as they drop out when multiplying with the above combination of $\epsilon$ poles.

The range on the plus distribution arises here from two considerations: The upper bound has its origin in one of the intermediate delta distributions, which only has support if $\omega_1>x_1$. The lower bound is added manually, to separate the quark-DPD from the antiquark-DPD.

For the quasi-diagram the gluon propagator is not cut, as the matrix element defining quasi-DPDs can be treated as time-ordered. The quasi-diagram results in  
%%%
\begin{align}\label{eq:diagrama}
    \tilde F^A
    &=
    -\pi \tilde p^z \mu_0^{2\epsilon}
    \int \df\omega_3 \, \Psi(\omega_3)
    \int\frac{\df^d k}{(2\pi)^d} \
    (\img g \gamma^\mu) \frac{\img}{\fsl{\tilde p}_1 - \fsl{k}} \tilde{\Gamma}_{q_1} 
    \frac{\img}{\fsl{\tilde p}_3 - \fsl{k}} (\img g \gamma^\nu) 
    \otimes \tilde{\Gamma}_{q_2}
    \\
    &\qquad\times
    \frac{-\img g_{\mu\nu}}{k^2} \,
    \delta\bigl[x_1 \tilde p^z - (\tilde p_1^z - k^z)\bigr]\,
    \delta\bigl[x_2 \tilde p^z - \tilde p_2^z\bigr]\,
    \delta\bigl[x_1 \tilde p^z - (\tilde p_3^z - k^z)\bigr]
    \nonumber
    \\[1.5ex]
    &=
    \alpha_s 
    \Psi(\omega_1)\,
    \delta\bigl(1-\tfrac{x_2}{\omega_2}\bigr) \ 
    \frac{\tilde{\Gamma}_{q_1} \otimes \tilde{\Gamma}_{q_2}}{\tilde p_1^z \ \tilde p_2^z} 
    \nn \\ & \quad \times
    \frac{\tilde\rho_{q_1}}{2}
    \biggl[
    \bigl(1-\tfrac{x_1}{\omega_1}\bigr)
    \biggl(\frac{1}{\epsilon_\text{ir}} + 2 
        + \ln\Bigl(\frac{\mu^2}{4(1-\frac{x_1}{\omega_1})^2 (\tilde p_1^z)^2}\Bigr)\biggr) \,
    \theta\bigl(\tfrac{x_1}{\omega_1}\bigr)\,
    \theta\bigl(1 - \tfrac{x_1}{\omega_1}\bigr)
    \nonumber\\
    &\qquad 
    -\text{sgn}\bigl(\tfrac{x_1}{\omega_1}\bigr) 
    \biggl(1 + \bigl(1-\tfrac{x_1}{\omega_1}\bigr)
        \ln\bigg|\frac{\frac{x_1}{\omega_1}}
            {1-\frac{x_1}{\omega_1}}\bigg|\biggr)
    \biggr]_+^{(-\infty,\infty)}
    \nonumber\\
    &\quad
    -\alpha_s
    \Psi(\omega_1) \,
    \delta\bigl(1 - \tfrac{x_1}{\omega_1}\bigr) \,
    \delta\bigl(1 - \tfrac{x_2}{\omega_2}\bigr) \, 
    \frac{\Gamma_{q_1}\otimes\Gamma_{q_2}}{p_1^+ p_2^+}\,
    \frac{\tilde\rho_{q_1}}{4} 
    \biggl(\frac{1}{\epsilon_\text{uv}}-\frac{1}{\epsilon_\text{ir}}\biggr) \, ,
\nn\end{align}
%%%
where we used~\eqref{eq:quasigamma} to relate the Dirac structures in the last term to the lightcone ones to highlight the similarity to~\eqref{B:LCa}. $\tilde{\rho}_{q_1}$ are again spin-dependent factors, given by
%%%
\begin{align}
    \tilde{\rho}_{q}=1-3\epsilon \, , \qquad
    \tilde{\rho}_{\Delta q} = 1+\epsilon \, , \qquad
    \tilde{\rho}_{\delta q}= 0 \, .
\end{align}
%%%
This time the $\mathcal{O}(\epsilon)$ contributions are relevant as they multiply an isolated $1/\epsilon$.
The intricate pattern of plus distribution ranges arises from the varying sign possibilities imposed by the absolute values in eq.~\eqref{eq:masterint}.

For both the lightcone and the quasi diagram the infinite normalization factor $\Psi(\omega_1)$ factors out (its argument is $\omega_1$ due to the interplay of the two delta distributions in the first line of eq.~\eqref{eq:diagrama}, which enforces $\omega_3=\omega_1$). Including the corresponding graph where the gluon connects to the other quark line, we obtain, in the notation of \eq{Delta_def},
%%%
\begin{align}
    \bigl[\Delta^A_{q_1 q_2}\bigr]_{q_1' q_2'}
    &=
    -2
    \delta_{q_1 q_1'}
    \delta_{q_2 q_2'}\,
    \delta(1-\tfrac{x_2}{\omega_2})\,
    \biggl\{
    \frac{\rho_{q_1}}{\epsilon_\text{uv}}
    \bigl[1-\tfrac{x_1}{\omega_1}\bigr]^{[0,1]}_+
    +
    \frac{\tilde{\rho}_{q_1}-\rho_{q_1}}{\epsilon_\text{ir}}
    \bigl[1-\tfrac{x_1}{\omega_1}\bigr]^{[0,1]}_+
    \\
    &\qquad
    +
    \tilde{\rho}_{q_1}
    \biggl[
    \bigl(1-\tfrac{x_1}{\omega_1}\bigr)
    \biggl[2 + \ln\Bigl(\frac{\mu^2}
        {4(1-\frac{x_1}{\omega_1})^2 (\tilde{p}_1^z)^2}\Bigr)\biggr]
    \biggr]^{[0,1]}_+
    \nonumber\\
    &\qquad
    -
    \tilde{\rho}_{q_1}
    \biggl[\text{sgn}\bigl(\tfrac{x_1}{\omega_1}\bigr)
    \biggl(1 + (1-x_1)\ln\bigg|
        \frac{\frac{x_1}{\omega_1}}{1-\frac{x_1}{\omega_1}}\bigg|\biggr)
    \biggr]^{(-\infty,+\infty)}_+\biggr\}
    +(1\leftrightarrow2)\,.
    \nonumber
\end{align}
%%%

\subsubsection*{Diagram B}
For diagram $B$ we need to implement the rapidity regulator. For the lightcone diagram we find
%%%
\begin{align}
    F^B
    &=
    -\pi p^+
    \mu_0^{2\epsilon}
    \int \df \omega_3 \, \Psi(\omega_3)
    \int\frac{\df^d k}{(2\pi)^d} \
    \Gamma_{q_1}\otimes (\img g \gamma^\mu) \frac{\img}{\fsl{p}_2 - \fsl{k}}
    \Gamma_{q_2} (\img g n_a^\nu) \frac{\img}{k^+ + \img\delta^+}
    \\
    &\quad\times
    (-g_{\mu\nu}) \ 2\pi \delta_+(k^2) \,
    \delta\bigl[x_1 p^+ - p_1^+\bigr] \,
    \delta\bigl[x_2 p^+ - (p_2^+ - k^+)\bigr] \,
    \delta\bigl[x_1 p^+ - p_3^+\bigr]
    \nonumber\\[1.5ex]
    &=
    -\alpha_s \,
    \Psi(\omega_1) \, \delta\bigl(1 - \tfrac{x_1}{\omega_1}\bigr) \, 
    \frac{\Gamma_{q_1}\otimes\Gamma_{q_2}}{p_1^+ p_2^+} \,
    \frac{1}{2} 
    \biggl(\frac{1}{\epsilon_\text{uv}}-\frac{1}{\epsilon_\text{ir}}\biggr)
    \biggl[\frac{\frac{x_2}{\omega_2}}{1-\frac{x_2}{\omega_2}}
    \biggr]_+^{[0,1]}
    \nonumber\\
    &\quad
    +\alpha_s
    \Psi(\omega_1) \,
    \delta\bigl(1 - \tfrac{x_1}{\omega_1}\bigr) \,
    \delta\bigl(1 - \tfrac{x_2}{\omega_2}\bigr) \,
    \frac{\Gamma_{q_1}\otimes\Gamma_{q_2}}{p_1^+ p_2^+} \,
    \frac{1}{2} 
    \biggl(\frac{1}{\epsilon_\text{uv}}-\frac{1}{\epsilon_\text{ir}}\biggr)
    \biggl[1 + \ln\biggl(\frac{\delta^+}{p_2^+}\biggr) + \frac{\img \pi}{2}\biggr]
\nn\,.\end{align}
%%%
For the color-summed DPD, the rapidity divergence ($\ln \delta^+$) of this diagram will cancel against that of diagram $E$. 
For the quasi-diagram we obtain,
%%%
\begin{align}
    \tilde{F}^B
    &=
    -\pi \tilde p^z \mu_0^{2\epsilon}
    \int \df\omega_3 \, \Psi(\omega_3)
    \int\frac{\df^d k}{(2\pi)^d} \, 
    \tilde{\Gamma}_{q_1} \otimes (\img g \gamma^\mu) \frac{\img}{\fsl{\tilde p}_2-\fsl{k}}
        \tilde{\Gamma}_{q_2} (\img g {\hat z}^\nu) \frac{\img}{k^z}
    \\
    &\qquad\times
    \frac{-\img  g_{\mu\nu}}{k^2}\, 
    \delta\bigl[x_1 \tilde p^z - \tilde p_1^z\bigr]\,
    \bigl(e^{-\img k^z \tilde \eta} -1 \bigr)\, \delta\bigl[x_2 \tilde p^z - (\tilde p_2^z - k^z)\bigr]\,
    \delta\bigl[x_1 \tilde p^z - \tilde p_3^z\bigr]
    \nonumber\\[1.5ex]
    &=
    \alpha_s
    \Psi(\omega_1)\,
    \delta\bigl(1-\tfrac{x_1}{\omega_1}\bigr) \,
    \frac{\tilde{\Gamma}_{q_1}\otimes\tilde{\Gamma}_{q_2}}{\tilde p_1^z \ \tilde p_2^z} 
    \nonumber\\
    &\qquad\times
    \frac{1}{2}
    \biggl[
    \frac{\frac{x_2}{\omega_2}}{1-\frac{x_2}{\omega_2}}
        \biggl(\frac{1}{\epsilon_\text{ir}}
            +\ln\Bigl(\frac{\mu^2}{4(1-\frac{x_2}
                {\omega_2})^2 (\tilde p_2^z)^2}\Bigr)\biggr)
        -1\biggr]_+^{(0,1)}
    \nonumber\\
    &\quad\qquad
    +\biggl[\text{sgn}\bigl(\tfrac{x_2}{\omega_2}\bigr)
        \frac{1}{1-\frac{x_2}{\omega_2}}
        \biggl(\frac{1}{2}-\frac{x_2}{\omega_2}\ln\bigg|
            \frac{\frac{x_2}{\omega_2}}{1-\frac{x_2}{\omega_2}}\bigg|\biggr)
    \biggr]_+^{(-\infty,\infty)}
    \nonumber\\
    &\quad
    +\alpha_s
    \Psi(\omega_1)\,
    \delta\bigl(1-\tfrac{x_1}{\omega_1}\bigr)\,
    \delta\bigl(1-\tfrac{x_2}{\omega_2}\bigr)\, 
    \frac{\tilde{\Gamma}_{q_1}\otimes\tilde{\Gamma}_{q_2}}{\tilde p_1^z \ \tilde p_2^z}
    \nonumber\\
    &\qquad\times
    4\img\pi^2\, \tilde p_2^z \mu_0^{2\epsilon}
    \int_{-\infty}^{\infty} \!\df y\,
    \frac{1 - e^{-\img y \tilde p_2^z \tilde \eta}}{y}
    \int_0^1 \! \df v\,
    \int\frac{\df^d \ell}{(2\pi)^d} \, 
    \frac{2\!-\!y\!-\!v}{\ell^4}\,
    \delta\bigl[\ell^z - (y\!-\!v)\tilde p_2^z\bigr] \, .
\nn\end{align}
%%%
While we were able to obtain a closed-form expression for the integral in the last term, it substantially simplifies if we first combine diagrams before performing the integral, which is all we need to calculate the matching. This is discussed at the end of this appendix. The contribution of diagram $B$ and its three sister topologies to the matching kernel is given by
%%%
\begin{align}
    \bigl[\Delta^B_{q_1 q_2}\bigr]_{q_1' q_2'}
    &=
    -4
    \delta_{q_1 q_1'}
    \delta_{q_2 q_2'}\,
    \delta\bigl(1-\tfrac{x_2}{\omega_2}\bigr)
    \biggl\{
    \biggl[\frac{\frac{x_1}{\omega_1}}{1-\frac{x_1}{\omega_1}}
        \biggl(\frac{1}{\epsilon_\text{uv}}
        +\ln\Bigl(\frac{\mu^2}{4(1-\frac{x_1}{\omega_1})^2 (\tilde p_1^z)^2}\Bigr)
        \biggr)
        -1\biggr]^{[0,1]}_+
    \nonumber\\
    &\qquad
    +\biggl[\text{sgn}\bigl(\tfrac{x_1}{\omega_1}\bigr) 
    \frac{1}{1-\frac{x_1}{\omega_1}}
        \biggl(\frac{1}{2}-\frac{x_1}{\omega_1}
        \ln\bigg|\frac{\frac{x_1}{\omega_1}}{1-\frac{x_1}{\omega_1}}\bigg|\biggr)
    \biggr]^{(-\infty,+\infty)}_+
    \nonumber\\
    &\qquad
    -
    \delta\bigl(1-\tfrac{x_1}{\omega_1}\bigr)\,
    \biggl(\frac{1}{\epsilon_\text{uv}}-\frac{1}{\epsilon_\text{ir}}\biggr)
    \biggl[1 + \ln\biggl(\frac{\delta^+}{p_1^+}\biggr)\biggr]
    \nonumber\\
    &\qquad
    +
    \delta\bigl(1-\tfrac{x_1}{\omega_1}\bigr)\,
    8\img\pi^2\, \tilde{p}_1^z \mu_0^{2\epsilon}
    \nonumber
    \\
    &\qquad\quad\times
    \int_{-\infty}^{\infty}\! \df y \,
    \frac{1-\cos(y \tilde{p}_1^z \tilde \eta)}{y}
    \int_0^1 \!\df v
    \int\frac{\df^d \ell}{(2\pi)^d} \, 
    \frac{2\!-\!y\!-\!v}{\ell^4}\,
    \delta\bigl[\ell^z - (y\!-\!v)\tilde{p}_1^z\bigr]\biggr\}
    \nonumber\\
    &\quad
    +(1\leftrightarrow2)\,,
\end{align}
%%%
where the appearance of the cosine is due to the combination of the two diagrams attaching to the same quark line.

\subsubsection*{Diagram C}
Diagram $C$ vanishes for the lightcone-DPD as both ends of the gluon line are connected to Wilson lines along the $n_a$ direction, leading to $n_a^2 = 0$. For the quasi-DPD it is given by, 
%%%
\begin{align}
    \tilde{F}^C
    &=
    -\pi \tilde p^z \mu_0^{2\epsilon}
    \int \df\omega_3 \ \Psi(\omega_3)
    \int\frac{\df^d k}{(2\pi)^d} \
    \tilde{\Gamma}_{q_1} \otimes \tilde{\Gamma}_{q_2} \
    \frac{\img}{k^z} (\img g {\hat z}^\mu) \frac{-\img  g_{\mu\nu}}{k^2} 
    (\img g {\hat z}^\nu) \frac{\img}{k^z}\,
    \delta\bigl[x_1 \tilde p^z - \tilde p_1^z\bigr]
    \nonumber\\
    &\quad\times
    \bigl(1 - e^{-\img k^z \tilde \eta}\bigr)
    \Bigl\{
    \delta\bigl[x_2 \tilde p^z - (\tilde p_2^z - k^z)\bigr]
    -e^{\img k^z \tilde \eta}\delta\bigl[x_2 \tilde p^z - \tilde p_2^z\bigr]
    \Bigr\}\,
    \delta\bigl[x_1 \tilde p^z - \tilde p_3^z\bigr]
    \nonumber\\[1.5ex]
    &=
    \alpha_s
    \Psi(\omega_1)\,
    \delta\bigl(1-\tfrac{x_1}{\omega_1}\bigr) \,
    \frac{\tilde{\Gamma}_{q_1} \otimes \tilde{\Gamma}_{q_2}}{\tilde p_1^z \ \tilde p_2^z} \, 
    \frac{1}{2}\Biggl[\bigg|\frac{1}{1-\frac{x_2}{\omega_2}}\bigg|\Biggr]_+^{(-\infty,\infty)}
    \nonumber\\
    &\quad
    +\alpha_s
    \Psi(\omega_1)\,
    \delta\bigl(1-\tfrac{x_1}{\omega_1}\bigr)\,
    \delta\bigl(1-\tfrac{x_2}{\omega_2}\bigr) \, 
    \frac{\tilde{\Gamma}_{q_1} \otimes \tilde{\Gamma}_{q_2}}{\tilde p_1^z \ \tilde p_2^z} \,
    \biggl[\frac{1}{\epsilon_\text{uv}} + 2
    +\ln\biggl(\frac{\mu^2 \tilde \eta^2}{b_0^2}\biggr)\biggr]
\,,\end{align}
%%%
where $b_0$ and the relation between $\mu_0$ and $\mu$ is given in \eq{mu0_b0}.
Dividing out $\Psi(\omega_1)$, the contribution of diagram $C$ to the matching kernel is
%%%
\begin{align}
    \bigl[\Delta^C_{q_1 q_2}\bigr]_{q_1' q_2'}
    &=
    -2
    \delta_{q_1 q_1'}
    \delta_{q_2 q_2'}\,
    \delta\bigl(1-\tfrac{x_2}{\omega_2}\bigr)
    \biggl\{
    \Biggl[\bigg|\frac{1}{1-\frac{x_2}{\omega_2}}\bigg|\Biggr]_+^{\mathrlap{(-\infty,\infty)}}
    +2
    \delta\bigl(1-\tfrac{x_1}{\omega_1}\bigr)\, 
    \biggl[\frac{1}{\epsilon_\text{uv}}+2
        +\ln\Bigl(\frac{\mu^2 \tilde \eta^2}{b_0^2}\Bigr)\biggr] 
    \biggr\}
    \nonumber
    \\
    &\quad
    +(1\leftrightarrow2)\,,
\end{align}
%%%
which also includes the related topology attaching to the other quark line.

\subsubsection*{Diagram D}
This is the quark self-energy diagram and is the same for the lightcone- and quasi-DPD, therefore
%%%
\begin{align}
    \bigl[\Delta^D_{q_1 q_2}\bigr]_{q_1' q_2'}
    &= 
    0\,.
\end{align}
%%%

\subsubsection*{Diagram E}
The lightcone diagram is given by
%%%
\begin{align}
    F^E
    &=
    -\pi p^+ \mu_0^{2\epsilon}
    \int \df\omega_3 \ \Psi(\omega_3)
    \int\frac{\df^d k}{(2\pi)^d} \ 
    (\img g \gamma^\mu) \frac{\img}{\fsl{p}_1 - \fsl{k}} \Gamma_{q_1}
    \frac{-\img}{k^+ - \img\delta^+} (\img g n_a^\nu)\otimes\Gamma_{q_2}
    \\
    &\quad\times
    \frac{-\img  g_{\mu\nu}}{k^2}\,
    \delta\bigl[x_1 p^+ - p_1^+\bigr]\,
    \delta\bigl[x_2 p^+ - p_2^+\bigr]\,
    \delta\bigl[x_1 p^+ - p_3^+\bigr]
    \nn \\[1.5ex]
    &=
    -\alpha_s
    \Psi(\omega_1)\,
    \delta\bigl(1-\tfrac{x_1}{\omega_1}\bigr)\,
    \delta\bigl(1-\tfrac{x_2}{\omega_2}\bigr)\, 
    \frac{\Gamma_{q_1}\otimes\Gamma_{q_2}}{p_1^+ p_2^+}\,
    \frac{1}{2}
    \biggl(\frac{1}{\epsilon_\text{uv}} - \frac{1}{\epsilon_\text{ir}}\biggr)
    \biggl[1 + \ln\biggl(\frac{\delta^+}{p_1^+}\biggr) - \frac{\img\pi}{2}\biggr] .
    \nonumber
\end{align}
%%%
The corresponding quasi-diagram yields,
%%%
\begin{align}
    \tilde{F}^E
    &=
    -\pi \tilde p^z \mu_0^{2\epsilon}
    \int \df\omega_3 \, \Psi(\omega_3)
    \int\frac{\df^d k}{(2\pi)^d} \
    (\img g \gamma^\mu) \frac{\img}{\fsl{\tilde p}_1 - \fsl{k}} \tilde{\Gamma}_{q_1}
    \frac{\img}{k^z} (\img g {\hat z}^\nu)\otimes\Gamma_{q_2}
    \\
    &\quad\times
    \frac{-\img  g_{\mu\nu}}{k^2}\,
    \Bigl\{\delta\bigl[x_1 \tilde p^z - \tilde p_1^z\bigr]
        - e^{-\img k^z \tilde \eta} \delta\bigl[x_1 \tilde p^z - (\tilde p_1^z - k^z)\bigr]\Bigr\}\,
    \delta\bigl[x_2 \tilde p^z - \tilde p_2^z\bigr] \,
    \delta\bigl[x_1 \tilde p^z - \tilde p_3^z\bigr]
    \nonumber\\[1.5ex]
    &=
    -\alpha_s
    \Psi(\omega_1)\,
    \delta\bigl(1-\tfrac{x_1}{\omega_1}\bigr) \,
    \delta\bigl(1-\tfrac{x_2}{\omega_2}\bigr) \, 
    \frac{\tilde{\Gamma}_{q_1}\otimes \tilde \Gamma_{q_2}}{\tilde p_1^z \ \tilde p_2^z}  
    \nn \\
    &\qquad\times
    4\img\pi^2\, \tilde p_1^z \mu_0^{2\epsilon}
    \int_{-\infty}^{\infty} \! \df y \, 
    \frac{1 - e^{-\img y \tilde p_1^z \tilde \eta}}{y}
    \int_0^1\! \df v\,
    \int\frac{\df^d \ell}{(2\pi)^d} \, 
    \frac{2\!-\!y\!-\!v}{\ell^4} \,
    \delta\bigl[\ell^z - (y\!-\!v)\tilde p_1^z\bigr] \,.
    \nonumber
\end{align}
%%%
The remaining integral can be carried out but again simplifies when first combined with other diagrams, as discussed at the end of this appendix.
The contribution of this diagram and its three sister topologies to the matching kernel is
%%%
\begin{align}
    \bigl[\Delta^E_{q_1 q_2}\bigr]_{q_1' q_2'}
    &=
    -4
    \delta_{q_1 q_1'}
    \delta_{q_2 q_2'}\,
    \delta\bigl(1-\tfrac{x_1}{\omega_1}\bigr)\,
    \delta\bigl(1-\tfrac{x_2}{\omega_2}\bigr)\,
    \biggl\{
    \biggl(\frac{1}{\epsilon_\text{uv}}-\frac{1}{\epsilon_\text{ir}}\biggr)
    \biggl[1 + \ln\biggl(\frac{\delta^+}{p_1^+}\biggr)\biggr]
    \\
    &\qquad
    -8\img\pi^2\, \tilde{p}_1^z \mu_0^{2\epsilon}
    \int_{-\infty}^{\infty}\! \df y \,
    \frac{1 \!-\! \cos(y \tilde{p}_1^z \tilde \eta)}{y}
    \int_0^1\! \df v
    \int\! \frac{\df^d \ell}{(2\pi)^d} \,
    \frac{2\!-\!y\!-\!v}{\ell^4} \,
    \delta\bigl[\ell^z - (y\!-\!v)\tilde{p}_1^z\bigr]
    \biggr\}
    \nonumber
    \\
    &\quad
    +(1\leftrightarrow2)\,.
\nn\end{align}
%%%

\subsubsection*{Diagram F}
Diagram $F$ vanishes for the lightcone-DPD. For the quasi-DPD it is given by
%%%
\begin{align}
    \tilde{F}^F
    &=
    -\pi \tilde p^z \mu_0^{2\epsilon}
    \int\! \df\omega_3 \, \Psi(\omega_3)
    \int\frac{\df^d k}{(2\pi)^d} \
    \tilde{\Gamma}_{q_1} \frac{\img}{k^z} (\img g {\hat z}^\mu) 
    \frac{\img}{k^z} (\img g {\hat z}^\nu)\otimes\tilde{\Gamma}_{q_2} \,
    \frac{-\img  g_{\mu\nu}}{k^2} 
    \\
    &\quad\times
    \delta\bigl[x_2 \tilde p^z - \tilde p_2^z\bigr] \,
    \delta\bigl[x_1 \tilde p^z - \tilde p_3^z\bigr] \,
    \Bigl\{e^{-\img k^z \tilde \eta}\delta\bigl[x_1 \tilde p^z - (\tilde p_1^z - k^z)\bigr]
    -\delta\bigl[x_1 \tilde p^z - \tilde p_1^z\bigr]\Bigr\}
    \nonumber\\[1.5ex]
    &=
    -\alpha_s
    \Psi(\omega_1)\,
    \delta\bigl(1-\tfrac{x_1}{\omega_1}\bigr)\,
    \delta\bigl(1-\tfrac{x_2}{\omega_2}\bigr)\, 
    \frac{\tilde{\Gamma}_{q_1}\otimes\tilde{\Gamma}_{q_2}}{\tilde p_1^z \tilde p_2^z} \, 
    \frac{1}{2}\biggl[\frac{1}{\epsilon_\text{uv}}
        + 2 + \ln\biggl(\frac{\mu^2 \tilde \eta^2}{b_0^2}\biggr)\biggr] \, .
\nn \end{align}
%%%
The $\Psi(\omega_1)$ factors out again, and the contribution of diagram $F$ and its three mirror topologies to the matching kernel is given by
%%%
\begin{align}
    \bigl[\Delta^F_{q_1 q_2}\bigr]_{q_1' q_2'}
    &=
    8
    \delta_{q_1 q_1'}
    \delta_{q_2 q_2'}\,
    \delta\bigl(1-\tfrac{x_1}{\omega_1}\bigr)\,
    \delta\bigl(1-\tfrac{x_2}{\omega_2}\bigr)\, 
    \biggl[\frac{1}{\epsilon_\text{uv}}+2
        +\ln\Bigl(\frac{\mu^2 \tilde{\eta}^2}{b_0^2}\Bigr)\biggr]\,.
\end{align}
%%%

\subsubsection*{Diagram G}

For diagram $G$ one does not need to use the modified partonic states of \sec{partonicstates}, so we can set $\Psi(\omega_3)=\delta(\omega_1-\omega_3)$ in the calculation of this diagram. The lightcone diagram results in
%%%
\begin{align}
    F^G
    &=
    -\pi p^+ \mu_0^{2\epsilon}
    \int\frac{\df^d k}{(2\pi)^d} \ 
    e^{\img \mathbf{k}_\perp \cdot \mathbf{b}_\perp} \ 
    (\img g \gamma^\mu) \frac{\img}{\fsl{p}_1 - \fsl{k}} \Gamma_{q_1}
    \otimes\Gamma_{q_2} \frac{\img}{\fsl{p}_4 - \fsl{k}} (\img g \gamma^\nu)
    \\
    &\quad\times
    (-g_{\mu\nu})\, 2\pi\delta_+(k^2)\,
    \delta\bigl[x_1 p^+ - (p_1^+ - k^+)\bigr]\,
    \delta\bigl[x_2 p^+ - p_2^+\bigr]\,
    \delta\bigl[x_1 p^+ - p_1^+\bigr]
    \nonumber\\[1.5ex]
    &=
    \alpha_s\,
    \delta(\omega_1-x_1)\,
    \delta(\omega_2-x_2)\, 
    \frac{1}{p_1^+ p_4^+}
    \biggl(g_\perp^{\mu\nu} 
        - 2\epsilon \frac{b^\mu b^\nu}{\mathbf{b}_\perp^2}\biggr)\, \gamma_\rho \gamma_\mu \Gamma_{q_1}
        \otimes\Gamma_{q_2} \gamma_\nu \gamma^\rho
    \nonumber\\
    &\quad\times
    \frac{1}{8}\biggl[\frac{1}{\epsilon_\text{ir}}
        +\ln\biggl(\frac{\mu^2 \mathbf{b}_\perp^2}{b_0^2}\biggr)\biggr]
    (\omega_1-x_1) \theta(x_1) \theta(\omega_1-x_1)
    \nonumber\\[1.5ex]
    &=0 \, .
\nn\end{align}
%%%
We start by noting that the corresponding quasi-diagram is UV and IR finite. We calculate it by first expanding in $1/(|\mathbf{b}_\perp| p^z)$ (usually we do it the other way around, but that is much more complicated here), leading to
%%%
\begin{align}
    \tilde{F}^G
    &=
    -\pi \tilde p^z \mu_0^{2\epsilon}
    \int \df\omega_3 \, \delta(\omega_3-\omega_1)
    \int\frac{\df^d k}{(2\pi)^d} \, 
    e^{\img \mathbf{k}_\perp \cdot \mathbf{b}_\perp} \, 
    (\img g\gamma^\mu)\frac{\img}{\fsl{\tilde p}_1-\fsl{k}}\tilde{\Gamma}_{q_1}
    \otimes\tilde{\Gamma}_{q_2}\frac{\img}{\fsl{\tilde p}_4-\fsl{k}}(\img g\gamma^\nu)
    \nonumber\\
    &\quad\times
    \frac{-\img  g_{\mu\nu}}{k^2}\,
    \delta\bigl[x_1 \tilde p^z - (\tilde p_1^z - k^z)\bigr]\,
    \delta\bigl[x_2 \tilde p^z - \tilde p_2^z\bigr]\,
    \delta\bigl[x_1 \tilde p^z - \tilde p_3^z\bigr]
    \nonumber\\[1.5ex]
    &=
    -\alpha_s\,
    \delta\bigl(1-\tfrac{x_1}{\omega_1}\bigr)\,
    \delta\bigl(1-\tfrac{x_2}{\omega_2}\bigr)\,
    \frac{1}{\tilde p_1^z \ \tilde p_2^z} \, 
    \gamma_{\perp\rho}\gamma_{\perp\mu}\tilde{\Gamma}_{q_1}
    \otimes\tilde{\Gamma}_{q_2}\gamma_{\perp\nu}\gamma_\perp^\rho
    \nonumber\\
    &\quad\times
    4\img\pi^2 \tilde p^z \mu_0^{2\epsilon}
    \frac{\partial}{\partial b_\mu}
    \frac{\partial}{\partial b_\nu}
    \int\!\frac{\df^d k}{(2\pi)^d}\,
    e^{\img \mathbf{k}_\perp \cdot \mathbf{b}_\perp} \, 
    \frac{1}{k^2 (k-\tilde p_1)^2 (k-\tilde p_4)^2}\,
    \delta(k^z)
    \,.
\end{align}
%%%
The precise expression for this diagram is ultimately not relevant for the calculation of the matching kernel, as the difference between the lightcone and quasi diagrams gets divided by the infinite normalization factor from the tree-level DPDs, so
%%%
\begin{align}
    \bigl[\Delta^G_{q_1 q_2}\bigr]_{q_1' q_2'}
    &=
    0\,.
\end{align}
%%%

\subsubsection*{Diagram H}
Since this diagram involves a gluon being connected to a Wilson line, its calculation requires the implementation of a rapidity regulator. The lightcone diagram yields
%%%
\begin{align}
    F^H
    &=
    -\pi p^+ \mu_0^{2\epsilon}
    \int \df\omega_3 \, \Psi(\omega_3)
    \int\frac{\df^d k}{(2\pi)^d} \, 
    e^{\img \mathbf{k}_\perp \cdot \mathbf{b}_\perp} \
    (\img g \gamma^\mu) \frac{\img}{\fsl{p}_1 - \fsl{k}} \Gamma_{q_1}
    \otimes
    \Gamma_{q_2} (\img g n_a^\nu) \frac{\img}{k^+ + \img\delta^+}
    \nonumber\\
    &\quad\times
    (-g_{\mu\nu}) \, 2\pi\delta_+(k^2)\,
    \delta\bigl[x_1 p^+ - (p_1^+ - k^+)\bigr]\,
    \delta\bigl[x_2 p^+ - p_2^+\bigr]\,
    \delta\bigl[x_1 p^+ - p_3^+\bigr]
    \nonumber\\[1.5ex]
    &=
    \alpha_s
    \Psi(x_1)\,
    \delta\bigl(1-\tfrac{x_2}{\omega_2}\bigr) \, 
    \frac{\Gamma_{q_1}\otimes\Gamma_{q_2}}{p_1^+ p_2^+}\,
    \frac{1}{2}\biggl[\frac{1}{\epsilon_\text{ir}} 
        +\ln\biggl(\frac{\mu^2 \mathbf{b}_\perp^2}{b_0^2}\biggr)\biggr]
    \biggl[\frac{\frac{x_1}{\omega_1}}{1-\frac{x_1}{\omega_1}}\,
    \theta\bigl(\tfrac{x_1}{\omega_1}\bigr)\,
    \theta\bigl(1 - \tfrac{x_1}{\omega_1}\bigr)
    \nonumber\\
    &\quad
    -\alpha_s
    \Psi(\omega_1) \,
    \delta\bigl(1-\tfrac{x_1}{\omega_1}\bigr) \,
    \delta\bigl(1-\tfrac{x_1}{\omega_1}\bigr) \, 
    \frac{\Gamma_{q_1}\otimes\Gamma_{q_2}}{p_1^+ p_2^+}\,
    \frac{1}{2}\biggl[\frac{1}{\epsilon_\text{ir}} 
        +\ln\biggl(\frac{\mu^2 \mathbf{b}_\perp^2}{b_0^2}\biggr)\biggr]
    \biggl[1 + \ln\biggl(\frac{\delta^+}{p_1^+}\biggr) + \frac{\img\pi}{2}\biggr] \, .
\end{align}
%%%
The calculation of the quasi diagram is more complicated because the divergences that arise when the gluon momentum goes to zero is regulated by both $\epsilon$ and $\tilde \eta$, 
%%%
\begin{align}
    \tilde{F}^H
    &=
    -\pi \tilde p^z \mu_0^{2\epsilon}
    \int \df\omega_3 \,\Psi(\omega_3)
    \int\frac{\df^d k}{(2\pi)^d} \, 
    e^{\img \mathbf{k}_\perp \cdot \mathbf{b}_\perp} \ 
    (\img g \gamma^\mu) \frac{\img}{\fsl{\tilde p}_1 - \fsl{k}} \tilde{\Gamma}_{q_1}
    \otimes\tilde{\Gamma}_{q_2} (\img g {\hat z}^\nu) \frac{\img}{k^z}
    \nonumber\\
    &\quad\times
    \frac{-\img  g_{\mu\nu}}{k^2} \,
    \delta\bigl[x_1 \tilde p^z - (\tilde p_1^z - k^z)\bigr]\,
    \delta\bigl[x_2 \tilde p^z - \tilde p_2^z\bigr]\,
    \delta\bigl[x_1 \tilde p^z - \tilde p_3^z\bigr]\,
    \bigl(e^{-\img k^z \tilde \eta} - 1\bigr)
    \nonumber\\[1.5ex]
    &=
    \alpha_s
    \Psi(x_1)\,
    \delta\bigl(1-\tfrac{x_2}{\omega_2}\bigr) \,
    \frac{\tilde{\Gamma}_{q_1}\otimes\tilde{\Gamma}_{q_2}}{\tilde p_1^z \, \tilde p_2^z} \,
    \frac{1}{2}
    \biggl[\frac{1}{\epsilon_\text{ir}} 
        +\ln\biggl(\frac{\mu^2 \mathbf{b}_\perp^2}{b_0^2}\biggr)\biggr]
    \biggl[
    \frac{\frac{x_1}{\omega_1}}{1-\frac{x_1}{\omega_1}}\,
    \theta\Bigl(\frac{x_1}{\omega_1}\Bigr) \theta\Bigl(1-\frac{x_1}{\omega_1}\Bigr)
    \biggr]_+
    \nonumber\\
    &\quad
    +\alpha_s
    \Psi(\omega_1)\,
    \delta\bigl(1-\tfrac{x_1}{\omega_1}\bigr)\,
    \delta\bigl(1-\tfrac{x_2}{\omega_2}\bigr)\, 
    \frac{\tilde{\Gamma}_{q_1}\otimes\tilde{\Gamma}_{q_2}}{\tilde p_1^z \, \tilde p_2^z}
    \\
    &\qquad\times
    4\img\pi^2\, \tilde p_1^z \mu_0^{2\epsilon}
    \int_{-\infty}^{\infty} \! \df y \,
    \frac{1-e^{-\img y \tilde p_1^z \tilde \eta}}{y}
    \int_0^1\! \df v
    \int\! \frac{\df^d \ell}{(2\pi)^d} \ 
    e^{\img \mathbf{\ell}_\perp \cdot \mathbf{b}_\perp} \, 
    \frac{2\!-\!y\!-\!v}{\ell^4}\, 
    \delta\bigl[\ell^z - (y\!-\!v)\tilde p_1^z\bigr] \, .
    \nonumber
\end{align}
%%%
The remaining integral can be carried out but again simplifies when first combined with other diagrams, as discussed at the end of this appendix.

A special comment concerning the wave function $\Psi$: In eq.~\eqref{eq:Delta_def} we must factor out $\Psi(\omega_1)$, and so far this has not been a problem: Either this arose naturally (e.g. in diagram A), or we encountered the combination $\Psi(x_1)\delta(1-\frac{x_1}{\omega_1})$, which is essentially the same. Here for the first time $\Psi(x_1)$ appears with a non-trivial function of $x_1$, naively precluding the extraction. However, the lightcone and quasi diagrams share the same non-trivial $x$-dependence. Therefore, after subtracting the two diagrams only a delta function term remains and we can factor out the normalization constant $\Psi(\omega_1)$ as before. The contribution of this diagram to the matching kernel is then given by
%%%
\begin{align}
    \bigl[\Delta^H_{q_1 q_2}\bigr]_{q_1' q_2'}
    &=
    -4
    \delta_{q_1 q_1'}
    \delta_{q_2 q_2'}\,
    \delta\bigl(1-\tfrac{x_1}{\omega_1}\bigr)\,
    \delta\bigr(1-\tfrac{x_2}{\omega_2}\bigr)\, 
    \biggr\{
    \biggl[\frac{1}{\epsilon_\text{ir}} 
        + \ln\biggl(\frac{\mu^2 \mathbf{b}_\perp^2}{b_0^2}\biggr)\biggr]
    \biggl[1 + \ln\biggl(\frac{\delta^+}{p_1^+}\biggr)\biggr]
    \nn \\
    & \qquad
    +
    8\img\pi^2\, \tilde{p}_1^z \mu_0^{2\epsilon}
     \int_{-\infty}^{\infty} \! \df y \,
    \frac{1-\cos(y \tilde{p}_1^z \tilde \eta)}{y}
    \int_0^1\! \df v
     \int  \frac{\df^d \ell}{(2\pi)^d} \,
    \nn \\ 
    & \qquad \quad \times
    e^{\img \mathbf{\ell}_\perp \cdot \mathbf{b}_\perp} \, 
    \frac{2-y-v}{\ell^4} \,
    \delta\bigl[\ell^z - (y-v)\tilde{p}_1^z\bigr] 
    \biggr\}
    +(1\leftrightarrow2)\,.
\end{align}
%%%

\subsubsection*{Diagram I}
This diagram vanishes for the lightcone-DPD as it involves a gluon line whose both ends are connected to Wilson lines in the $n_a$ direction. For the quasi-DPD this diagram gives
%%%
\begin{align}
    \tilde{F}^I
    &=
    -\pi \tilde p^z \mu_0^{2\epsilon}
    \int \df\omega_3\, \Psi(\omega_3)
    \int\frac{\df^d k}{(2\pi)^d} \, 
    e^{\img \mathbf{k}_\perp \cdot \mathbf{b}_\perp} \ 
    \frac{\img}{k^z} (\img g {\hat z}^\mu) \tilde{\Gamma}_{q_1}\otimes
    \tilde{\Gamma}_{q_2} (\img g {\hat z}^\nu) \frac{\img}{k^z} \
    \frac{-\img  g_{\mu\nu}}{k^2}
    \\
    &\quad\times
    \bigl(1 - e^{-\img k^z \tilde \eta}\bigr)
    \Bigl\{\delta\bigl[x_1 \tilde p^z - (\tilde p_1^z - k^z)\bigr] 
        - e^{\img k^z \tilde \eta}\, \delta\bigl[x_1 \tilde p^z - \tilde p_1^z\bigr]\Bigr\}\,
    \delta\bigl[x_2 \tilde p^z - \tilde p_2^z\bigr]\,
    \delta\bigl[x_1 \tilde p^z - \tilde p_3^z\bigr]
    \nonumber\\[1.5ex]
    &=
    \alpha_s\,
    \delta\bigl(1-\tfrac{x_1}{\omega_1})\,
    \delta\bigl(1-\tfrac{x_2}{\omega_2})\, 
    \frac{\tilde{\Gamma}_{q_1}\otimes\tilde{\Gamma}_{q_2}}{\tilde p_1^z \ \tilde p_2^z} \, 
    \biggl[\ln\biggl(\frac{\tilde \eta^2}{b_\perp^2}\biggr) + 2
        - \frac{\pi \tilde \eta}{b_\perp}\biggr]
    \nonumber
\,.\end{align}
%%%
The contribution to the matching kernel is then
%%%
\begin{align}
    \bigl[\Delta^I_{q_1 q_2}\bigr]_{q_1' q_2'}
    &=
    -4
    \delta_{q_1 q_1'}
    \delta_{q_2 q_2'}\,
    \delta\bigl(1-\tfrac{x_1}{\omega_1}\bigr)\,
    \delta\bigl(1-\tfrac{x_2}{\omega_2}\bigr)\,
    \biggl[\ln\Bigl(\frac{\tilde{\eta}^2}{\mathbf{b}_\perp^2}\Bigr) 
        +2 -\frac{\pi \tilde \eta}{|\mathbf{b}_\perp|}\biggr]
    +(1\leftrightarrow2)\,.
\end{align}
%%%

\subsubsection*{Diagram J}
For this diagram we can set $\Psi(\omega_3)=\delta(\omega_1-\omega_3)$ from the beginning, as no squared delta functions show up here. The lightcone diagram can be calculated directly and results in
%%%
\begin{align}
    F^J
    &=
    -\pi p^+ \mu_0^{2\epsilon}
    \int \df\omega_3 \, \Psi(\omega_3)
    \int\frac{\df^d k}{(2\pi)^d} \, 
    e^{\img \mathbf{k}_\perp \cdot \mathbf{b}_\perp} \ 
    (\img g \gamma^\mu) \frac{\img}{\fsl{p}_1 - \fsl{k}} \Gamma_{q_1}\otimes
    (\img g \gamma^\nu) \frac{\img}{\fsl{p}_2 + \fsl{k}} \Gamma_{q_2}
    \nonumber\\
    &\quad\times
    \frac{-\img  g_{\mu\nu}}{k^2} \,
    \delta\bigl[x_1 p^+ - (p_1^+ - k^+)\bigr]\,
    \delta\bigl[x_2 p^+ - (p_2^+ + k^+)\bigr]\,
    \delta\bigl[x_1 p^+ - p_3^+\bigr]
    \nonumber\\[1.5ex]
    &=
    -\alpha_s \,
    \delta\bigl(1-\tfrac{x_1}{\omega_1}\bigr) \,
    \delta\bigl(1-\tfrac{x_2}{\omega_2}\bigr) \,
    \frac{\gamma_{\perp\rho} \gamma_{\perp\mu} \Gamma_{q_1}\otimes
    \gamma_{\perp}^\rho \gamma_{\perp\nu} \Gamma_{q_2}}{p_1^+ p_2^+}
    \biggl(g_\perp^{\mu\nu}
    -2\epsilon \frac{b^\mu b^\nu}{\mathbf{b}_\perp^2} \biggr) \,
    \nonumber\\
    &\quad\times
    \frac{1}{8}\biggl[\frac{1}{\epsilon_\text{ir}}
        + \ln\biggl(\frac{\mu^2 \mathbf{b}_\perp^2}{b_0^2}\biggr)\biggr]
    \nonumber
\end{align}
%%%
As for diagram G, the calculation of the quasi diagram is more complicated and so we first expanding in $1/(|\mathbf{b}_\perp| p^z)$, leading to
%%%
\begin{align}
    \tilde{F}_c
    &=
    -\pi \tilde p^+ \mu_0^{2\epsilon}
    \int \df\omega_3 \, \Psi(\omega_3)
    \int\frac{\df^d k}{(2\pi)^d} \, 
    e^{\img \mathbf{k}_\perp \cdot \mathbf{b}_\perp} \ 
    (\img g \gamma^\mu)\frac{\img}{\fsl{\tilde p}_1-\fsl{k}}\tilde{\Gamma}_{q_1}\otimes
    (\img g \gamma^\nu) \frac{\img}{\fsl{\tilde p}_2 + \fsl{k}} \tilde{\Gamma}_{q_2}
    \nonumber\\
    &\quad\times
    \frac{-\img  g_{\mu\nu}}{k^2} \,
    \delta\bigl[x_1 \tilde p^z - (\tilde p_1^z - k^z)\bigr] \,
    \delta\bigl[x_2 \tilde p^z - (\tilde p_2^z + k^z)\bigr] \,
    \delta\bigl[x_1 \tilde p^z - \tilde p_3^z\bigr]
    \nonumber\\[1.5ex]
    &=
    -\alpha_s \,
    \delta\bigl(1-\tfrac{x_1}{\omega_1}\bigr) \,
    \delta\bigl(1-\tfrac{x_2}{\omega_2}\bigr) \, 
    \frac{ \gamma_\rho \gamma_\mu \tilde{\Gamma}_{q_1}\otimes
    \gamma^\rho \gamma_\nu \tilde{\Gamma}_{q_2}}{\tilde p_1^z \ \tilde p_2^z} \,
    \biggl(\frac{1}{8} \, 
    \frac{1}{\epsilon_\text{ir}}\, g_\perp^{\mu\nu} 
    +\mathcal{O}(\eps^0)\biggr) \, .
    \nonumber
\end{align}
%%%
Because the lightcone and quasi diagram share the same IR poles, their difference is finite. Dividing by the infinite normalization constant $\Psi(\omega_1)$ yields
%%%
\begin{align}
    \bigl[\Delta^J_{q_1 q_2}\bigr]_{q_1' q_2'}
    &=
    0\,.
\end{align}
%%%

\subsubsection*{Diagram K}
For lightcone diagram $K$ we have 
%%%
\begin{align}
    F^K
    &=
    -\pi p^+ \mu_0^{2\epsilon}
    \int \df\omega_3 \, \Psi(\omega_3)
    \int\frac{\df^d k}{(2\pi)^d} \,
    e^{\img \mathbf{k}_\perp \cdot \mathbf{b}_\perp} \ 
    (\img g \gamma^\mu) \frac{\img}{\fsl{p}_1 - \fsl{k}} \Gamma_{q_1}\otimes
    \Gamma_{q_2} \frac{-\img}{-k^+ - \img\delta^+} (-\img g n_a^\nu)
    \nonumber\\
    &\quad\times
    \frac{-\img  g_{\mu\nu}}{k^2} \,
    \delta\bigl[x_1 p^+ - (p_1^+ - k^+)\bigr] \,
    \delta\bigl[x_2 p^+ - (p_2^+ + k^+)\bigr] \,
    \delta\bigl[x_1 p^+ - p_3^+\bigr]
    \nonumber\\[1.5ex]
    &=
    -\alpha_s \Psi(x_1) \,
    \delta(1-x_1-x_2) \, 
    \frac{\Gamma_{q_1}\otimes\Gamma_{q_2}}{p_1^+ p_2^+} \, 
    \frac{1}{2}
    \biggl[\frac{1}{\epsilon_\text{ir}}
        +\ln\biggl(\frac{\mu^2 \mathbf{b}_\perp^2}{b_0^2}\biggr)\biggr]
    \biggl[\frac{\omega_2\, \frac{x_1}{\omega_1}}{1-\frac{x_1}{\omega_1}}\,
    \theta\Bigl(\frac{x_1}{\omega_1}\Bigr) \theta\Bigl(1-\frac{x_1}{\omega_1}\Bigr) \biggr]_+
    \nonumber\\
    &\quad
    +\alpha_s
    \Psi(\omega_1)\,
    \delta\bigl(1-\tfrac{x_1}{\omega_1}\bigr)\,
    \delta\bigl(1-\tfrac{x_2}{\omega_2}\bigr)\, 
    \frac{\Gamma_{q_1}\otimes\Gamma_{q_2}}{p_1^+ p_2^+} \, 
    \frac{1}{2}
    \biggl[\frac{1}{\epsilon_\text{ir}}
        +\ln\biggl(\frac{\mu^2 \mathbf{b}_\perp^2}{b_0^2}\biggr)\biggr]
    \biggl[1 + \ln\biggl(\frac{\delta^+}{p_1^+}\biggr) + \frac{\img\pi}{2}\biggr] \, .
\end{align}
%%%
For the quasi diagram we have, 
%%%
\begin{align}
    \tilde{F}^K
    &=
    -\pi \tilde p^z \mu_0^{2\epsilon}
    \int \df\omega_3\, \Psi(\omega_3)
    \int\frac{\df^d k}{(2\pi)^d} \, 
    e^{\img \mathbf{k}_\perp \cdot \mathbf{b}_\perp} \ 
    (\img g \gamma^\mu) \frac{\img}{\fsl{\tilde p}_1 - \fsl{k}} \tilde{\Gamma}_{q_1}
    \otimes\tilde{\Gamma}_{q_2} \frac{\img}{k^z} (\img g {\hat z}^\nu)
    \\
    &\quad\times
    \frac{-\img  g_{\mu\nu}}{k^2}\, 
    \delta\bigl[x_1 \tilde p^z - (\tilde p_1^z - k^z)\bigr]\,
    \Bigl\{\delta\bigl[x_2 \tilde p^z - (\tilde p_2^z + k^z)\bigr]
        - e^{-\img k^z \tilde \eta}\, \delta\bigl[x_2 \tilde p^z - \tilde p_2^z\bigr]\Bigr\}\,
    \delta\bigl[x_1 \tilde p^z - \tilde p_3^z\bigr]
    \nonumber\\[1.5ex]
    &=
    -\alpha_s \Psi(x_1)\,
    \delta(1-x_1-x_2) \, 
    \frac{\tilde{\Gamma}_{q_1}\otimes\tilde{\Gamma}_{q_2}}{\tilde p_1^z \, \tilde p_2^z} \, 
    \frac{1}{2}
    \biggl[\frac{1}{\epsilon_\text{ir}}
        +\ln\biggl(\frac{\mu^2 \mathbf{b}_\perp^2}{b_0^2}\biggr)\biggr]
    \biggl[\frac{\omega_2\, \frac{x_1}{\omega_1}}{1-\frac{x_1}{\omega_1}}\,
    \theta\Bigl(\frac{x_1}{\omega_1}\Bigr)\theta\Bigl(1-\frac{x_1}{\omega_1}\Bigr)\biggr]_+
    \nonumber\\
    &\quad
    -\alpha_s
    \Psi(\omega_1) \,
    \delta\bigl(1-\tfrac{x_1}{\omega_1}\bigr) \,
    \delta\bigl(1-\tfrac{x_2}{\omega_2}\bigr) \,
    \frac{\tilde{\Gamma}_{q_1}\otimes\tilde{\Gamma}_{q_2}}{\tilde p_1^z\, \tilde p_2^z} \,
    \nonumber\\
    &\qquad\times
    4\img\pi^2\, \tilde p_1^z \mu_0^{2\epsilon}
    \int_{-\infty}^{\infty} \! \df y \, 
    \frac{1-e^{-\img y \tilde p_1^z \tilde \eta}}{y}
    \int_0^1 \! \df v
    \int\frac{\df^d \ell}{(2\pi)^d} \, 
    e^{\img \mathbf{\ell}_\perp \cdot \mathbf{b}_\perp} \, 
    \frac{2\!-\!y\!-\!v}{\ell^4} \,
    \delta\bigl[\ell^z - (y\!-\!v)\tilde p_1^z\bigr] \, .
\nn\end{align}
%%%
Upon subtracting the lightcone and quasi diagrams, the non-trivial $x$-dependence cancels between the two and $\Psi(\omega_1)$ can be factored out, to yield
%%%
\begin{align}
    \bigl[\Delta^K_{q_1 q_2}\bigr]_{q_1' q_2'}
    &=
    4
    \delta_{q_1 q_1'}
    \delta_{q_2 q_2'} \,
    \delta\bigl(1-\tfrac{x_1}{\omega_1}\bigr)\,
    \delta\bigl(1-\tfrac{x_2}{\omega_2}\bigr)\,
    \biggl\{
    \biggl[\frac{1}{\epsilon_\text{ir}}
        +\ln\biggl(\frac{\mu^2 \mathbf{b}_\perp^2}{b_0^2}\biggr)\biggr]
    \biggl[1 + \ln\biggl(\frac{\delta^+}{p_1^+}\biggr)\biggr]
    \\
    &\quad
    +8\img\pi^2 \tilde{p}_1^z \mu_0^{2\epsilon}
    \int_{-\infty}^{\infty} \! \df y \, 
    \frac{1-\cos(y \tilde{p}_1^z \tilde \eta)}{y}
    \int_0^1 \! \df v \,
    \int\frac{\df^d \ell}{(2\pi)^d} 
    \nn \\ & \qquad \times
    e^{\img \mathbf{\ell}_\perp \cdot \mathbf{b}_\perp} \, 
    \frac{2 - y - v}{\ell^4} \,
    \delta\bigl[\ell^z - (y - v)\tilde{p}_1^z\bigr]
    \biggr\}
    +(1\leftrightarrow2)\,.
\nn\end{align}
%%%

\subsubsection*{Diagram L}
For the lightcone-DPD, diagram $L$ vanishes due to the gluon line being connected to two Wilson lines in the $n_a$ direction. For the quasi diagram we find
%%%
\begin{align}
    \tilde{F}^L
    &=
    -\pi \tilde p^z \mu_0^{2\epsilon}
    \int \df\omega_3 \, \Psi(\omega_3)
    \int\frac{\df^d k}{(2\pi)^d} \, 
    e^{\img \mathbf{k}_\perp \cdot \mathbf{b}_\perp} \ 
    \tilde{\Gamma}_{q_1} \frac{\img}{k^z} (\img g {\hat z}^\mu) 
    \otimes\tilde{\Gamma}_{q_2} \frac{-\img}{k^z} (\img g {\hat z}^\nu)
    \frac{-\img  g_{\mu\nu}}{k^2} 
    \\
    &\qquad\times
    \delta\bigl[x_1 \tilde p^z - \tilde p_3^z\bigr]
    \Bigl\{\delta\bigl[x_1 \tilde p^z - (\tilde p_1^z - k^z)\bigr] 
        - e^{-\img k^z \tilde \eta} \,\delta\bigl[x_1 \tilde p^z - \tilde p_1^z\bigr]\Bigr\}
    \nonumber\\
    &\qquad\times        
    \Bigl\{\delta\bigl[x_2 \tilde p^z - (\tilde p_2^z + k^z)\bigr] 
        - e^{+\img k^z \tilde \eta} \, \delta\bigl[x_2 \tilde p^z - \tilde p_2^z\bigr]\Bigr\}
    \nonumber\\[1.5ex]
    &=
    -\alpha_s
    \Psi(\omega_1) \,
    \delta\bigl(1-\tfrac{x_1}{\omega_1}\bigr) \,
    \delta\bigl(1-\tfrac{x_2}{\omega_2}\bigr) \, 
    \frac{\tilde{\Gamma}_{q_1}\otimes\tilde{\Gamma}_{q_2}}{\tilde p_1^z \ \tilde p_2^z} \,
    \biggl[\ln\biggl(\frac{\tilde \eta^2}{b_\perp^2}\biggr)+2-\frac{\pi \tilde \eta}{b_\perp}\biggr]
    \, .
\end{align}
%%%
Dividing out $\Psi(\omega_1)$, we find that the contribution of diagram $L$ to the matching kernel is given by
%%%
\begin{align}
    \bigl[\Delta^L_{q_1 q_2}\bigr]_{q_1' q_2'}
    &=
    4
    \delta_{q_1 q_1'}
    \delta_{q_2 q_2'} \,
    \delta\bigl(1-\tfrac{x_1}{\omega_1}) \,
    \delta\bigl(1-\tfrac{x_2}{\omega_2}) \,
    \biggl[\ln\Bigl(\frac{\tilde{\eta}^2}{\mathbf{b}_\perp^2}\Bigr) 
        + 2 - \frac{\pi \tilde \eta}{b_\perp}\biggr]
    +(1\leftrightarrow2)\,.
\end{align}
%%%

\subsubsection*{Remaining integrals in $\Delta^B, \Delta^H, \Delta^E, \Delta^K$}

For the color-summed DPD, the remaining integrals cancel between $\Delta^B$ and $\Delta^E$. For the color-correlated DPDs, this cancellation no longer holds, due to the different color factors in table~\ref{tab:colorfactors}. In that case, the combination of all four diagrams leads to the following integral,
%%%
\begin{align}
    \mathcal{I}(\tilde p^z)
    &=
    \tilde p^z \mu_0^{2\epsilon}\int_{-\infty}^{\infty} \! \df y \, 
    \frac{1-\cos(y \tilde p^z \tilde \eta)}{y}
    \int_0^1 \! \df v\,
    \int\frac{\df^d \ell}{(2\pi)^d} \, 
    (1-e^{\img \mathbf{\ell}_\perp \cdot \mathbf{b}_\perp}) \ 
    \frac{2\!-\!y\!-\!v}{\ell^4} \,
    \delta\bigl[\ell^z - (y\!-\!v)\tilde p^z\bigr] .
\end{align}
%%%
Note that this integral does not contain any IR divergences due to factor of $1-e^{\img \mathbf{b}_\perp\cdot\mathbf{l}_\perp}$. However, we cannot set $d=4$ because the term proportional to $y$ in the numerator of the momentum integral results in a UV divergence for $y\rightarrow\pm\infty$. However, this term is finite as $y\rightarrow0$ and so we can take $\tilde \eta\rightarrow\infty$ in this term, leading to trivial integrals over $y$ and $v$. Splitting off this term, we can rewrite the above integral as
%%%
\begin{align}
    \mathcal{I}(\tilde p^z)
    &=
    \tilde p^z \int_{-\infty}^{\infty} \! \df y \, 
    \frac{1-\cos(y \tilde p^z \tilde \eta)}{y}
    \int_0^1 \! \df v \, (2-v)
    \int\frac{d^4 \ell}{(2\pi)^4} \, 
    \frac{1-e^{\img \mathbf{\ell}_\perp \cdot \mathbf{b}_\perp}}{\ell^4} \,
    \delta\bigl[\ell^z - (y\!-\!v)\tilde p^z\bigr]
    \nonumber\\
    &\quad
    -\mu_0^{2\epsilon}
    \int\frac{\df^d \ell}{(2\pi)^d} \,
    \frac{1-e^{\img \mathbf{\ell}_\perp \cdot \mathbf{b}_\perp}}{\ell^4} \, .
\end{align}
%%%

The integral in the second line is straightforward to compute. To calculate the momentum integral in the first line we first perform a Wick rotation, integrate over all components of the momentum perpendicular to the plane spanned by $\mathbf{b}_\perp$ and $\hat{z}$, and lastly integrate over the component of $\ell$ that is parallel to $\mathbf{b}_\perp$. The result is 
%%%
\begin{align}
    \mathcal{I}(\tilde p^z)
    &=
    \frac{\img}{16\pi^2}
    \int_{-\infty}^{\infty} \! \df y \,
    \frac{1-\cos(y \tilde p^z \tilde \eta)}{y}
    \int_0^1 \! \df v \, \frac{2-v}{|y-v|} 
    \bigl(1-e^{-|y-v|b_\perp \tilde p^z}\bigr)
    \\
    &\quad
    -\frac{\img}{16\pi^2}\biggl(\frac{1}{\epsilon_\text{uv}} + L_b\biggr) \,,
\nn \end{align}
%%%
with $L_b$ defined in \eq{LbLp}.
To perform the remaining integral over $y$ and $v$, we first change the integration bound of the $y$ integral to $[0,\infty)$ by symmetrizing the integrand. The resulting integrand is finite as $y\rightarrow0$ and so we can drop the regulator $\tilde \eta$, leading to
%%%
\begin{align}
    \mathcal{I}(\tilde p^z)
    &=
    \frac{\img}{16\pi^2}
    \int_0^{\infty} \! \df y \, 
    \frac{1}{y} 
    \int_0^1 \! \df v \, (2-v)
    \biggl(\frac{1-e^{-|y-v|b_\perp \tilde p^z}}{|y-v|}
    -\frac{1-e^{-|y+v|b_\perp \tilde p^z}}{|y+v|}\biggr)
    \\
    &\quad
    -\frac{\img}{16\pi^2}\biggl(\frac{1}{\epsilon_\text{uv}} + L_b\biggr) \, .
\nn\end{align}
%%%
The remaining integrals over $y$ and $v$ can then be performed. For large $|\mathbf{b}_\perp| \tilde p^z$ the result can be written as
%%%
\begin{align}
    \mathcal{I}(\tilde p^z)
    &=
    -\frac{\img}{16\pi^2}
    \biggl(
    \frac{1}{\epsilon_\text{uv}}
    +2L_b
    -L_b L_p
    -2
    +L_p
    -\frac{1}{2}L_b^2
    -\frac{1}{2}L_p^2\biggr) \,,
\end{align}
%%%
where
%%%
\begin{align} \label{eq:LbLp}
    L_b
    &= \ln \biggl(\frac{\mu^2 \mathbf{b}_\perp^2}{b_0^2}\biggr)
    \,, \qquad
    L_p
    =
    \ln\biggl(\frac{4(\tilde p^z)^2}{\mu^2}\biggr)
\,.\end{align}
%%%

\bibliographystyle{JHEP}
\bibliography{quasi_dpds.bib}

\end{document}